%% file: DNDK_to_pions_second_draft_to_PRD.tex
\documentclass[aps,prd,twocolumn,showpacs,superscriptaddress,floatfix]{revtex4-1}  

\usepackage{graphicx}  
\usepackage{dcolumn}   
\usepackage{bm}        
\usepackage{amssymb}   
\usepackage{multirow}

\usepackage{amsmath}
\usepackage{lineno}
\usepackage{mathtools}
\usepackage{subfigure}
\usepackage{mdwlist}
\usepackage{pdfpages}
\usepackage{makecell}
\usepackage{feynmp}
\usepackage{enumerate}
\usepackage{pst-pdf}
\usepackage{diagbox}

\usepackage{graphicx}
\usepackage{epstopdf}

\begin{document}


\title{Search for dinucleon decay into pions at Super-Kamiokande}
\input{authors-20140520}

\date{\today}

\begin{abstract}
A search for dinucleon decay into pions with the Super-Kamiokande detector has been performed with an exposure of 282.1 kiloton-years. Dinucleon decay is a process that violates baryon number by two units. We present the first search for dinucleon decay to pions in a large water Cherenkov detector. The modes $^{16}$O$(pp) \rightarrow$ $^{14}$C$\pi^{+}\pi^{+}$, $^{16}$O$(pn) \rightarrow$ $^{14}$N$\pi^{+}\pi^{0}$, and $^{16}$O$(nn) \rightarrow$ $^{14}$O$\pi^{0}\pi^{0}$ are investigated. No significant excess in the Super-Kamiokande data has been found, so a lower limit on the lifetime of the process per oxygen nucleus is determined. These limits are: $\tau_{pp\rightarrow\pi^{+}\pi^{+}} > 7.22 \times 10^{31}$ years, $\tau_{pn\rightarrow\pi^{+}\pi^{0}} > 1.70 \times 10^{32}$ years, and $\tau_{nn\rightarrow\pi^{0}\pi^{0}} > 4.04 \times 10^{32}$ years. The lower limits on each mode are about two orders of magnitude better than previous limits from searches for dinucleon decay in iron.
\end{abstract}

\pacs{13.30.Eg, 14.20.Dh, 29.40.Ka}


\maketitle


\section{Introduction}

Baryon number seems to be a conserved quantity in order to explain the stability of matter, yet baryon number must be violated to explain the predominance of matter over antimatter in the universe~\cite{Sakharov:1967dj}. Single nucleon decay processes that violate baryon number by one unit but conserve $\Delta|B-L|$ (where B is baryon number and L is lepton number) have been studied extensively at Super-Kamiokande~\cite{Abe:2014mwa, Abe:2013lua, Nishino:2012ipa, Regis:2012sn, Kobayashi:2005pe} in the context of Grand Unified Theories (GUTs) that predict an unstable proton. Lower limits on proton decay modes such as $p\rightarrow e^{+}\pi^{0}$ of $ 1.2 \times 10^{34}$ years~\cite{Raaf:2012pva} have ruled out the simplest GUTs, such as minimal $SU(5)$~\cite{Georgi:1974sy}. Other processes, such as those that violate B by two units, are of great theoretical interest although not necessarily predicted by GUT-inspired models. Searches for such processes constitute an important subset of searches for baryon number violation, both in their own right and given the continued non-observation of single nucleon decay. Discovery of such a process would be a spectacular revelation of new physics, while non-observation provides important constraints on the theories which predict them. Dinucleon decay -- the simultaneous decay of two bound nucleons into leptons or mesons -- constitutes such a process. In this paper we search for the processes $^{16}$O$(pp) \rightarrow$ $^{14}$C$\pi^{+}\pi^{+}$, $^{16}$O$(pn) \rightarrow$ $^{14}$N$\pi^{+}\pi^{0}$, and $^{16}$O$(nn) \rightarrow$ $^{14}$O$\pi^{0}\pi^{0}$ (abbreviated throughout to $pp\rightarrow\pi^{+}\pi^{+}$, $pn\rightarrow\pi^{+}\pi^{0}$, and $nn\rightarrow\pi^{0}\pi^{0}$).

\par Processes violating baryon number by two units can occur in a broad class of models~\cite{Arnold:2012sd}. The experimentally accessible processes of this kind are dinucleon decay and neutron-antineutron oscillation (n-nbar). The latter process may also be pursued in experiments with free neutrons~\cite{Babu:2013yww}, whereas dinucleon decay requires the close overlap of bound nucleons. The tree-level Feynman diagram for n-nbar and dinucleon decay to pions is the same, with two spectator quarks in the latter case. An example of a dinucleon decay diagram from one of the models discussed in \cite{Arnold:2012sd} is shown in Fig. 1. The expected time scale for dinucleon decay to pions can be related to the time scale for n-nbar, if the same operator is dominant for each process. This is discussed further in Appendix A.

\begin{figure}[h]
\includegraphics[width=80mm]{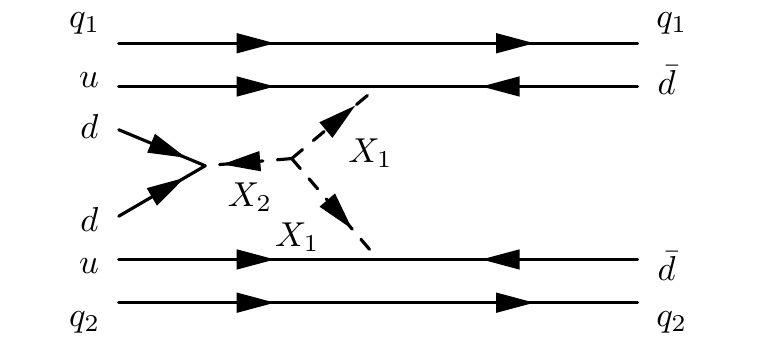}        

\caption{Feynman diagram for dinucleon decay to pions from one of the models discussed in~\cite{Arnold:2012sd}, with spectator quarks $q_{1}, q_{2} \in (u,d)$. The $X_1$ and $X_2$ scalar particles belong to $(\bar{6},1,-\frac{1}{3})$ and $(\bar{6},1,\frac{2}{3})$ representations of the Standard Model $SU(3) \otimes SU(2) \otimes U(1)$ gauge group.}
\end{figure}

\par One dinucleon decay search has been performed before at Super-Kamiokande, namely a search for decay to charged kaons~\cite{Litos:2014}. The Frejus experiment, which used iron nuclei, performed an extensive search covering dinucleon decay final states~\cite{Berger:1991fa}. The search in \cite{Berger:1991fa} includes the modes analyzed in this paper. The more exclusive limits in this paper relative to those in \cite{Berger:1991fa} reflect the large exposure of Super-Kamiokande. Our limit calculation accounts for systematic uncertainties, particularly those due to intranuclear pion interactions. This is the first search for dinucleon decay into pions in a water Cherenkov detector.

\par This paper proceeds as follows. Section II describes some basic features of the Super-Kamiokande experiment. Section III describes relevant aspects of our Monte-Carlo simulation, for both dinucleon decay signal and atmospheric neutrino background. Section IV describes standard data selection and reconstruction algorithms used in Super-Kamiokande. Section V describes the analysis of all three modes, first discussing some features of boosted decision trees that are relevant for the $pp\rightarrow\pi^{+}\pi^{+}$ and $pn\rightarrow\pi^{+}\pi^{0}$ searches. Sections VI and VII describe our results and conclusion.


\section{The Super-Kamiokande Detector}

Super-Kamiokande (hereafter ``SK'') is a large, cylindrical water Cherenkov detector. Full details of the detector can be found in \cite{Fukuda:2002uc}; here, we note only the most basic features.

\par The detector contains two optically-separated regions: the inner detector (ID) and outer detector (OD). Inward-facing, 20-inch photomultiplier tubes (PMTs) collect Cherenkov light in the ID, while light from penetrating particles (typically cosmic-ray muons) is detected by 8-inch, outward-facing PMTs in the OD. The fiducial volume of the ID is defined as the cylindrical volume with surfaces two meters inward from the ID wall. This corresponds to 22.5 ktons, and $7.5 \times 10^{33}$ oxygen nuclei. 

\newcommand{\specialcell}[2][c]{%
  \begin{tabular}[#1]{@{}c@{}}#2\end{tabular}}


\par SK has had four data-taking periods. It began taking data in April 1996, with 11,146 PMTs covering 40\% of the ID surface. The first data-taking period, called SK-I, continued until July 2001, totaling 1489.2 livetime days, or 91.7 kton-years. An accident occurred in November 2001, which caused the implosion of about half of the ID PMTs. The remaining 5,182 PMTs were re-distributed uniformly on the surface of the ID, providing 19\% photo-coverage. To prevent such accidents in the future, each PMT has been enclosed in acrylic and fiber-reinforced plastic cases. The second data-taking period, which used this decreased photo-coverage, lasted from December 2002 until October 2005, and is called SK-II. This lasted 798.6 livetime days, or 49.2 kton-years. Replacement PMTs were produced and the photo-coverage was returned to 40\% in 2006. The third data-taking period, from July 2006 until September 2008, is called SK-III, comprising 518.1 livetime days (31.9 kton-years). The detector was upgraded in the summer of 2008, with improved electronics that record all PMT hit information without dead time~\cite{Nishino:upgrade}~\cite{Abe:2013gga}. This is the current configuration of the detector, called SK-IV. In this paper, SK-IV data are used up to March 2014, which comprises 1775.6 livetime days (109.3 kton-years). The total SKI-IV dataset is 282.1 kton-years. 


\section{Simulation}

Dinucleon decay signal and atmospheric neutrino background Monte Carlo simulations (MC) are generated to estimate signal efficiency, expected background rates, and systematic uncertainties. Since each of the four SK data-taking periods have different detector configurations, separate sets of MC for both the dinucleon decay signal and the atmospheric neutrino background are used for each period. The dinucleon decay MC is based on single nucleon decay MC used for previous SK analyses~\cite{Abe:2014mwa, Abe:2013lua, Nishino:2012ipa, Regis:2012sn, Kobayashi:2005pe}. The atmospheric neutrino MC is the same MC sample used in other SK analyses, as described in detail in \cite{Ashie:2005ik}.

\subsection*{Dinucleon decay}

As dinucleon decay requires the close overlap of two nucleons, we restrict attention to the oxygen nucleus for the decaying nucleons. Fermi momentum, correlation between nucleons, nuclear binding energy, and pion-nucleon interactions are taken into account. This last effect is especially prominent, given the dense nuclear material the pions traverse.
\par
\begin{figure}[h!]
\includegraphics[width=80mm]{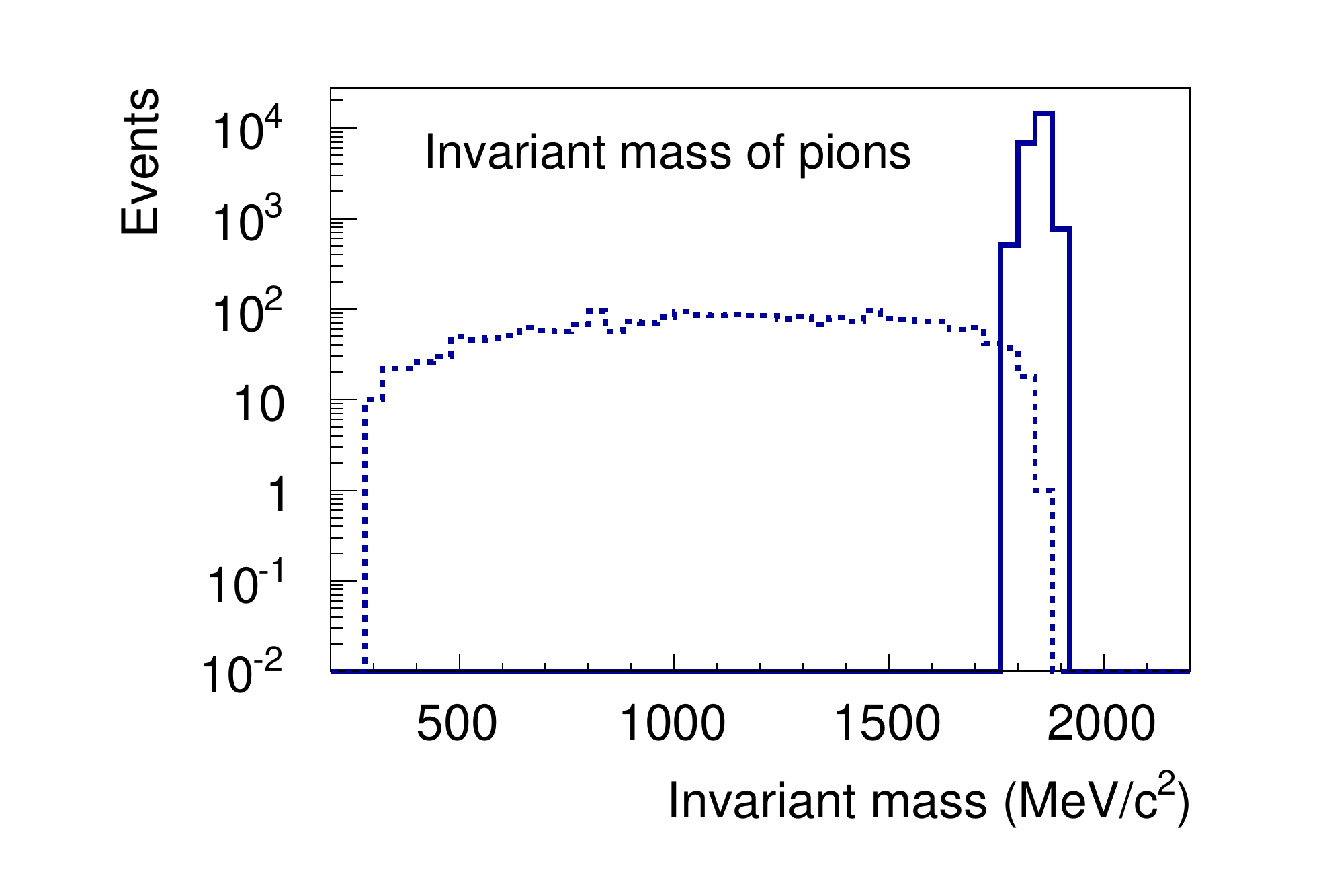}
\caption{Invariant mass of the pions from $pp\rightarrow\pi^{+}\pi^{+}$, before nuclear effects. Plots for $pn\rightarrow\pi^{+}\pi^{0}$ and $nn\rightarrow\pi^{0}\pi^{0}$ look similar. The dashed histogram shows the spectrum for events with a correlated nucleon.}
\label{figure:minv_nonuc}
\end{figure}

 Fermi momentum and nuclear binding energy are taken from measured electron-$^{12}$C scattering~\cite{Nakamura:1976}. Nuclear binding energy is taken into account by modifying the nucleon masses. Reference~\cite{Yamazaki:1999} estimates that 10\% of single nucleons that undergo decay have wavefunctions that are correlated with other nucleons in the nucleus; we assume the same level of correlation for dinucleon decay pairs. Such correlated decay events typically have a lower total invariant mass, due to the momentum of the correlated nucleon, which is below Cherenkov threshold (Fig.~\ref{figure:minv_nonuc}). The positions of the decaying nucleons in $^{16}$O are approximated using the Woods-Saxon nuclear density model~\cite{Woods:1954zz}. Dinucleon decay leaves the remaining nucleus in an excited state from which it de-excites by the emission of gamma rays. However, the de-excitation gamma rays are on the order of a few MeV, while the pions are on the order of a GeV, so the extra light coming from the gamma rays is negligible relative to the pions. The main difference in dinucleon decay simulation from single-nucleon decay is the available phase space. For dinucleon decay, the initial total momentum and energy are summed over both nucleons (or three nucleons, if there is a correlated decay). 

\par Pion-nucleon interactions in the nucleus are simulated using NEUT~\cite{Hayato:2002sd}. The interactions modeled include quasielastic scattering, in which the incoming and outgoing pion have the same charge, but there is some energy loss; charge exchange (e.g. $\pi^{+} + n \rightarrow \pi^{0} + p$); absorption (e.g. $\pi^{+} + n \rightarrow p$); and hadron production (e.g. $\pi^{+} + p \rightarrow \pi^{+} + \pi^{0} + p$). Quasielastic scattering is relevant at all pion momenta, as is charge exchange, although it has a lower overall cross-section. Absorption has a sizable cross-section in the pion momentum range 100-300 MeV/$c$ range, but is negligle elsewhere. Hadron production is relevant for pion momentum greater than 500 MeV/$c$, and is therefore relevant for most pions immediately after the decay. With these effects taken into account, less than half of the generated events in each mode escape the nucleus in the original pion configuration (Table~\ref{table:trupi}). 

\par After the pions leave the nucleus, their propagation and Cherenkov light emission is passed through a custom detector simulation based on GEANT3~\cite{CERN:1994}. Charged pions are susceptible to further secondary interactions in the water, while neutral pions decay before such interactions. Propagation of charged pions through the water below 500 MeV/$c$ is simulated by custom code based on \cite{Nakahata:1986}, while charged pions above this momentum are handled by GCALOR~\cite{Albanese:1980np}.



\begin{table}[h]
\setlength{\tabcolsep}{3pt}
\begin{center}
$\mathbf{pp}\boldsymbol{\rightarrow}\boldsymbol{\pi^{+}}\boldsymbol{\pi^{+}}$ \\
    \begin{tabular}{r | r r r r}

\hline \hline
\backslashbox{{\bf{$\pi^{0}$}}}{{\bf{$\pi^{+}$}}} & 0 & 1 & 2 & 3\\ \hline
0 & --- & 11.8\% & \bf{45.1\%} &  2.4\% \\ 
1 &  2.7\% & 21.8\% &  5.2\% & --- \\ 
2 &  2.8\% &  3.0\% & --- & --- \\ 

\hline \hline
\end{tabular}

\vspace{5mm}

$\mathbf{pn}\boldsymbol{\rightarrow}\boldsymbol{\pi^{+}}\boldsymbol{\pi^{0}}$ \\

\begin{tabular}{r | r r r r}

\hline \hline
\backslashbox{{\bf{$\pi^{0}$}}}{{\bf{$\pi^{+}$}}} & 0 & 1 & 2 & 3\\ \hline
0 & --- & 6.8\% & 12.3\% & 1.4\% \\ 
1 & 7.1\% & \bf{46.8}\% & 3.7\% & --- \\ 
2 & 10.5\% & 4.1\% & --- & --- \\ 

\hline \hline
\end{tabular}

\vspace{5mm}

$\mathbf{nn}\boldsymbol{\rightarrow}\boldsymbol{\pi^{0}}\boldsymbol{\pi^{0}}$ \\

\begin{tabular}{r | r r r} 

\hline \hline
\backslashbox{{\bf{$\pi^{0}$}}}{{\bf{$\pi^{+}$}}} & 0 & 1 & 2\\ \hline
0 & --- & 2.6\% & 3.5\% \\ 
1 & 11.5\% & 25.1\% & 2.8\% \\ 
2 & \bf{43.0\%} & 4.2\% & --- \\ 

\hline \hline
\end{tabular}

\caption{\protect \small Number of pions for $pp\rightarrow\pi^{+}\pi^{+}$ (top), $pn\rightarrow\pi^{+}\pi^{0}$ (middle), and $nn\rightarrow\pi^{0}\pi^{0}$ (bottom) after nuclear effects. Percentages indicate the percentage of events with the specified number of $\pi^{+}$'s (columns) and number of $\pi^{0}$'s (rows). Entries with a ``---'' indicate a small contribution ($<$0.1\%). The number of pions prior to nuclear effects is shown in bold typeface.}
\label{table:trupi}
\end{center}
\end{table}

\par Monte carlo events are generated in a broadened fiducial volume (FV), greater than one meter from the ID wall as opposed to two meters which defines the FV. This allows us to account for events that migrate across the FV boundary when reconstructed. The final detection efficiency is defined as the ratio of events surviving all selection criteria to the total number of events generated in the fiducial volume. For each SK period, we generate a total of 25,000 events. Three times this number of events is generated in modes for which we use a multivariate search method. This is because the multivariate method requires three separate samples, as described later. 

\subsection*{Atmospheric neutrinos}
The atmospheric neutrino MC sample is based on the Honda atmospheric neutrino flux~\cite{Honda:2006qj}, and the NEUT neutrino interaction and nuclear effects simulator~\cite{Hayato:2002sd}. The dominant neutrino interaction background for the dinucleon decay searches comes from charged-current single pion production (CC$1\pi$, $\nu + N \rightarrow \ell + N' + \pi$ for nucleons $N,N'$ and charged lepton $\ell$), as well as both charged and neutral-current deep inelastic scattering (CCDIS and NCDIS, $\nu + N \rightarrow \ell + N' +$ hadrons, where $\ell$ can be charged or neutral). Pions in $CC1\pi$ interactions are produced following the model of Rein and Sehgal \cite{Rein:1980wg}. This is the dominant hadron production interaction for hadronic invariant mass $W$ below about $2$ $GeV/c^{2}$. Deep inelastic scattering is treated differently depending on $W$. For $1.3$ $GeV/c^{2}$ $\leq$ $W$ $\leq$ $2.0$ $GeV/c^{2}$, only pions are considered as outgoing hadrons, and the number of pions is two or greater to differentiate it from $CC1\pi$. For $W$ $>$ $2.0$ $GeV/c^{2}$, other mesons are also considered, such as $\eta$ mesons and kaons. After leaving the nucleus, particles are propagated through the water using the same tools described above for dinucleon decay.
\par The atmospheric neutrino samples correspond to an exposure of 500 years for each SK period. Event rates are weighted to include the effect of neutrino oscillations. Final event rates after all selection criteria have been applied are normalized to the corresponding SKI-IV detector livetime.


\section{\label{sec:redrec}Reduction and Reconstruction}

This search uses the fully-contained (FC) dataset. Fully-contained events are defined as events in which all charged particles start and stop in the ID. Exiting and entering particles, particularly cosmic ray muons, are excluded by requiring minimal activity in the OD. The threshold for a triggered event in the FC dataset is 5.7 MeV of visible energy in SK-I, III, and IV, and 8 MeV of visible energy in SK-II. Visible energy is defined as the energy to produce the observed light in the event if it were produced by a single electron. The majority of triggered events are cosmic ray muons and low energy radioactivity from material surrounding the detector. A dedicated set of event selection algorithms is applied to eliminate the majority of these backgrounds, such that all but a fraction of remaining events come from atmospheric neutrinos~\cite{Ashie:2005ik}. 
\par Events that remain are passed through a reconstruction program. The reconstruction for each event determines an overall event vertex and the number of Cherenkov rings, as well as the direction, momentum, and particle classification of each ring. Events are assumed to originate from a single vertex, and the distribution of observed charge is used to find the first Cherenkov ring. Additional rings are found using a Hough transform method~\cite{Davies:1997}. Full details on our event reconstruction can be found in \cite{Ashie:2005ik} and \cite{Shiozawa:1999sd}.
\par Each Cherenkov ring is classified as ``$e$-like'' or ``$\mu$-like'', referring to electrons and muons. This is a conventional way of denoting whether a ring displays the characteristics of electromagnetic showers or not. Showering rings are produced by electrons and photons, but not protons, muons, or charged mesons. The classification of each ring is determined by its light pattern. Electromagnetic showers produce fuzzy rings due to scattering and pair production, while heavier particles produce rings with sharper edges. 
\par The momentum of each ring is determined based on the particle type, and the number of photo-electrons, referred to as charge, within a 70 degree half-angle cone of the ring direction. The charge is corrected for light attenuation and PMT angular acceptance. The overall charge in the cone is corrected for variations in the PMT gains.
\par In addition to the standard ring-finding program, an algorithm described in detail in \cite{Barszczak:2005sf} is used to detect particle tracks. Unlike the standard ring-finder, this algorithm does not search for ring edges, but rather tests particle hypotheses as either electrons or photons (specifically, two photons from a $\pi^{0}$ decay) using only the light pattern. It has been used in atmospheric neutrino~\cite{Wendell:2010md} and long-baseline~\cite{Abe:2013xua} oscillation analyses to reject neutral current single $\pi^{0}$ events. It is used in the $pn \rightarrow \pi^{+}\pi^{0}$ search to identify a $\pi^{0}$ from the decay and use its reconstructed mass as a discriminating variable for the analysis.
\par After the primary event, electrons from stopping muons (Michel electrons) are found by searching for clusters of hits detected at the same time. The time window for such clusters extends to 20 $\mu$s after the main event. In the SK-I, II, and III data-taking periods, there was an impedance mismatch in the electronics which caused signal reflection around 1000 ns after the main event. Thus, the time period 800-1200 ns after the main event was excluded from the Michel electron search. The improved electronics of SK-IV eliminated this signal reflection, and as a result the tagging efficiency has improved from 80\% to 96\% for $\mu^{+}$ decays and 63\% to 83\% for $\mu^{-}$ decays between SK-I-III and SK-IV. Hence, event selections that place an upper limit on the number of Michel electrons reduce the event rates more in SK-IV than in SK-I-III.
\par After the reconstruction, a final set of selection criteria is applied to isolate fully-contained fiducial volume events (FCFV):

\begin{itemize*} \itemsep0pt \parsep0pt \parskip0pt
\item No significant OD activity: the largest OD hit cluster contains less than 10 hits (SK-I) or 16 hits (SK-II-IV).
\item The total visible energy is greater than 30 MeV.
\item The reconstructed event vertex must be at least 2 m from the inner detector wall. This is the fiducial volume cut as defined in Section II.
\end{itemize*}

\noindent After the FCFV selection, the fraction of events coming from sources other than atmospheric neutrinos is estimated to be $< 1\%$. In the next section, all analyses start with the FCFV events.


\section{Analysis}

The experimental signature of a dinucleon decay to pions event, for all three modes considered, is two back-to-back pions, each forming at least one Cherenkov ring of the correct type. The characterization of such events is made difficult due to the interactions of charged pions in both the residual oxygen nucleus and the water. Interactions in the water lead to truncated Cherenkov light signatures, which typically lead to poor momentum reconstruction. Thus, for modes involving at least one $\pi^{+}$ ($pp\rightarrow\pi^{+}\pi^{+}$ and $pn\rightarrow\pi^{+}\pi^{0}$), simple selection criteria based on kinematic variables such as the total momentum and invariant mass that are used in other nucleon decay analyses do not give optimal sensitivity. For these modes, a multivariate method is applied to statistically identify events with some, but perhaps not all, of the expected characteristics that differentiate signal and background. Specifically, a boosted decision tree (BDT) method is used. The method is implemented in the ROOT-based TMVA~\cite{Hocker:2007ht} analysis library. This was also used in the dinucleon decay analysis in \cite{Litos:2014}.
\par For $nn \rightarrow \pi^{0}\pi^{0}$, the situation is more straightforward, as the $\pi^{0}$'s that escape the nucleus decay to energetic photons before interaction in the water, and the electromagnetic showers are well-reconstructed. For this mode, simple kinematic selection criteria are sufficient for the analysis.


\subsection*{Boosted Decision Trees}
\label{sec:BDT}

Here we describe some relevant details of the BDT, applicable for both $pp \rightarrow \pi^{+}\pi^{+}$ and $pn\rightarrow \pi^{+}\pi^{0}$ searches. Similar to other multivariate methods, a BDT takes a set of input variables -- in this case, reconstructed information for both dinucleon decay signal and atmospheric neutrino background -- with the goal of producing a single discriminating output variable that separates signal and background. To get to this single variable, three stages are required. They are referred to as training, testing, and analysis, and each uses separate sets of signal and background MC. We describe the process briefly. 

\par Boosted decision trees are sets of individual decision trees, simple binary trees formed by selection cuts on single variables. Trees are built using the training sample as follows. Starting with the full training sample, a selection cut on a single variable is made that best separates signal and background. For the analyses in this paper, this optimal separation is determined by the Gini index $g = p(1-p)$, where $p$ is the purity of the sample after the selection. Other separation types exist, though they don't lead to significantly different performance~\cite{Hocker:2007ht}. The sample then splits off into two samples, or nodes, based on this selection. Each of the daughter nodes does the same thing, separately finding the variable that best separates between signal and background, and so forth until some stopping criterion is reached (such as a minimum number of events remaining on a node). Terminal nodes are declared signal-like or background-like based on the purity of signal events in the node.
\par The training process continues over many such trees (typicially 1000 or so) using the same training sample, with the prescription that events in subsequent trees are re-weighted, or ``boosted.'' Several boosting algorithms exist, and are described in \cite{Hocker:2007ht}. Upon re-weighting, the best set of discriminating variables may change. Variables can also be used more than once in a given tree.

\par Once the training sample has been processed, additional datasets may pass through the trained set of trees. On an event-by-event basis, the final BDT output is given by 
\begin{equation} Y_{BDT} = \frac{1}{N_{tree}}\sum_{i=1}^{N_{tree}}{w_{i}h_{i}\left(x\right)}. \end{equation} \label{eq:BDT}

 \noindent Here $N_{tree}$ is the size of the ensemble of trees, $w_i$ is the weight of each tree in the ensemble, determined by the boosting method, $h_{i}$ is the ``performance'' of each event for the $i^{th}$ tree (whether it ended on a signal or background-like node) and $x$ is the set of input variables. Variables are given a ranking, or ``importance,'' a numerical value that roughly corresponds to the number of times the variable is used in the training process relative to the other variables in the set.
\par After the BDT is trained, two additional, statistically independent sets of signal and background MC are necessary. The {\em test} samples are used to determine the optimal boost method, as well as some additional parameters. The {\em analysis} samples are used to obtain the final sensitivity. Unlike in the training and testing process, no information is given to the BDT regarding the composition of the analysis samples as either signal or background. Both the test and analysis samples are also important in detecting training bias in the BDT. This bias is a situation in which the BDT has been trained to model statistical fluctuations in the training sample, as opposed to modeling from general trends in the input variables. Such overtraining can lead to different performance between MC samples, even if they are statistically similar.
\par Table \ref{table:BDT_samples} quantifies the signal and background MC samples used for each stage. Separate trees are trained, tested, and analyzed for each SK period. Real data are processed through the trained BDTs as a final step.

\begin{table}[h!]
\setlength{\tabcolsep}{4pt}

\begin{center}
\begin{tabular}{c | ccc } \hline\hline
& training & testing & analysis \\ \hline
\specialcell{dinuc.\\ decay} & 25000 ev. & 25000 ev. & 25000 ev. \\ \hline
\specialcell{atm-$\nu$ } & \specialcell{$\sim$750,000 ev. \\ (150 yr.) } & \specialcell{ $\sim$750,000 ev. \\ (150 yr.) \\ } & \specialcell{$\sim$1,000,000 ev. \\ (200 yr.) } \\ 
\hline \hline
\end{tabular}
\end{center}
\caption{Independent MC sample sizes used for each stage of the BDT for both the $pp \rightarrow \pi^{+}\pi^{+}$ and $pn\rightarrow\pi^{+}\pi^{0}$ searches, prior to any event selection. Dinucleon decay sample sizes show only the number of events, while the atmospheric neutrino sample sizes show the approximate number of events and the equivalent number of years. The same statistics are used for SKI-IV.}
\label{table:BDT_samples}
\end{table}


\subsection*{$\mathbf{pp}\boldsymbol{\rightarrow}\boldsymbol{\pi^{+}}\boldsymbol{\pi^{+}}$}

The interactions of $\pi^{+}$'s in the nucleus and the water complicate the event signature for $pp\rightarrow\pi^{+}\pi^{+}$. A set of selection criteria is applied to keep $pp\rightarrow\pi^{+}\pi^{+}$ events, and reduce the atmospheric neutrino background. The key characteristics are that the event has two ``back-to-back" rings, as the ring-finding algorithm is still good at reconstructing the ring direction even with the poor momentum resolution, and that the rings are identified as $\mu$-like. Due to interactions in the water, the $\pi^{+}$'s can simulate showering effects and therefore be mis-identified as $e$-like. However, the converse -- in which rings produced by electrons or photons are misidentified as $\mu$-like -- is rare. The following selection criteria are applied:

\begin{enumerate}
\item[(A1)] There is more than one Cherenkov ring.
\item[(A2)] The two most energetic rings are $\mu$-like. These are assumed to correspond to the $\pi^{+}$ candidates.
\item[(A3)] The angle between the two most energetic rings is greater than 120 degrees, a minimal requirement for ``back-to-back'' pions.
\item[(A4)] The total visible energy is less than 1600 MeV, approximately the maximum visible energy expected from a dinucleon decay event $(2M_{p} - 2M_{\pi^{+}})$, where $M_{p}$ is the proton and $M_{\pi^{+}}$ the pion mass.
\end{enumerate}        

\begin{table}[h!]
\setlength{\tabcolsep}{4pt}        
\begin{center}
\begin{tabular}{ l | r r r r} \hline \hline
& SK-I & SK-II & SK-III & SK-IV \\ \hline
Eff.(\%) & 11.2$\pm$0.2 & 10.5$\pm$0.2 & 12.0$\pm$0.2 & 12.1$\pm$0.2 \\
Bkg. & 33$\pm$0.9 & 17$\pm$0.5 & 13$\pm$0.4 & 45$\pm$1.2 \\
data & 27 & 14 & 8 & 43 \\

\hline \hline
\end{tabular}
\caption {Efficiencies, expected backgrounds, and data events after selection criteria are applied in the $pp\rightarrow\pi^{+}\pi^{+}$ search. Background rates are scaled to the appropriate SK period livetime. Statistical errors are shown.}
\label {pp_precut_stats}
\end{center}
\end{table}

\begin{figure*}[t!]
        
\subfigure{\includegraphics[width=80mm]{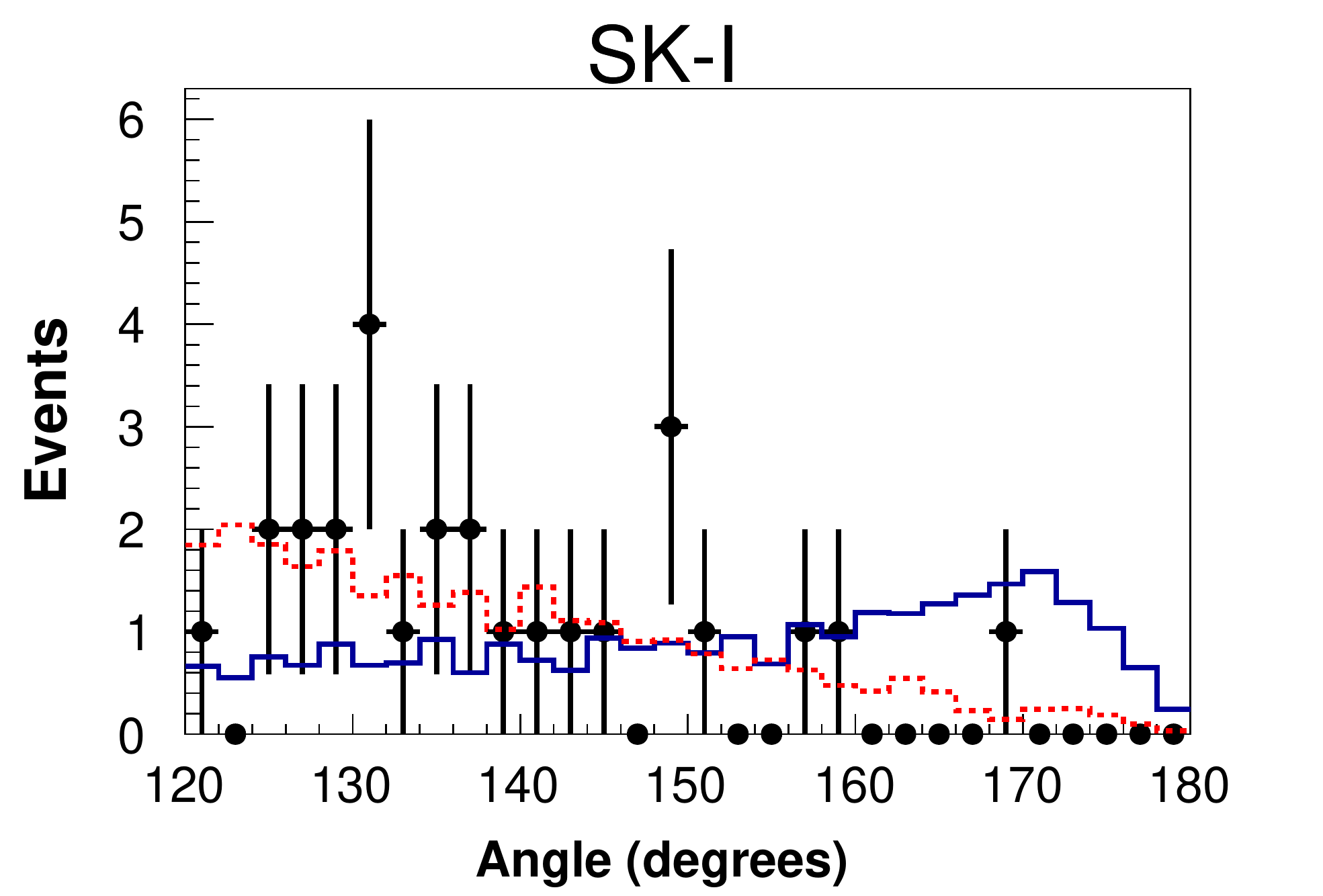}}
\subfigure{\includegraphics[width=80mm]{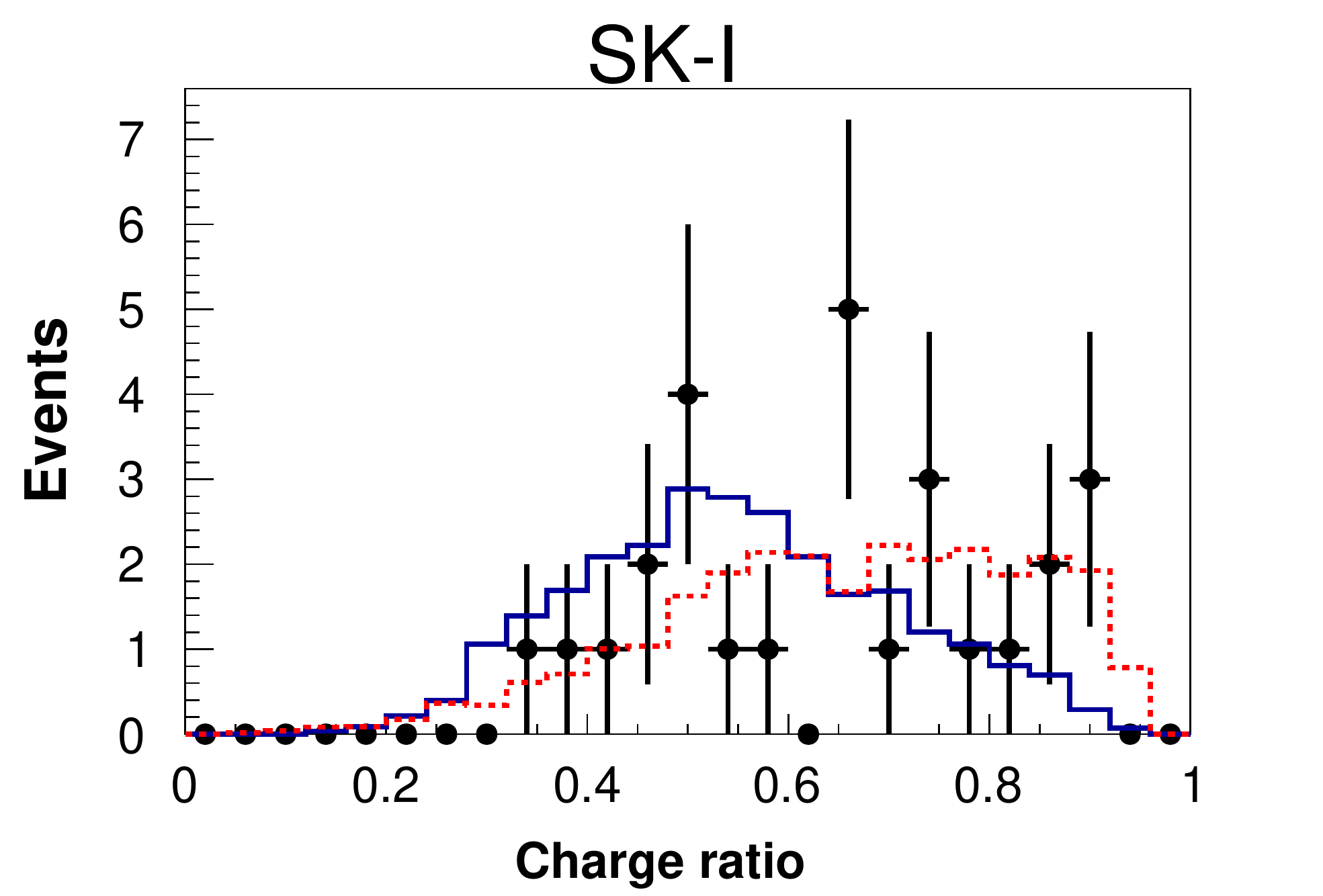}}
\subfigure{\includegraphics[width=80mm]{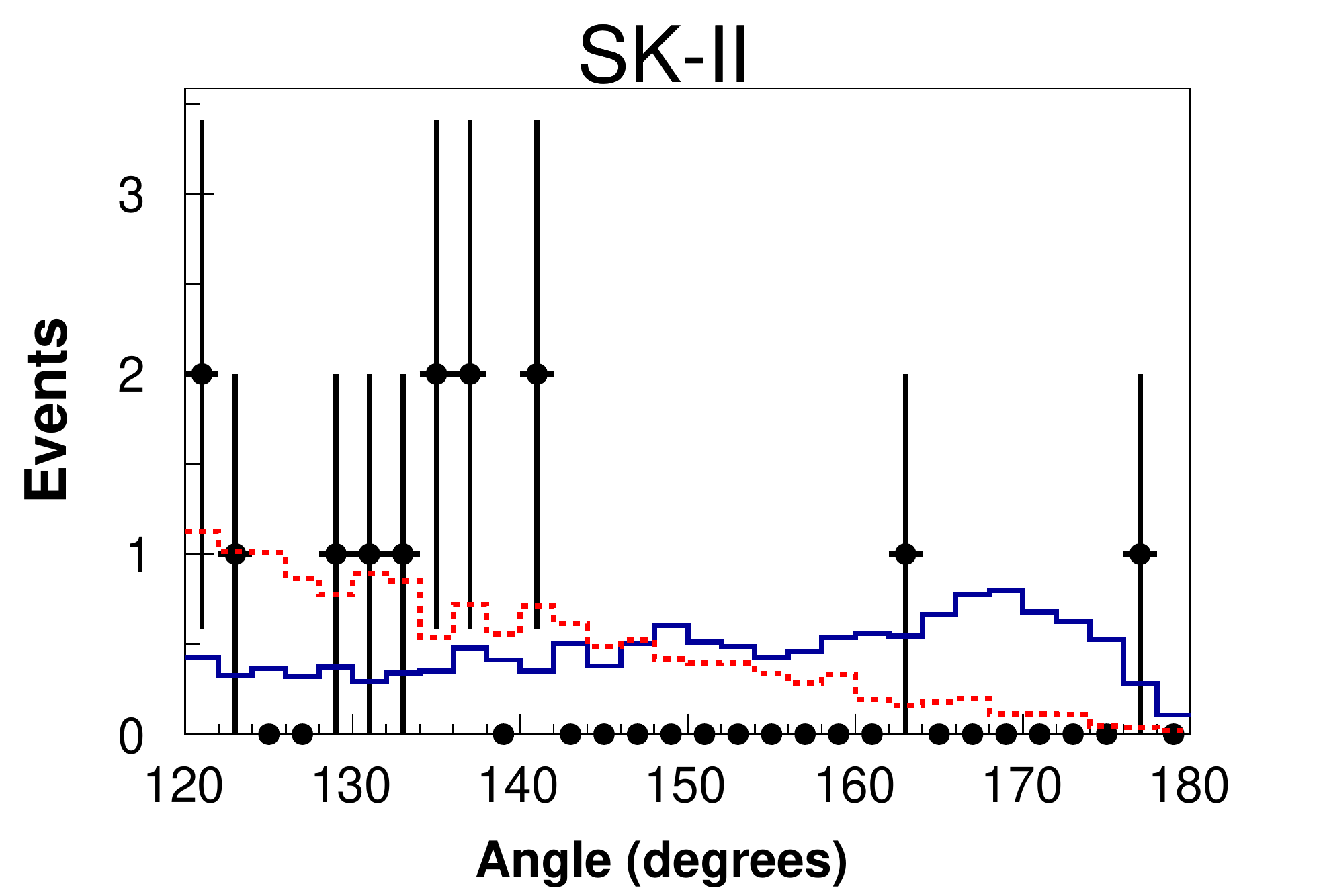}}
\subfigure{\includegraphics[width=80mm]{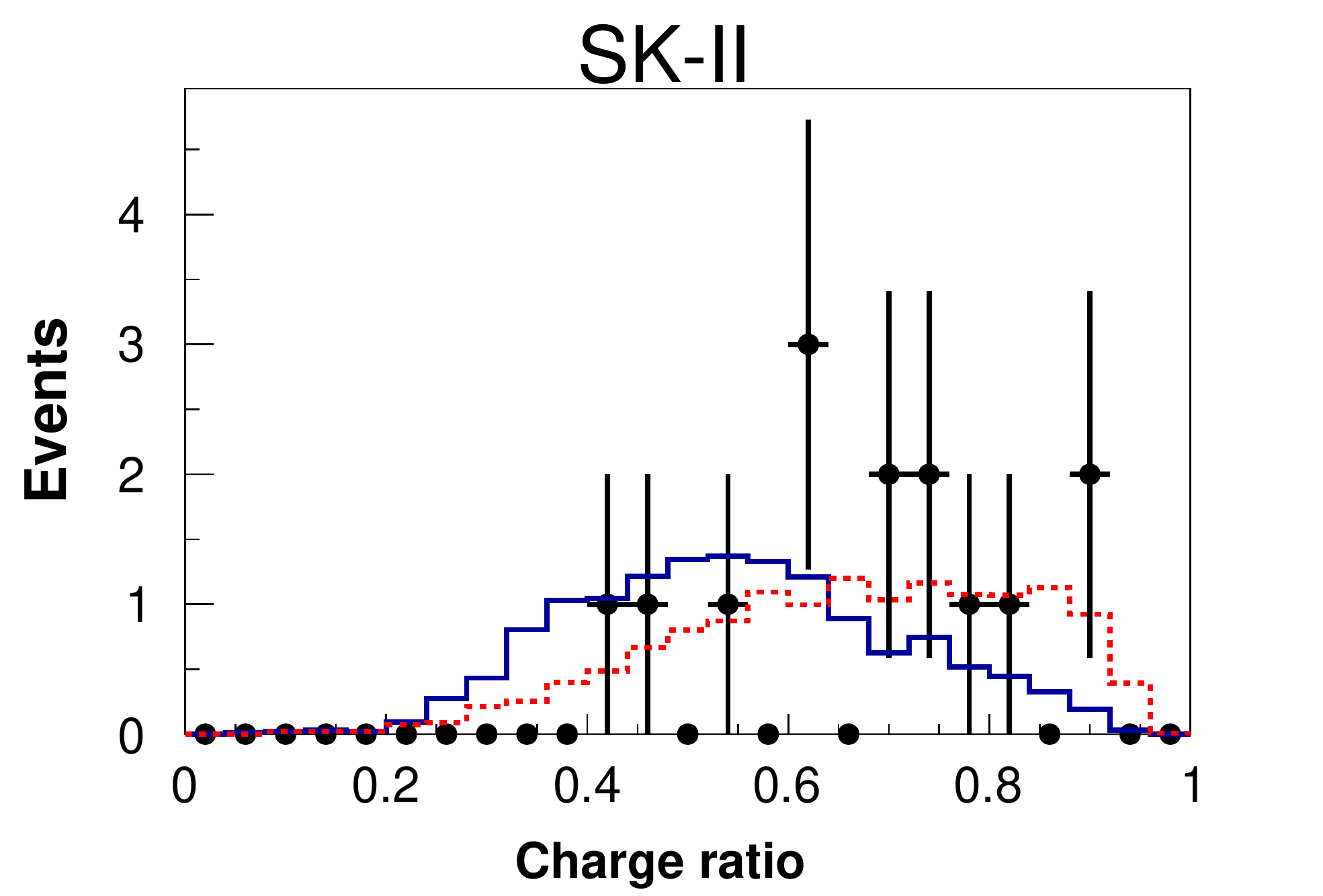}}
\subfigure{\includegraphics[width=80mm]{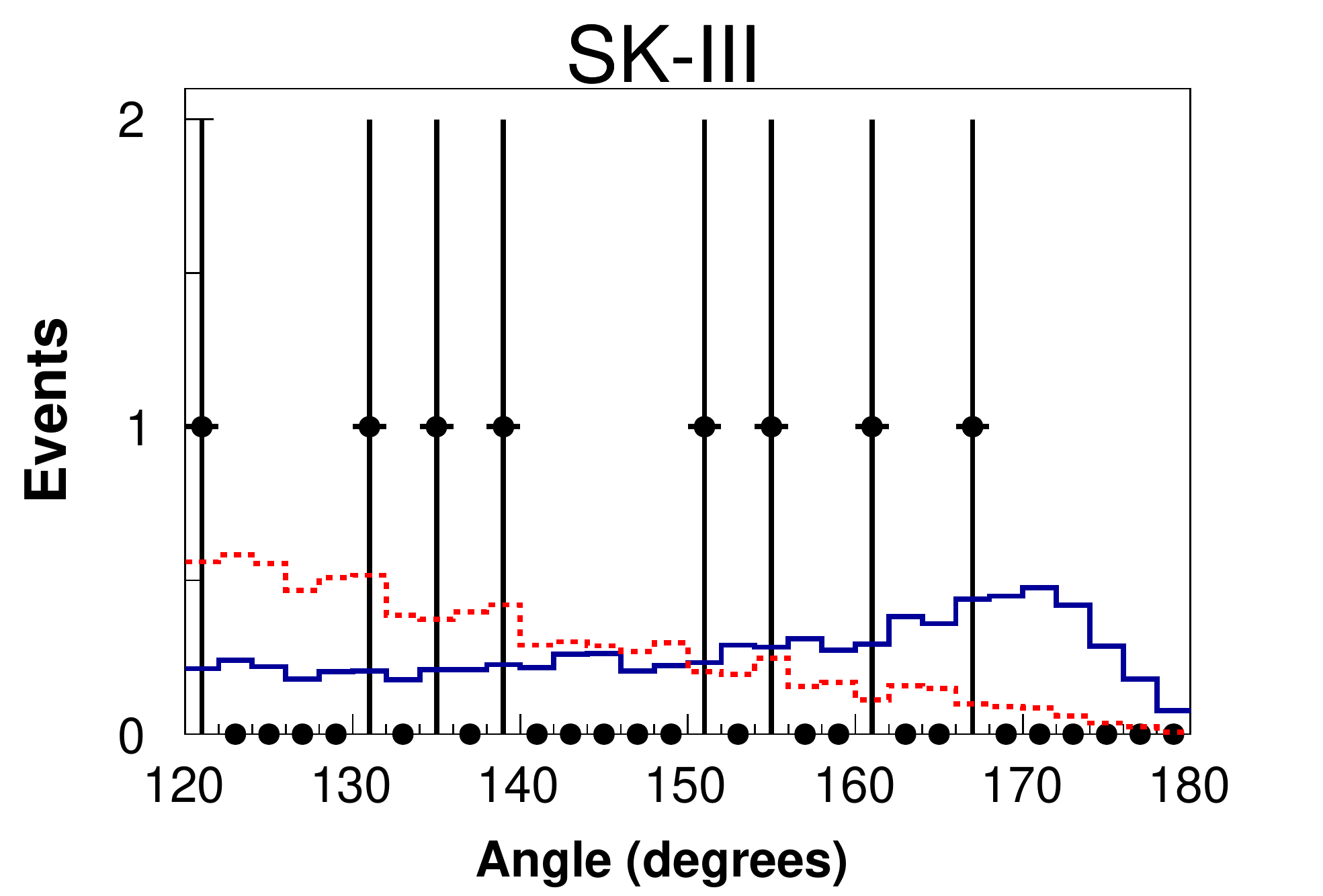}}
\subfigure{\includegraphics[width=80mm]{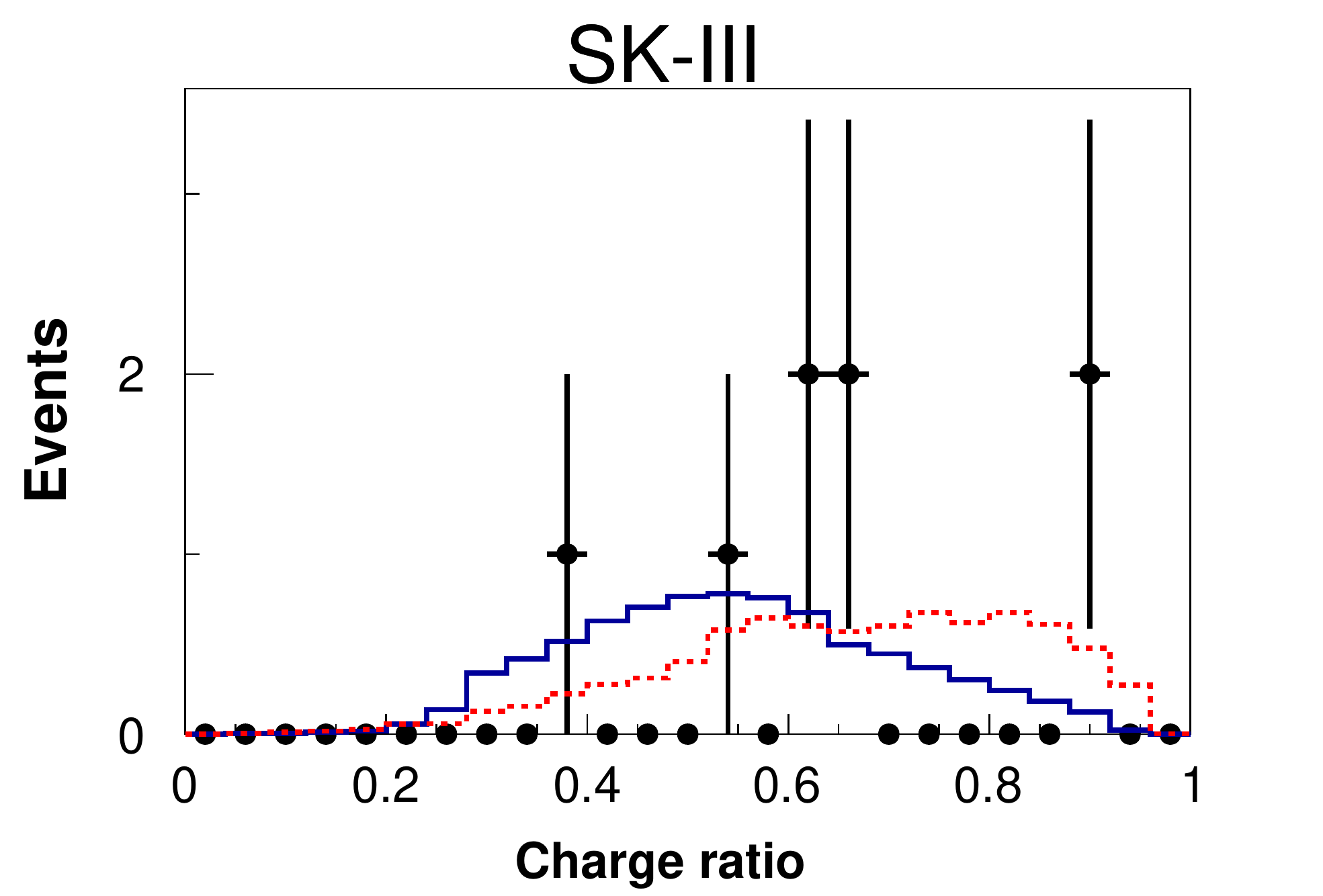}}
\subfigure{\includegraphics[width=80mm]{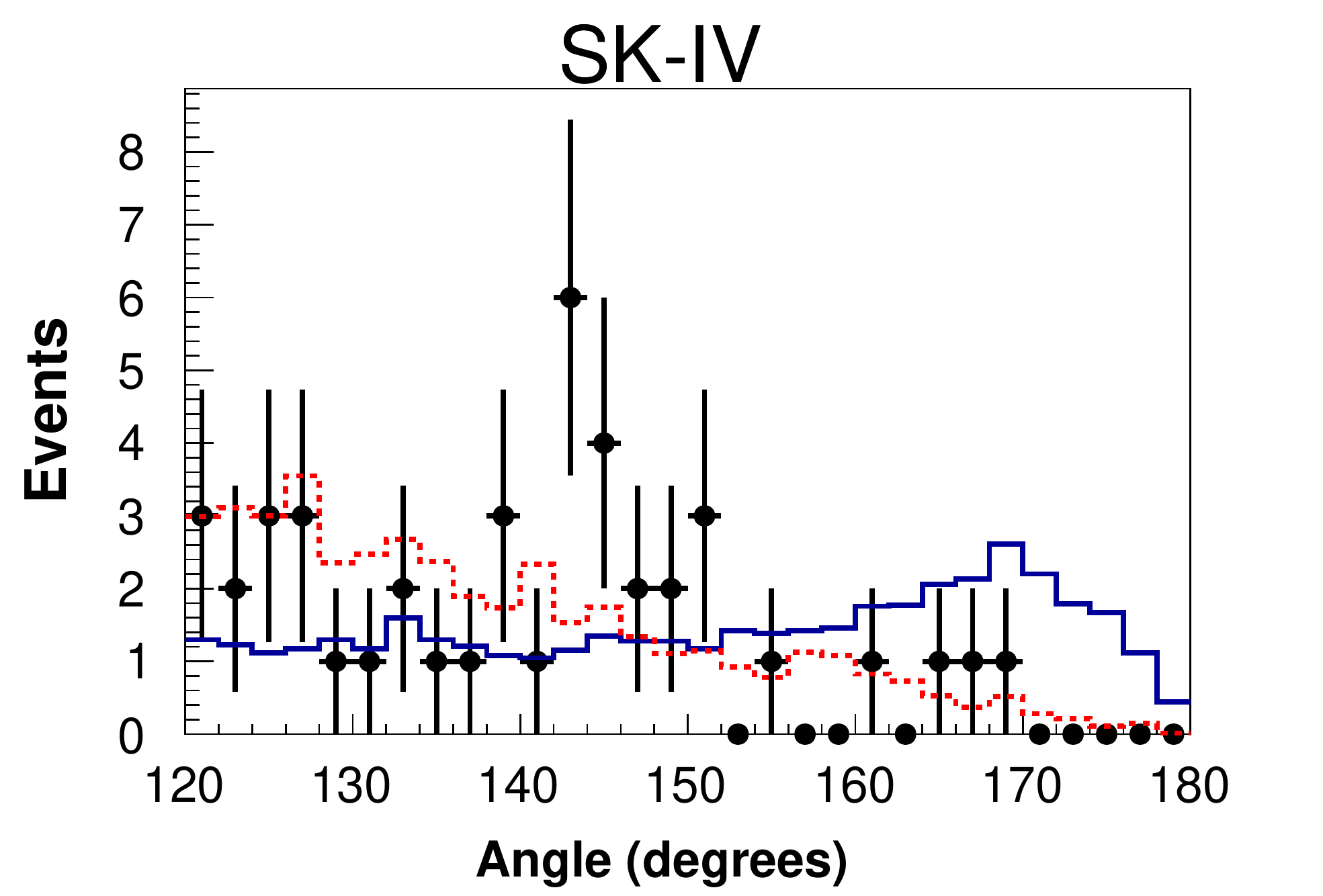}}
\subfigure{\includegraphics[width=80mm]{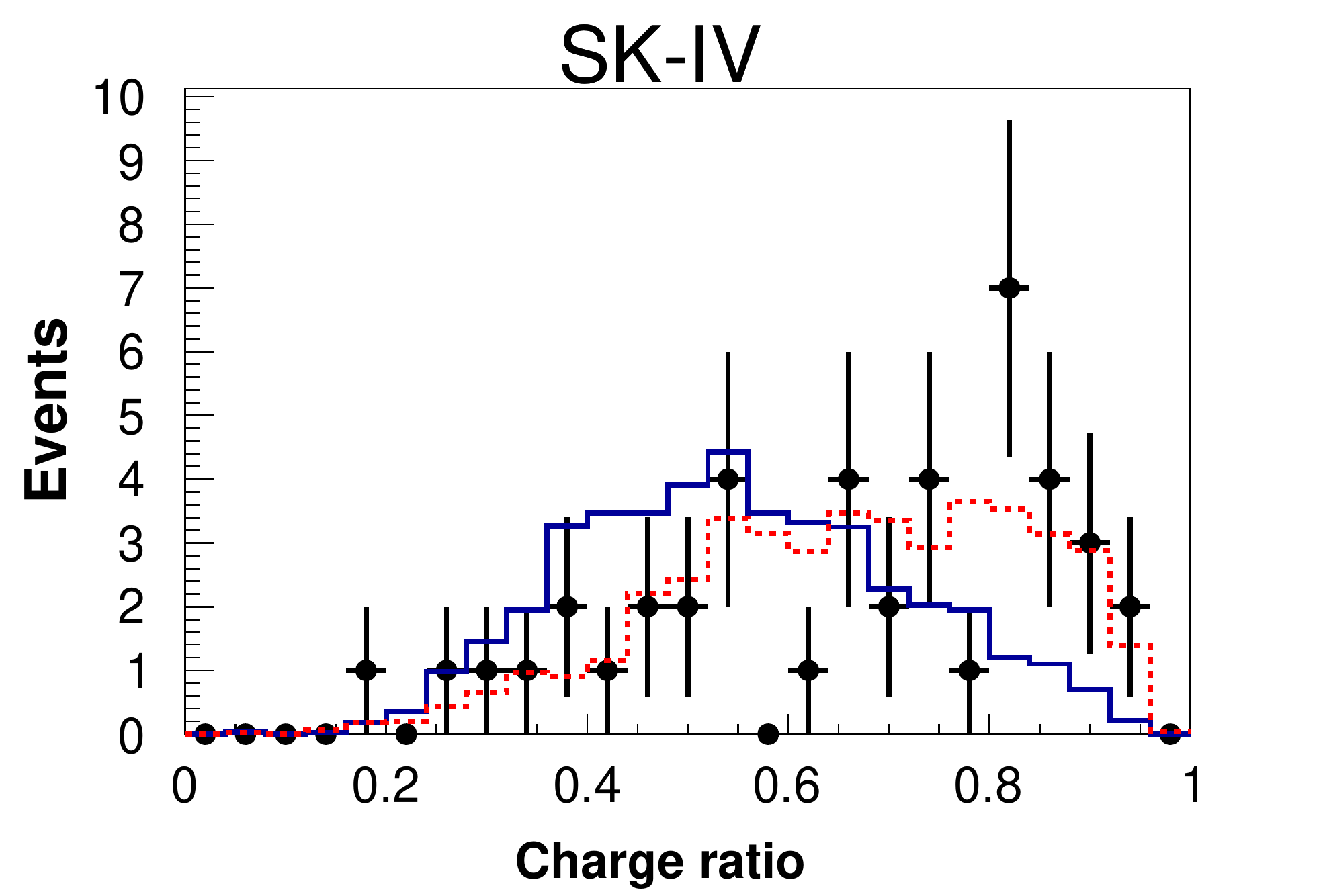}}
\caption{The angle between $\mu$-like rings and the ratio of charge carried by the most-energetic ring, for the $pp\rightarrow\pi^{+}\pi^{+}$ search, for signal MC (solid histogram), atmospheric neutrino MC (dashed histogram), and data (crosses). Distributions are normalized to the number of events in the data. Details of each variable are described in the text.}
\label{fig:pp_bestvars}
\end{figure*}

\par The signal efficiency, expected background, and data events remaining after the initial selection criteria is shown in Table~\ref{pp_precut_stats}. After applying these criteria, we input the remaining events into the BDT. The following set of input variables is used, ordered by their relative importance, as described in the BDT section above:

\begin{enumerate} \itemsep0pt \parsep0pt \parskip0pt
\item[(a1)]{The angle between the two most energetic $\mu$-like rings. It is expected to peak around a perfect ``back-to-back" angle of 180 degrees for the signal, and have a falling spectrum for increasing angle in the background.}
\item[(a2)]{The ratio of charge carried by the highest-energy ring to the total charge of all rings. This is expected to be smaller for signal, since the pions are expected to have roughly equal energy. By contrast, we expect most of the energy from a charged-current (CC) atmospheric neutrino event to be in the charged lepton ring.}
\item[(a3)]{The total visible energy. This has a broad distribution up to 1600 MeV for the signal, but peaks at a lower value for the background around 200 MeV.}
\item[(a4)]{The maximum distance between a Michel electron vertex and the primary event vertex. We expect this to be larger for energetic muons than for pions, since the muons penetrate water without sudden hadronic interactions. Muons above Cherenkov threshold sometimes appear in the signal (about 9\% of the time), but are much more common in the background, due to charged-current $\nu_{\mu}$ interactions such as single pion production.}
\item[(a5)]{The maximum angle between any $\mu$-like ring direction vector, and any Michel electron vertex vector (as pointed to from the primary vertex). Similar to (a1), this is designed to look for ``back-to-back" events, and is an additional check on the topology of the $\mu$-like rings.}
\item[(a6)]{The magnitude of the vector sum of corrected charge associated with all rings, where the corrections are applied for various factors, as described in Section IV. This is similar to the total momentum, since the corrected charge associated with each ring is used to determine its momentum. Due to typically poor momentum resolution from hadronic energy losses, it is preferable to use this variable instead of the total momentum.}
\item[(a7)] {The number of Cherenkov rings. Though we are trying to isolate two pions, they typically interact in the water, often producing visible secondary tracks, so that the number of rings distribution peaks at three. By contrast, the Cherenkov ring spectrum for the background falls with increasing ring number.}
\item[(a8)]{The number of Michel electrons. Naively we would expect two Michel electrons from two charged pions. However, the pions could undergo charge exchange or absorption, leading to zero or one Michel electrons. Though the signal and background distributions both peak at one decay electron, this variable is relatively independent of most of the other variables, which depend on Cherenkov ring-counting.}
\item[(a9)]{The number of $\mu$-like rings. Similar to the total number of rings, this variable specifically seeks scatters that produce a secondary charged pion.}
\end{enumerate}

The relative importance of all input variables is shown in Table~\ref{pp_ranking}. Fig.~\ref{fig:pp_bestvars} shows plots of the two most discriminating variables, the angle between $\pi^{+}$ candidates and the ratio of charge carried by the most energetic ring. The number of events in the data is small after the selection criteria is applied, but the data and atmospheric neutrino MC are consistent.

\par
\begin{figure*}[t!]
\subfigure{\includegraphics[width=80mm]{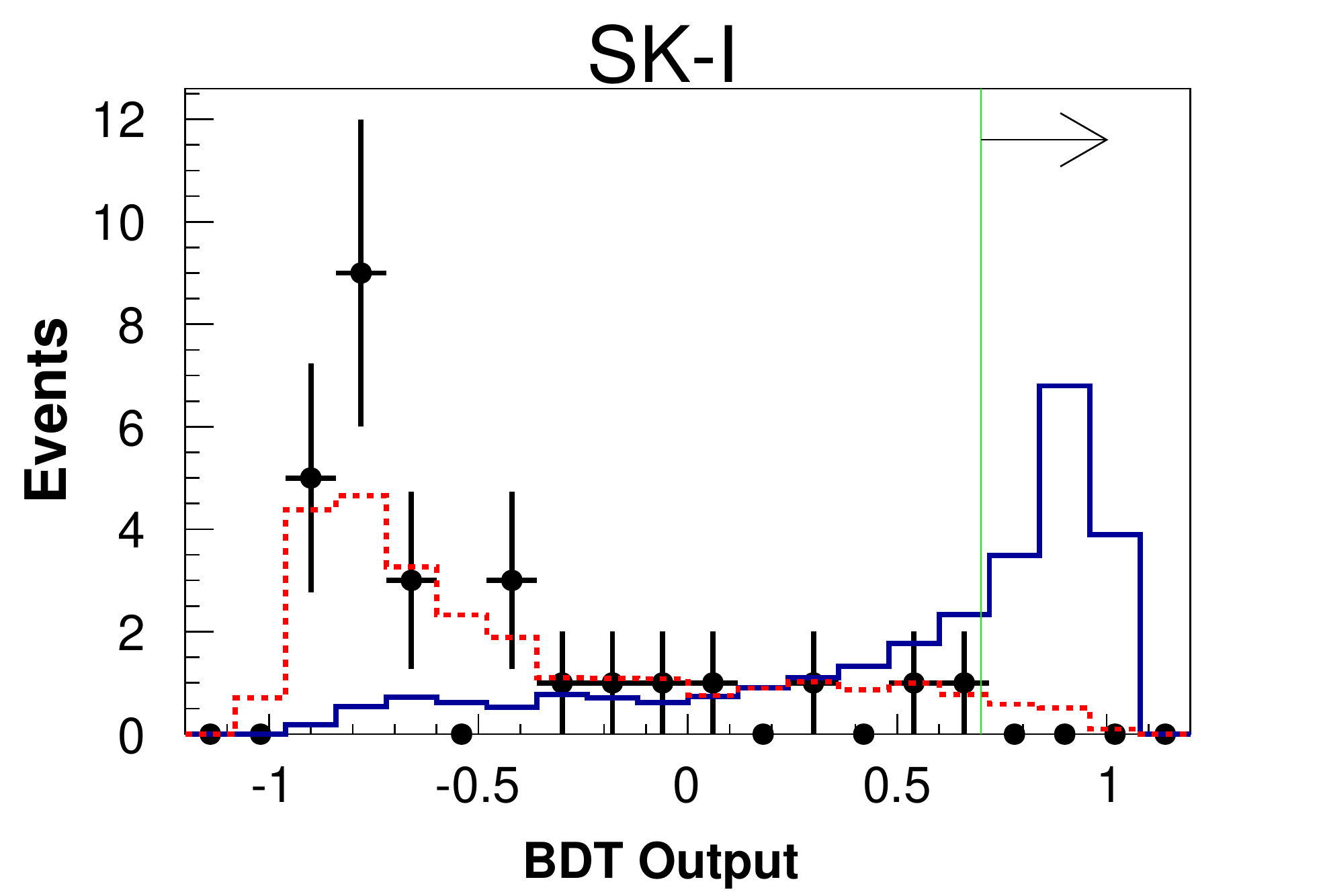}}
\subfigure{\includegraphics[width=80mm]{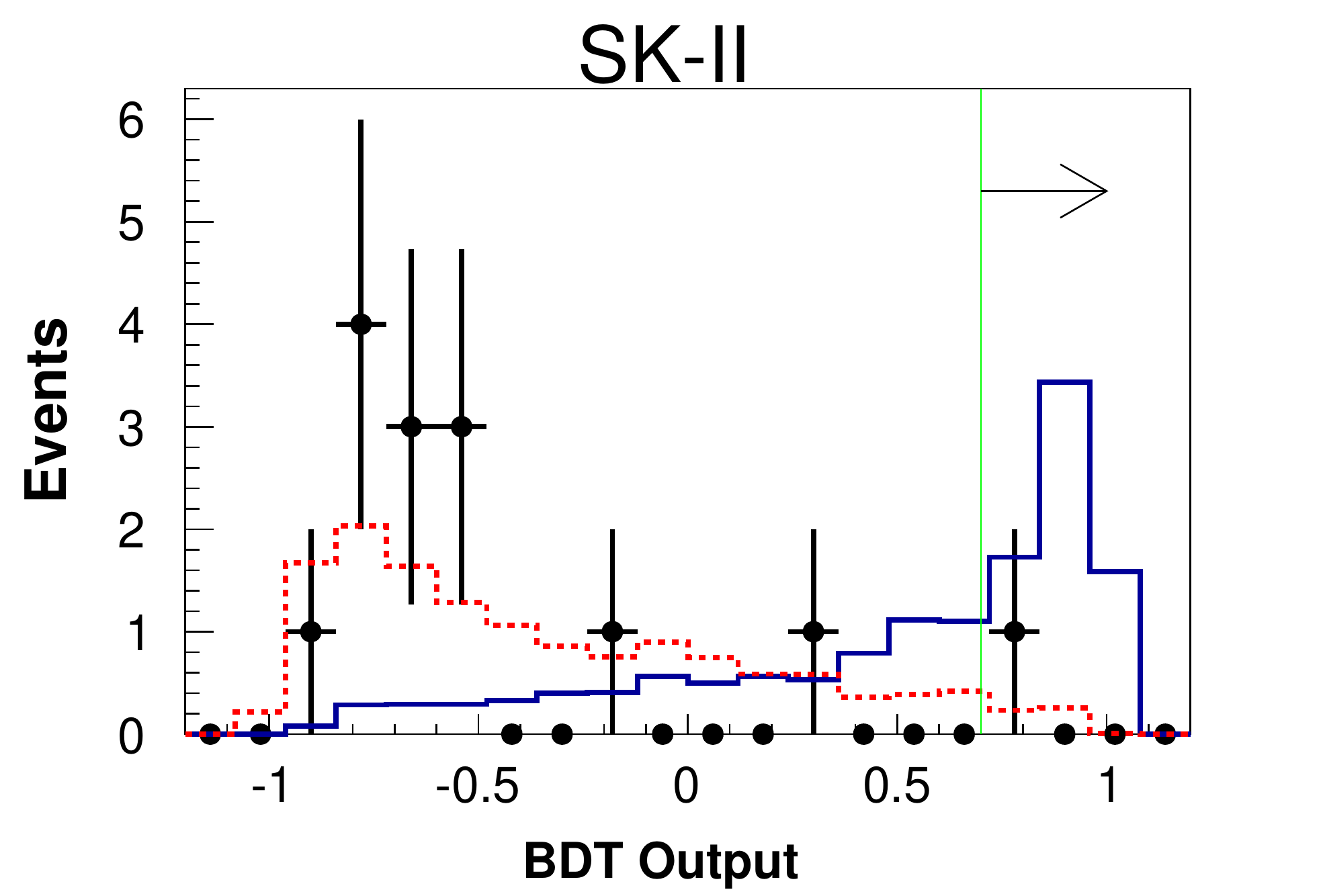}}
\subfigure{\includegraphics[width=80mm]{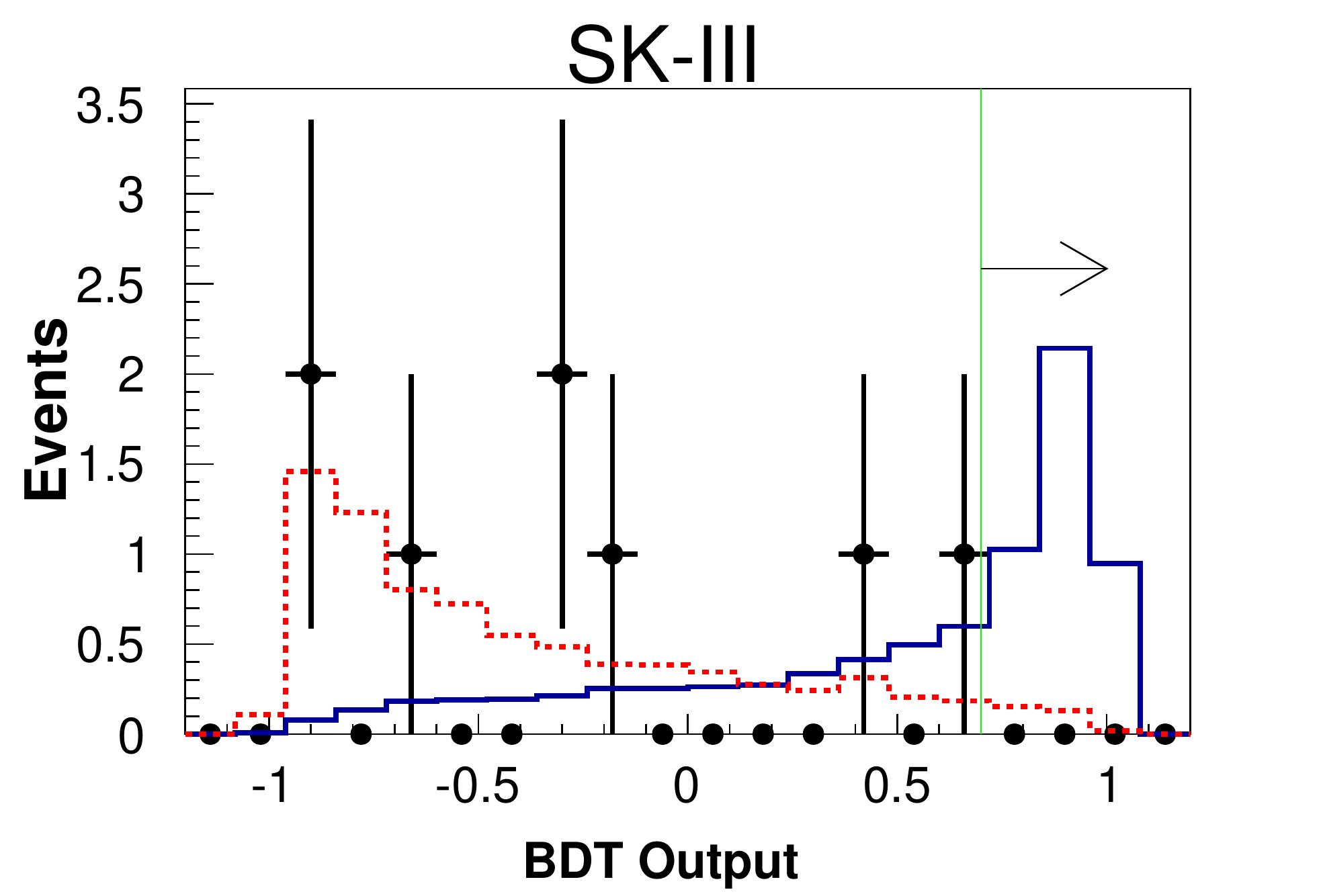}}
\subfigure{\includegraphics[width=80mm]{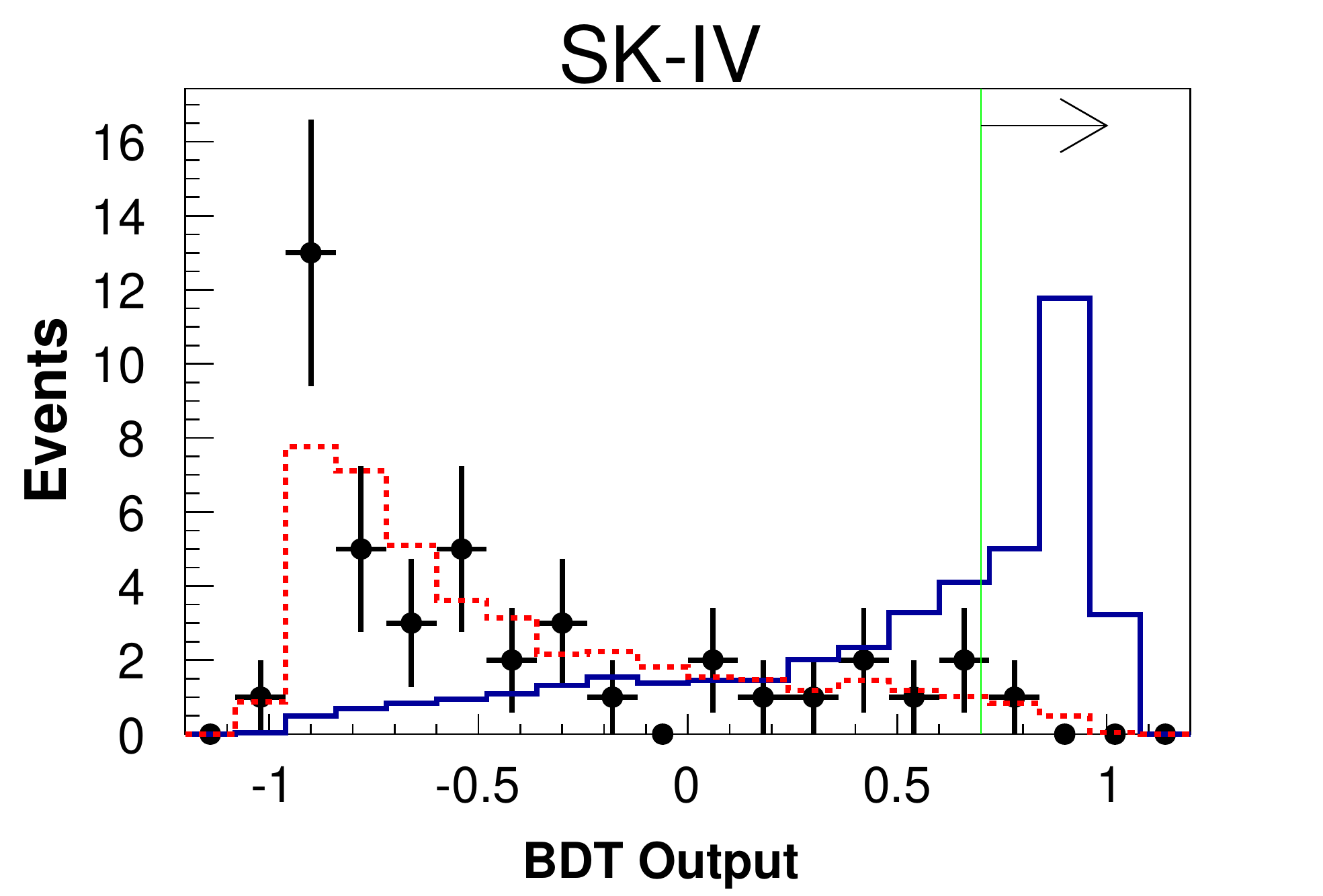}}
\caption{BDT output for $pp\rightarrow\pi^{+}\pi^{+}$ signal MC (solid), atmospheric neutrino background MC (dashed), and data (crosses). The vertical green line indicates the BDT cut value, and the arrow indicates that only events to the right of the cut are kept.}
\label{fig:BDTG}
\end{figure*}

 The final distributions of the BDT output for all SK periods are shown in Fig.~\ref{fig:BDTG}. For $pp \rightarrow \pi^{+}\pi^{+}$, the gradient boost~\cite{Hocker:2007ht} method is used. The final efficiencies, backgrounds, and candidate data events are shown in Table~\ref{table:pp_stats}. SK-II has a slightly lower efficiency than the other SK periods, as fewer events pass selection criterion (A2). 

\begin{table}[h!]
\setlength{\tabcolsep}{6pt}        
  \begin{center}
    \begin{tabular}{lr}
      \hline \hline
      Variable &  Importance \\
      \hline 
      Angle between $\mu$-like rings & 0.16\\
      \specialcell{Ratio of charge carried by \\ most energetic ring} & 0.15\\
      Visible energy & 0.15 \\
      \specialcell{Max. distance to \\ Michel vertex} & 0.13 \\
      \specialcell{Max. angle between $\mu$-like \\ ring and Michel vertex} & 0.13 \\
      \specialcell{Magnitude of vector sum \\of corrected charge} & 0.12 \\
      Number of rings & 0.071 \\
      Number of Michel electrons & 0.055 \\
      Number of $\mu$-like rings & 0.045 \\                        
      
      \hline \hline
    \end{tabular}
  \caption{\protect \small 
Relative importance of each variable in the  \\ $pp\rightarrow \pi^{+}\pi^{+}$ search, averaged across SK periods.
}
  \label{pp_ranking}
  \end{center}
\end{table}

\begin{table}[h!]
\setlength{\tabcolsep}{4pt}
\begin{center}
\begin{tabular}{l | r r r r } \hline\hline
  & SK-I & SK-II & SK-III & SK-IV\\ \hline
Eff. (\%) & 6.1$\pm$0.2 & 5.3$\pm$0.2 & 6.4$\pm$0.2 & 5.8$\pm$0.2\\
Bkg. (MT-yr) & 17.8$\pm$1.8 & 14.3$\pm$1.6 & 17.4$\pm$1.7 & 14.2$\pm$1.6 \\
Bkg. (SK live.) & 1.6 & 0.70 & 0.56 & 1.6 \\

Candidates & 0 & 1 & 0 & 1\\
\hline \hline
\end{tabular}
\end{center}
\caption{Efficiency, expected background events, and candidate data events for the SKI-IV $pp\rightarrow\pi^{+}\pi^{+}$ search. Statistical errors are quoted for efficiency and background. Background is quoted both for the appropriate SK livetime, and per megaton-year.} 
\label{table:pp_stats}
\end{table}

\begin{table}[h!]

\setlength{\tabcolsep}{5pt}

\begin{center}
\begin{tabular}{r | r r r r} \hline\hline

mode & SK-I & SK-II & SK-III & SK-IV\\ \hline



CC1$\pi$  & 45$\pm$7\% & 42$\pm$7\% & 51$\pm$7\% & 42$\pm$7\%\\ 
CCDIS  & 30$\pm$5\% & 31$\pm$6\% & 24$\pm$5\% & 34$\pm$7\%\\ 
NCDIS  & 10$\pm$4\% & 9$\pm$4\% & 6$\pm$3\% & 13$\pm$5\%\\ 
CCQE  & 9$\pm$3\% & 9$\pm$3\% & 10$\pm$3\% & 10$\pm$4\%\\ 
NC1$\pi$  & 5$\pm$3\% & 8$\pm$4\% & 8$\pm$3\% & 0+1\% \\ 

\hline \hline
\end{tabular}
\caption{Atmospheric neutrino interaction modes in the remaining background for the $pp\rightarrow \pi^{+}\pi^{+}$ search by percentage, as estimated by atmospheric neutrino MC.}
\label{table:pp_bkg}
\end{center}
\end{table}

\begin{table*}[t!]
\setlength{\tabcolsep}{6pt}        

    \begin{center}      
     \begin{tabular}{l|rrrrrr}
      \hline \hline
      Variable & \multicolumn{2}{c}{Candidate} & \multicolumn{2}{c}{Signal} & \multicolumn{2}{c}{Background} \\ 
      & SK-II & SK-IV & SK-II & SK-IV & SK-II & SK-IV\\
      \hline
      Angle between $\mu$-like rings (degrees) & 177.7 & 151.8 & 152.3 & 152.1 & 138.3 & 138.1 \\
      Ratio of charge carried by most energetic ring (dimensionless) & 0.60 & 0.46 & 0.55 & 0.55 & 0.66 & 0.66 \\
      Visible energy (MeV) & 400.0 & 745.6 & 680.5 & 693.7 & 440.2 & 407.3 \\
      Max. distance to Michel vertex (cm) & 1245.6 & 1420.1 & 440.7 & 424.8 & 523.0 & 459.2 \\
      Max. angle between $\mu$-like ring and Michel vertex (degrees) & 166.3 & 111.9 & 94.2 & 109.0 & 93.2 & 110.9  \\
      Magnitude of vector sum of corrected charge (corrected p.e.) & 306.9 & 1799.5 & 601.3 & 1310.0 & 448.7 & 797.5 \\
      Number of rings (dimensionless) & 3 & 3 & 2.89 & 2.95 & 2.39 & 2.40 \\
      Number of Michel electrons (dimensionless) & 1 & 2 & 0.91 & 1.09 & 1.00 & 1.21 \\
      Number of $\mu$-like rings (dimensionless) & 2 & 2 & 2.14 & 2.16 & 2.05 & 2.05 \\                        
      \hline \hline
     \end{tabular}
\caption{Candidate variable values, and mean variable values for signal and background distributions, for each of the variables used in the BDT for the $pp\rightarrow \pi^{+}\pi^{+}$ search. Values are given for SK-II and SK-IV, for which single candidate events were found.}
\label{table:pp_var_breakdown}  
\end{center}
\end{table*}

\par The final cut on the BDT output was determined using only signal and background MC, not the real data. The cut takes two basic considerations into account: (1) the cut maximizes $\epsilon/\sqrt{b}$, where $\epsilon$ is signal efficiency and $b$ is the background rate; and (2) the cut is not too close to the signal peak, since variations are propagated through the BDT to estimate systematics, and the part of the distribution near the peak changes the most under these variations. For (2), 0.7 was the closest value to the peak that we considered. As the efficiency (background) distribution increases (decreases) nearly monotonically for all SK periods, it was found that a cut at 0.7 satisfied (1) while obeying the constraint of (2).

\begin{figure}[h!]
\centering

\setlength\fboxsep{0pt}
\setlength\fboxrule{0pt}

\fbox{\includegraphics[trim ={7mm 65mm 10mm 3mm},clip=true, width=80mm]{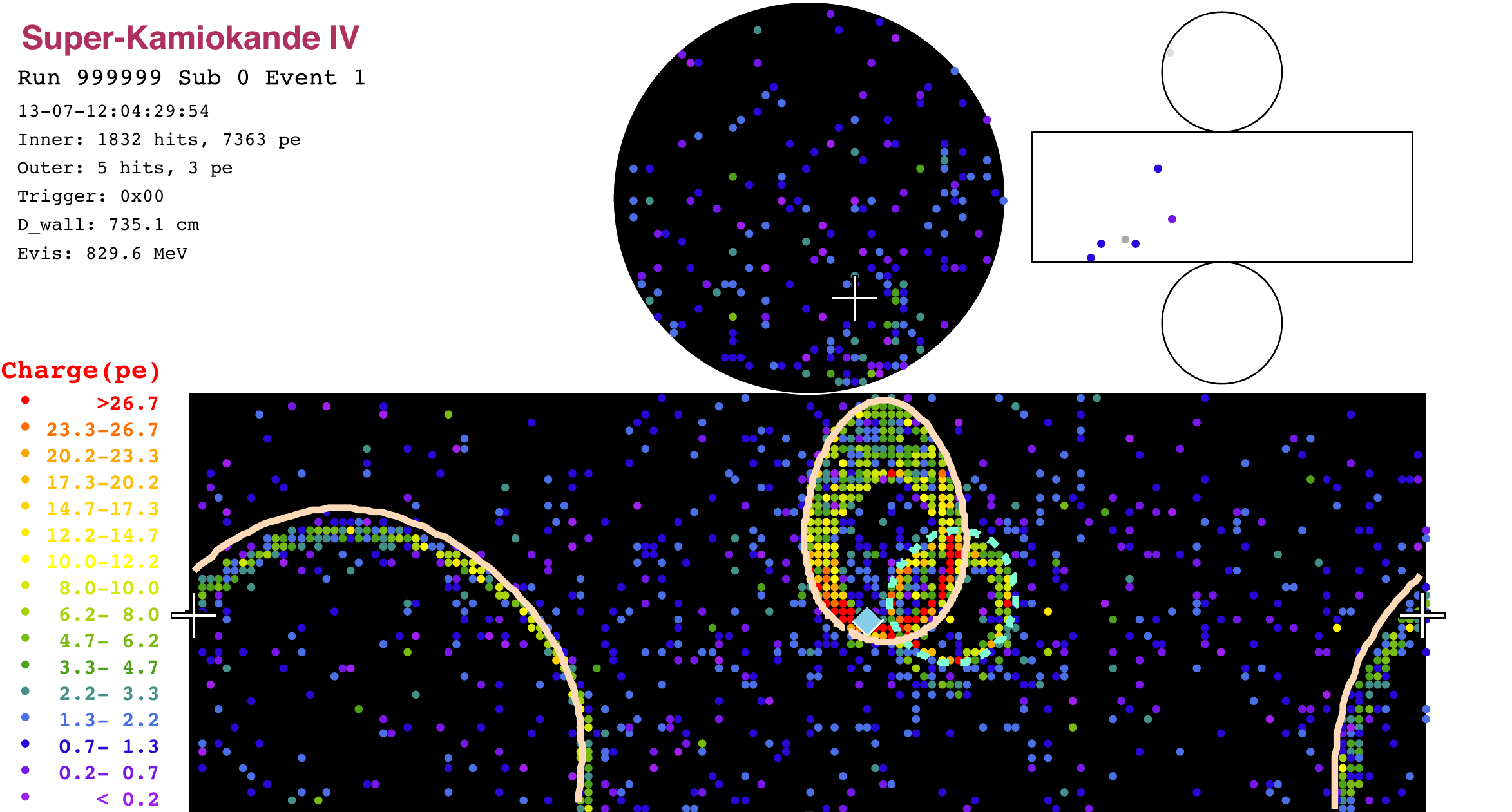}}
\linebreak
\fbox{\hspace{1.5em}\includegraphics[trim ={100mm 3mm 88mm 132mm},clip=true, width=22mm]{pp_pippip_ev1_notru_rotated.pdf}}

\caption{A $pp \rightarrow \pi^{+}\pi^{+}$ MC event. The intensity scale indicates the amount of charge deposited on the PMTs. The main display shows the ID, while the small display on the top right shows the OD. The white $+$'s indicate the horizontal and vertical location of the reconstructed vertex, projected onto the detector wall. Orange rings are fit as $\mu$-like, and cyan rings as $e$-like. Solid rings indicate that the ring's opening angle is near the ultra-relativistic limit in water of $\theta_{c}=cos^{-1}\left(1/n\right) \approx 41$ degrees, while dashed or dotted rings are more collapsed. The two solid orange rings correspond to true $\pi^{+}$'s, and have an angle of 164 degrees between them. The dotted cyan ring is fit as $e$-like, but is actually a hard scatter of one of the $\pi^{+}$'s.}
\label{figure:pp_mc_disp}
\end{figure}

\begin{figure}[h!]
\centering

\setlength\fboxsep{0pt}
\setlength\fboxrule{0pt}

\fbox{\includegraphics[trim ={7mm 65mm 10mm 3mm},clip=true, width=80mm]{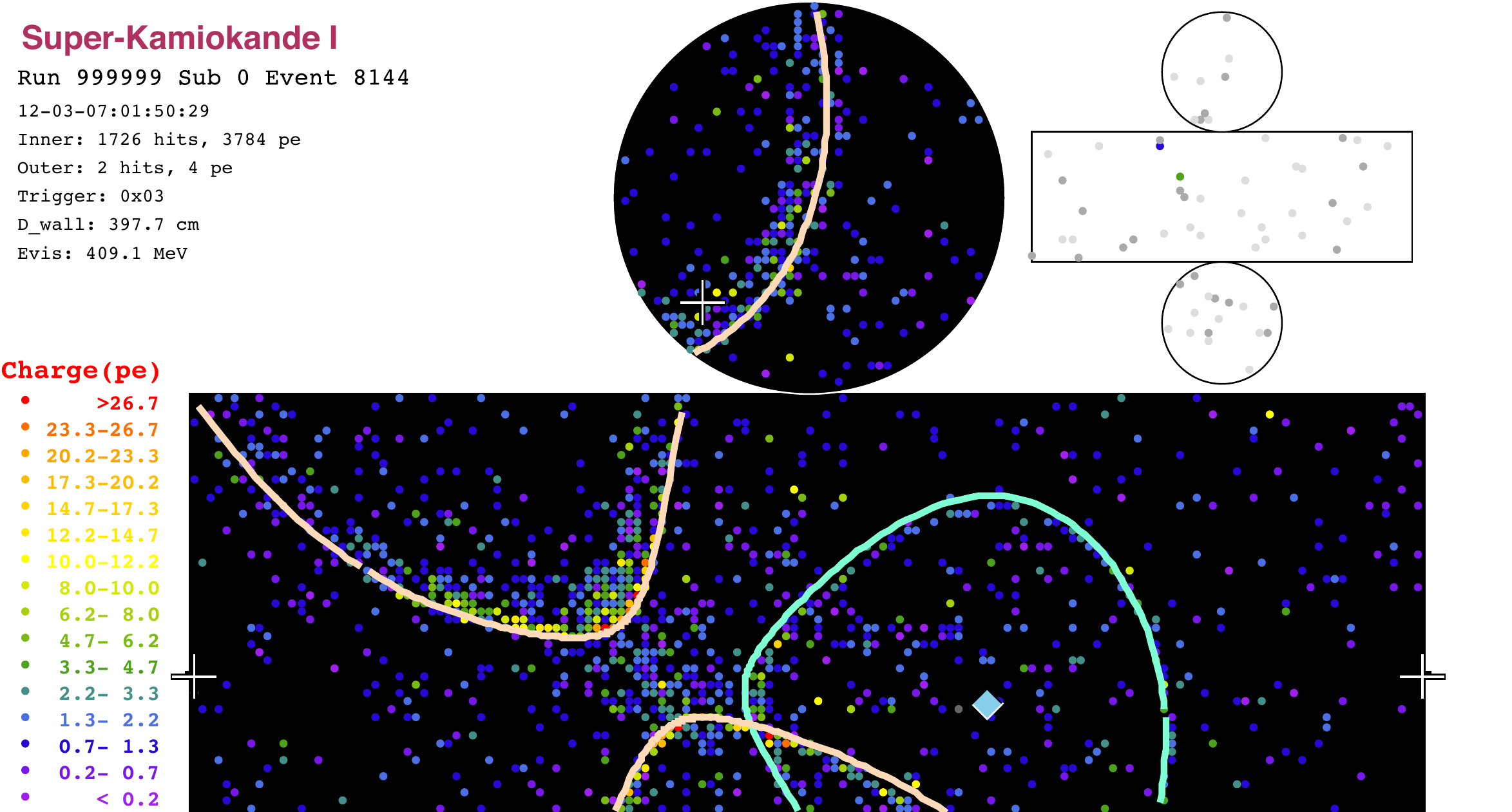}}
\linebreak
\fbox{\hspace{1.5em}\includegraphics[trim ={100mm 3mm 88mm 132mm},clip=true, width=22mm]{sk1_pp_pippip_bkg_file330_ev8144.pdf}}

\caption{ Surviving atmospheric neutrino MC event for the $pp\rightarrow\pi^{+}\pi^{+}$ search. The interaction type is $\nu_{\mu}$ charged-current single $\pi^{+}$ production. The uppermost $\mu$-like ring corresponds to a $\mu^{-}$, while the smaller $\mu$-like ring is a $\pi^{+}$. The $e$-like ring is mis-fit, as it comes from a hard scatter of the $\pi^{+}$. The angle between the two $\mu$-like rings is 170 degrees, thereby mimicking the main event signature. The momentum of the muon $\mu$-like ring is 402 MeV/$c$, and the momentum of the $\pi^{+}$ $\mu$-like ring is 301.3 MeV/$c$ }
\label{figure:pp_atm_surv}
\end{figure}

\begin{figure}[h!]
\begin{center}

\setlength\fboxsep{0pt}
\setlength\fboxrule{0pt}

\fbox{\includegraphics[trim ={7mm 65mm 10mm 3mm},clip=true, width=85mm]{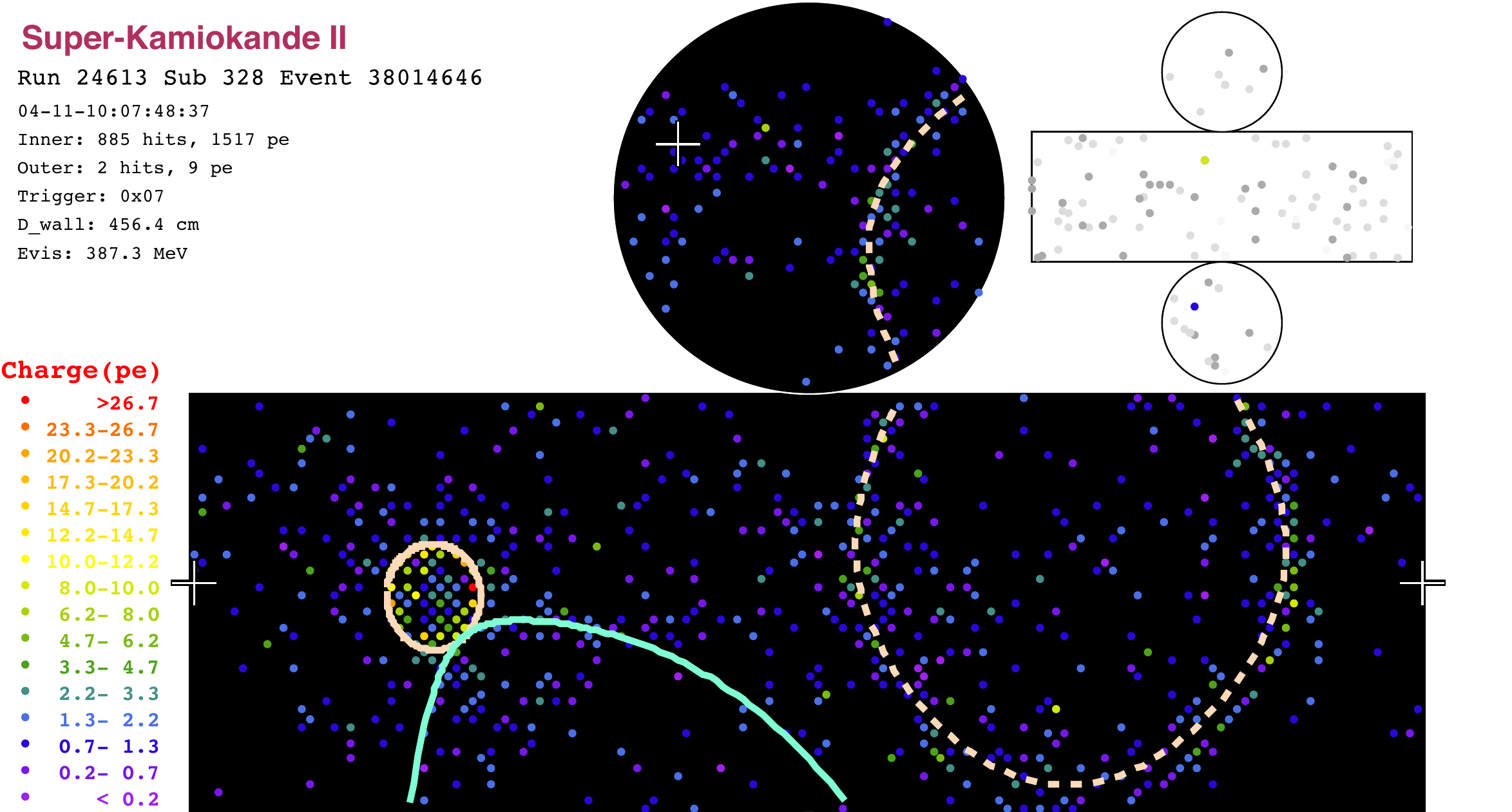}}
\linebreak
\fbox{\hspace{1.5em}\includegraphics[trim ={100mm 3mm 88mm 132mm},clip=true, width=22mm]{sk2_candidate_24613.pdf}}

\caption{$pp \rightarrow \pi^{+}\pi^{+}$ candidate event in SK-II. The two $\mu$-like rings are strikingly back-to-back, at almost exactly 180 degrees to each other. The $e$-like ring could be due to a hard scatter. There is one Michel electron, not shown. The $\mu$-like rings have momenta of 399.0 MeV/$c$ (solid) and 280.5 MeV/$c$ (dashed), and the $e$-like ring has momentum 42.0 MeV/$c$. The visible energy is 387.0 MeV, the total momentum 124.2 MeV/$c$, and the invariant mass 710.9 MeV/$c^2$.}
\label{figure:pp_cand_sk2}
\end{center}
\end{figure}

\begin{figure}[h!]
\centering

\setlength\fboxsep{0pt}
\setlength\fboxrule{0pt}

\fbox{\includegraphics[trim ={7mm 65mm 10mm 3mm},clip=true, width=80mm]{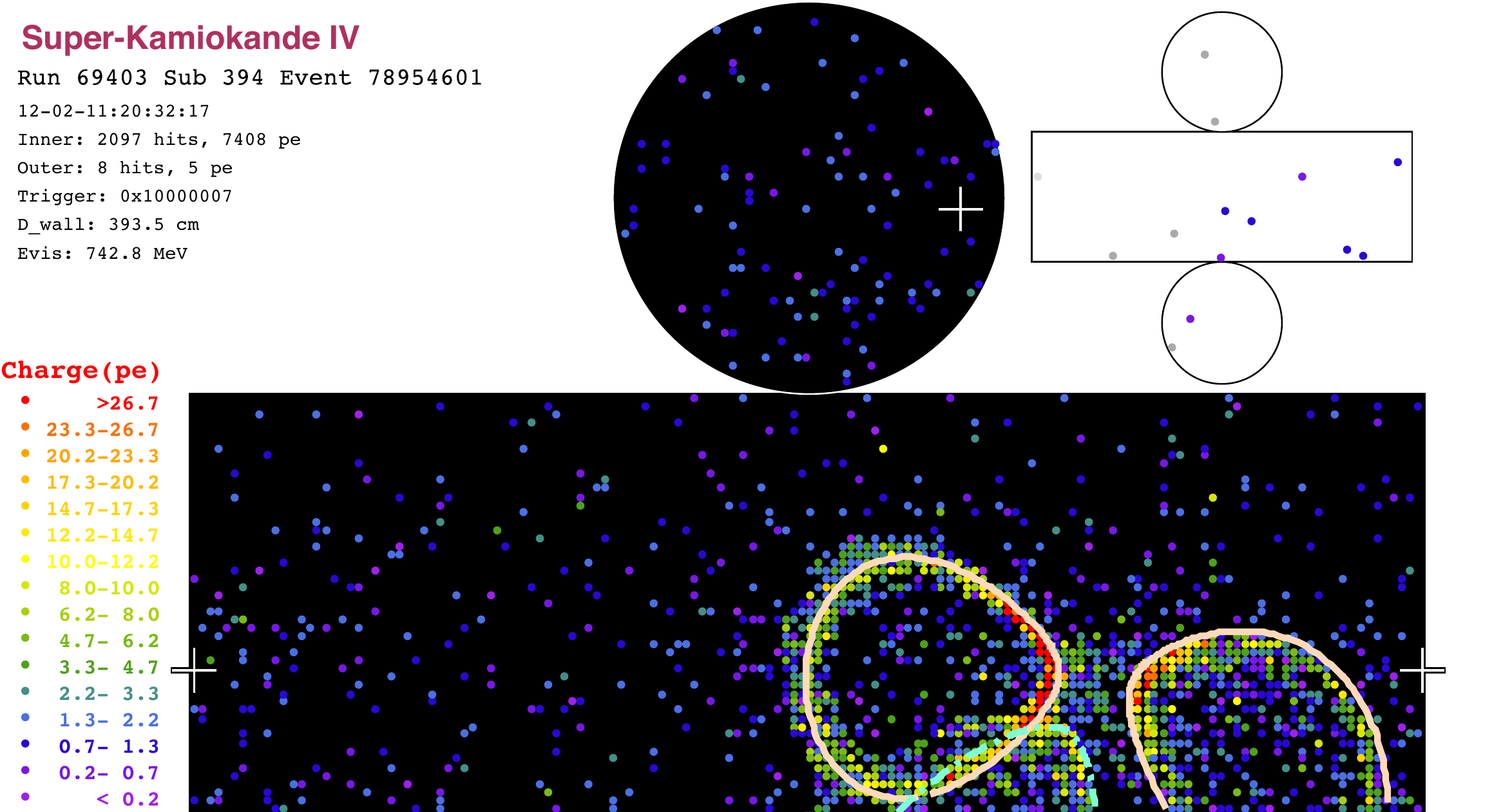}}
\linebreak
\fbox{\hspace{1.5em}\includegraphics[trim ={100mm 3mm 88mm 132mm},clip=true, width=22mm]{sk4_candidate_69403.pdf}}

\caption{$pp\rightarrow \pi^{+}\pi^{+}$ candidate event in SK-IV. The two $\mu$-like rings are at an angle of 151 degrees to each other, with momenta of 541.4 and 389.5 MeV/$c$. The dashed $e$-like ring could be due to a hard scatter. There are 2 Michel electrons, not shown. The visible energy is 742.8 MeV, the total momentum is 445.8 MeV/$c$, and the invariant mass is 1022.3 MeV/$c^2$.}
\label{figure:pp_cand_sk4}
\end{figure}

\par The largest component ($\sim$40\%) of the remaining atmospheric background is from charged current single pion production (CC1$\pi$). In this mode, typically a $\nu_{\mu}$ will produce a muon and a charged pion at a large angle, thus passing the $\pi^{+}$ candidate angle requirement. Other contributing background interaction modes include charged-current deep inelastic scattering (CCDIS), neutral-current deep inelastic scattering (NCDIS), charged-current quasi-elastic (CCQE) and neutral-current single pion production (NC1$\pi$). Table~\ref{table:pp_bkg} quantifies the neutrino interaction modes of the remaining background. The percentages agree within statistical fluctuations across SK periods.

\par A $pp \rightarrow \pi^{+}\pi^{+}$ MC event is shown in Fig.~\ref{figure:pp_mc_disp}. An atmospheric neutrino event surviving all selection criteria, including the final BDT cut, is shown in Fig.~\ref{figure:pp_atm_surv}. Two candidate events were found, one in SK-II (Fig.~\ref{figure:pp_cand_sk2}) and another in SK-IV (Fig.~\ref{figure:pp_cand_sk4}). No events are observed in SK-I or SK-III. Table~\ref{table:pp_var_breakdown} shows values of the BDT input variables for each candidate, compared to the means of the signal and background distributions. While both candidate events appear signal-like in some variables, particularly the angle between $\mu$-like rings (both candidates) and energy variables (SK-IV candidate), the candidates are consistent with the expected background rate for each SK period (Table~\ref{table:pp_stats}), as well as the total background rate of 4.5 expected across SKI-IV. Thus, we conclude that there is no evidence for $pp \rightarrow \pi^{+}\pi^{+}$ in the data.


\subsection*{$\mathbf{pn}\boldsymbol{\rightarrow}\boldsymbol{\pi^{+}}\boldsymbol{\pi^{0}}$}

There are many similarities between the $pn \rightarrow \pi^{+}\pi^{0}$ search and the $pp\rightarrow\pi^{+}\pi^{+}$ search. The main difference is that the $\pi^{0}$ provides a new set of discriminatory variables, and some of the multivariate input variables only apply to the reconstructed $\pi^{0}$. 
\par First, we apply a set of selection criteria:

\begin{enumerate}
\item[(B1)] There is more than one Cherenkov ring.
\item[(B2)] There is at least one $e$-like and one $\mu$-like ring. The $\pi^{+}$ is assumed to correspond to the most energetic $\mu$-like ring, while the $\pi^{0}$ is assumed to correspond to the $e$-like ring (if one $e$-like ring is found) or the two most energetic $e$-like rings (if more than one is found).
\item[(B3)] The number of Michel electrons is no more than one.
\item[(B4)] The ``reduced" visible energy, defined as the total visible energy minus the energy of the $\pi^{0}$ as defined in the previous criterion, is less than 800 MeV. This is similar to criterion (A4) used in the $pp\rightarrow\pi^{+}\pi^{+}$ search - the maximum visible energy from the $\pi^{+}$ is $M_{proton}-M_{\pi^{+}}\approx 800$ MeV.
\item[(B5)] The angle between the reconstructed $\pi^{+}$ and $\pi^{0}$ is greater than 120 degrees.
\end{enumerate}

\begin{table}[h!]
\setlength{\tabcolsep}{4pt}        
\begin{center}
\begin{tabular}{ c | r r r r} \hline \hline
& SK-I & SK-II & SK-III & SK-IV \\ \hline
Eff. (\%) & 21.0$\pm$0.3 & 21.9$\pm$0.3 & 21.6$\pm$0.3 & 21.1$\pm$0.3 \\
Bkg. & 132$\pm$1.8 & 69$\pm$1.0 & 48$\pm$0.6  & 147$\pm$2.0 \\
data & 136 & 66 & 45 & 171 \\

\hline \hline
\end{tabular}
\caption {Efficiencies, expected backgrounds, and data events after selection criteria in the $pn\rightarrow\pi^{+}\pi^{0}$ search. Background rates are scaled to the appropriate SK period livetime. Statistical errors are shown. }
\label {pn_precut_stats}
\end{center}
\end{table}

\par The signal efficiency, expected background, and data events remaining after the selection criteria is shown in Table~\ref{pn_precut_stats}. About a third of the primary $\pi^{0}$'s in the $pn\rightarrow\pi^{+}\pi^{0}$ events surviving the selection criteria are reconstructed as a single $e$-like ring. This is mainly because the typically high momentum of the $\pi^{0}$'s leads to a small opening angle between the photons in the subsequent $\pi^{0}\rightarrow\gamma\gamma$ decay, so that the rings are not distinct.
\par Events surviving the selection criteria are passed into a BDT. The following set of input variables is used, ordered according to variable importance:

\begin{enumerate}
\item[(b1)] The momentum of the $\pi^{0}$ candidate. The $\pi^{0}$ momentum peaks near 1 GeV for the signal, which is typically well-reconstructed. The background peaks at a much lower value of $\sim 200$ MeV.
\item[(b2)] The angle between the reconstructed $\pi^{+}$ and $\pi^{0}$. This peaks near 180 degrees for the signal.
\item[(b3)] The momentum of the $\pi^{+}$ candidate. The momentum resolution for charged pions from dinucleon decay is typically poor. Despite this, it still peaks at a higher value in the signal than the background (470 MeV/$c$ vs. 390 MeV/$c$), and is relatively independent of the better-reconstructed $\pi^{0}$ momentum.
\item[(b4)] The invariant mass of the $\pi^{0}$. This can be estimated for both one and two-ring $\pi^{0}$ hypotheses, by means of the specialized algorithm referred to in Section~\ref{sec:redrec}.
\item[(b5)] \label{itm:fifth} The ratio of charge carried by the highest-energy ring to the total charge of all rings.
\item[(b6)] \label{itm:sixth} The total visible energy.
\item[(b7)] \label{itm:seventh} The number of Michel electrons.
\end{enumerate}

\begin{figure*}[t!]
\subfigure{\includegraphics[width=80mm]{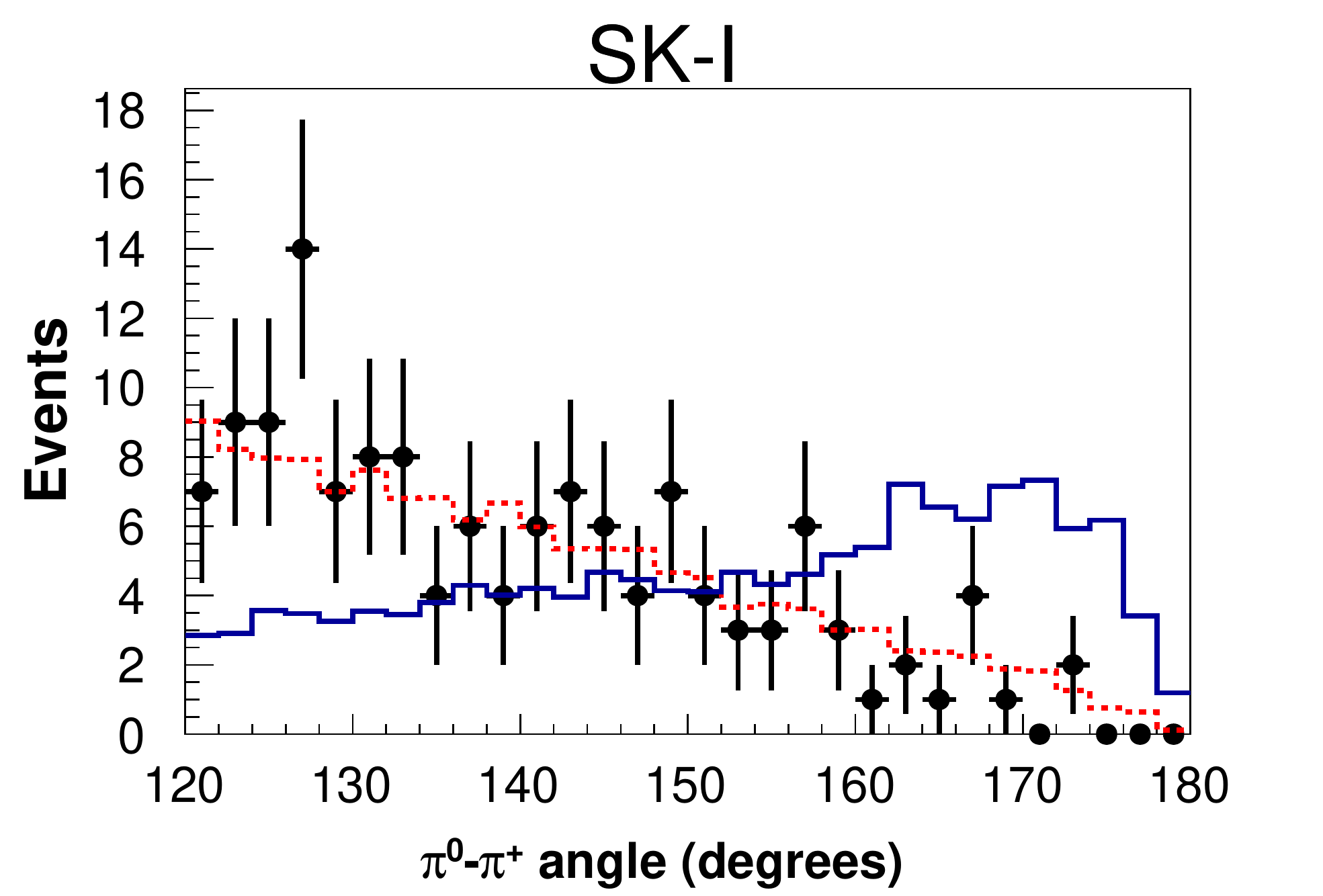}}
\subfigure{\includegraphics[width=80mm]{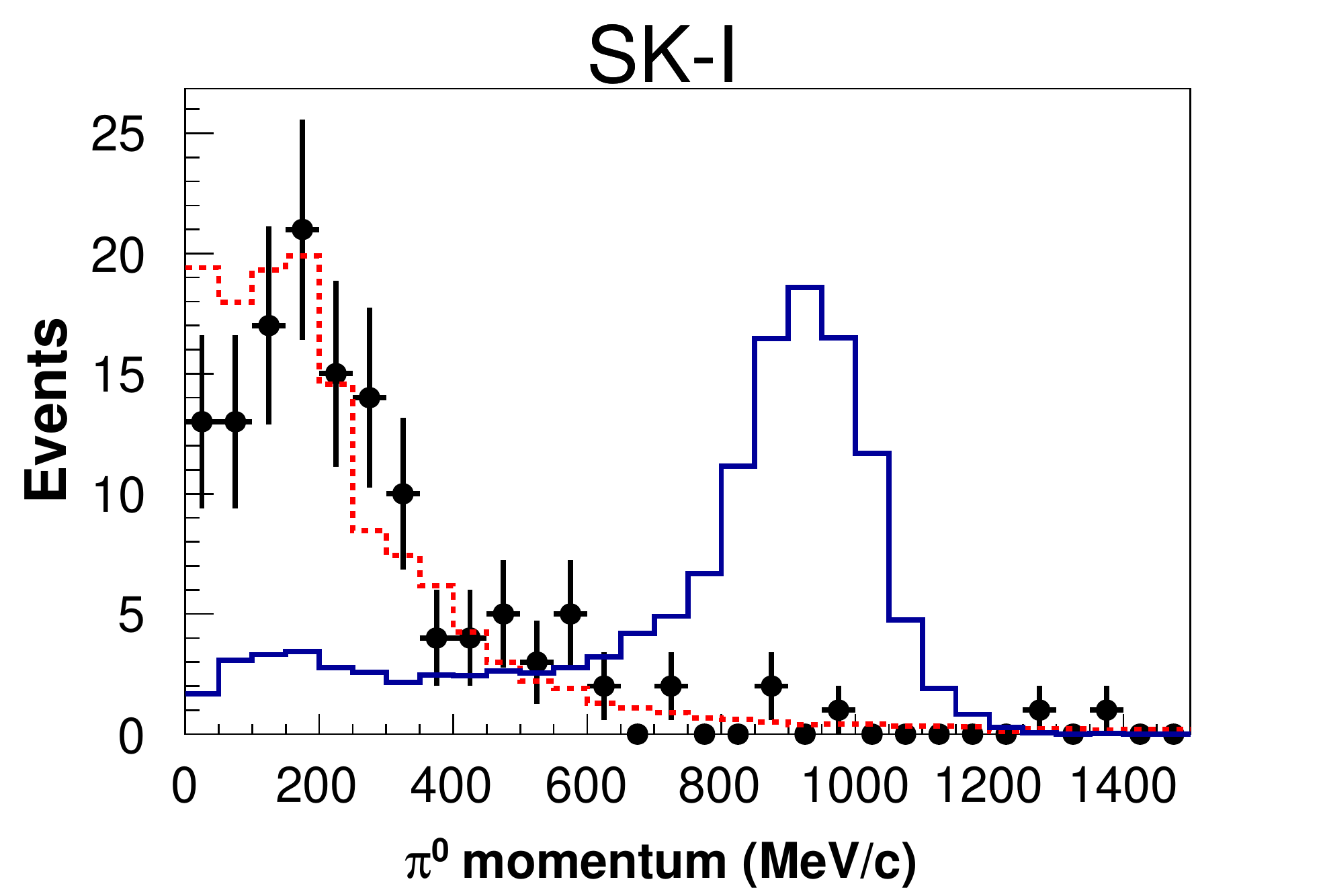}}
\subfigure{\includegraphics[width=80mm]{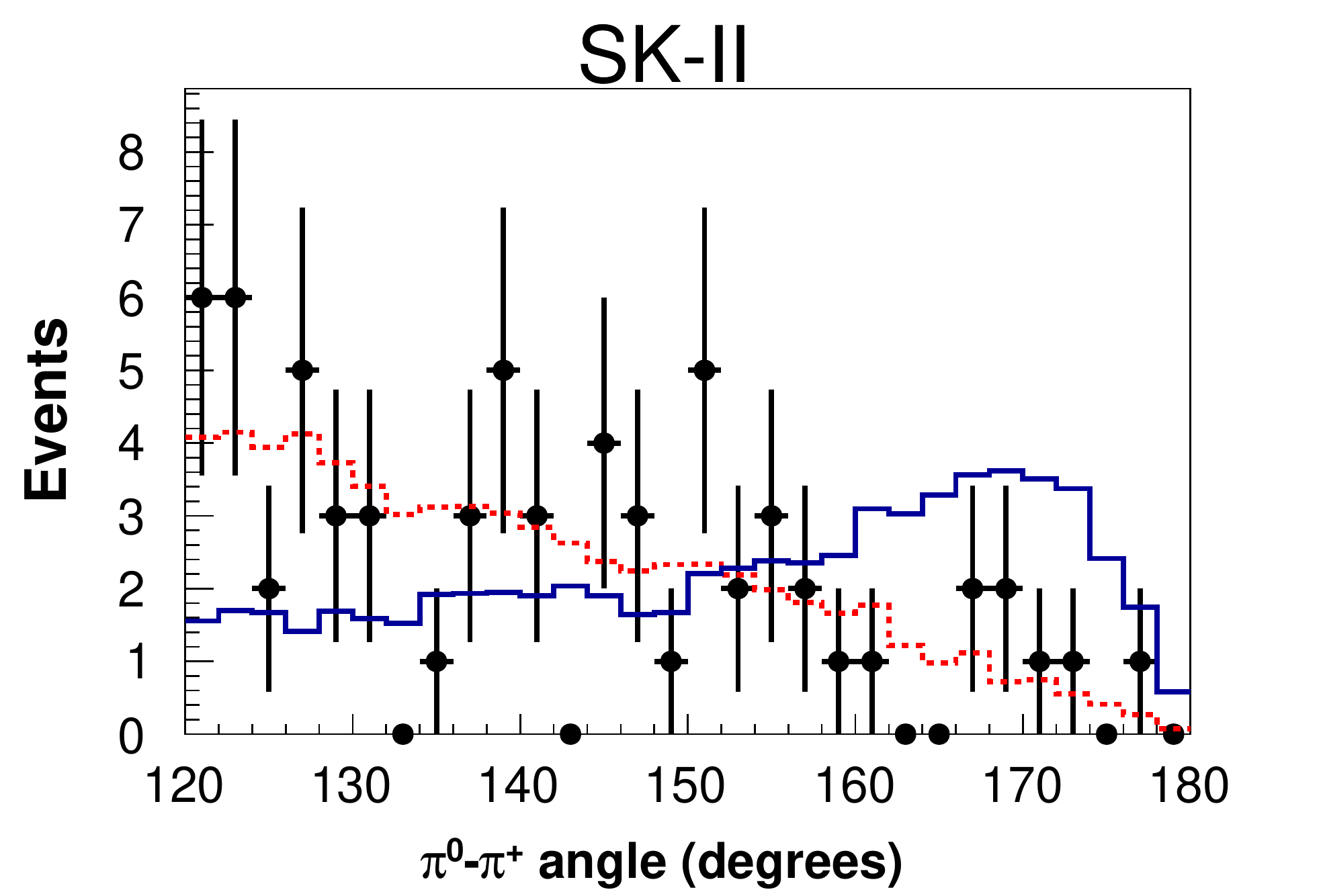}}
\subfigure{\includegraphics[width=80mm]{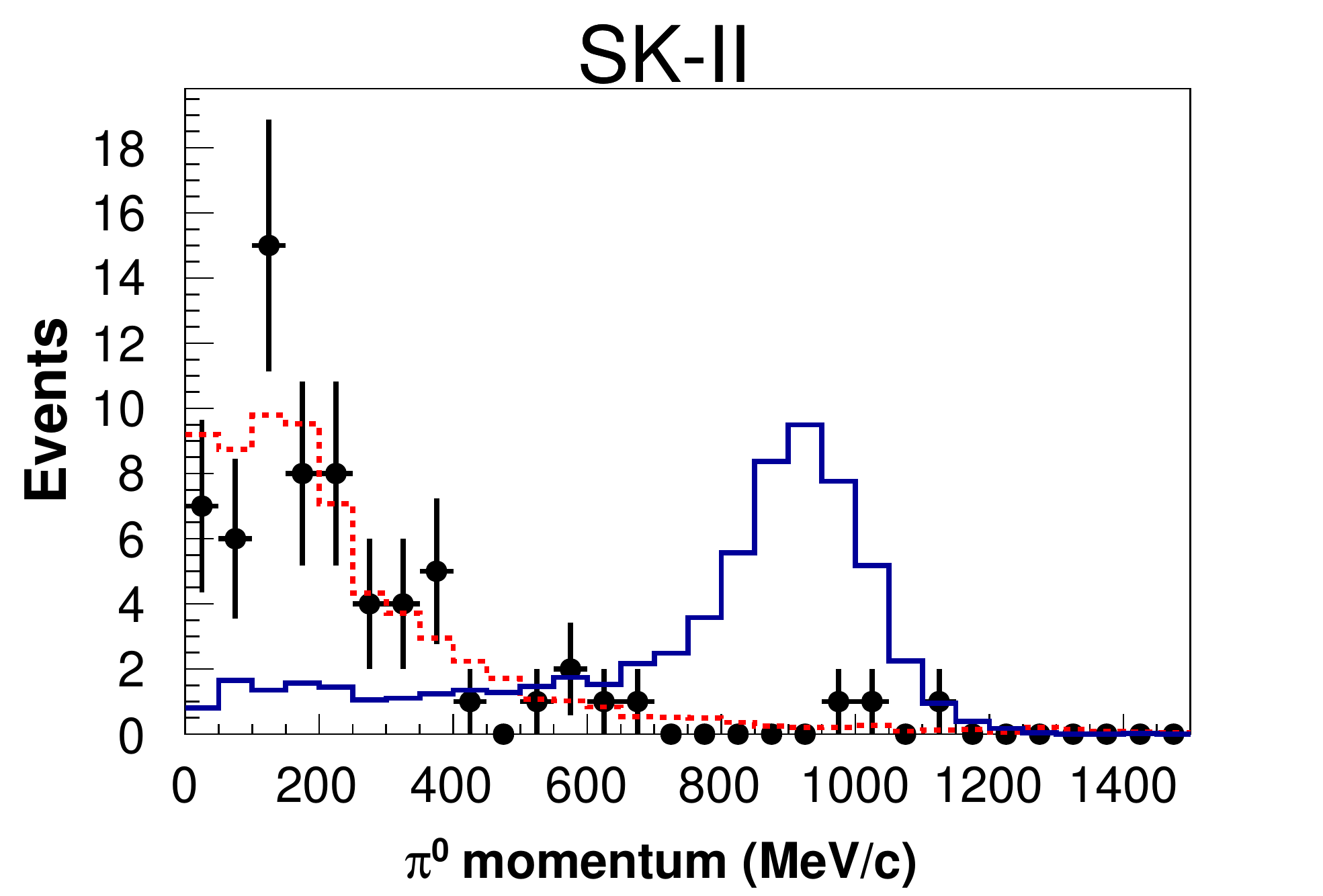}}
\subfigure{\includegraphics[width=80mm]{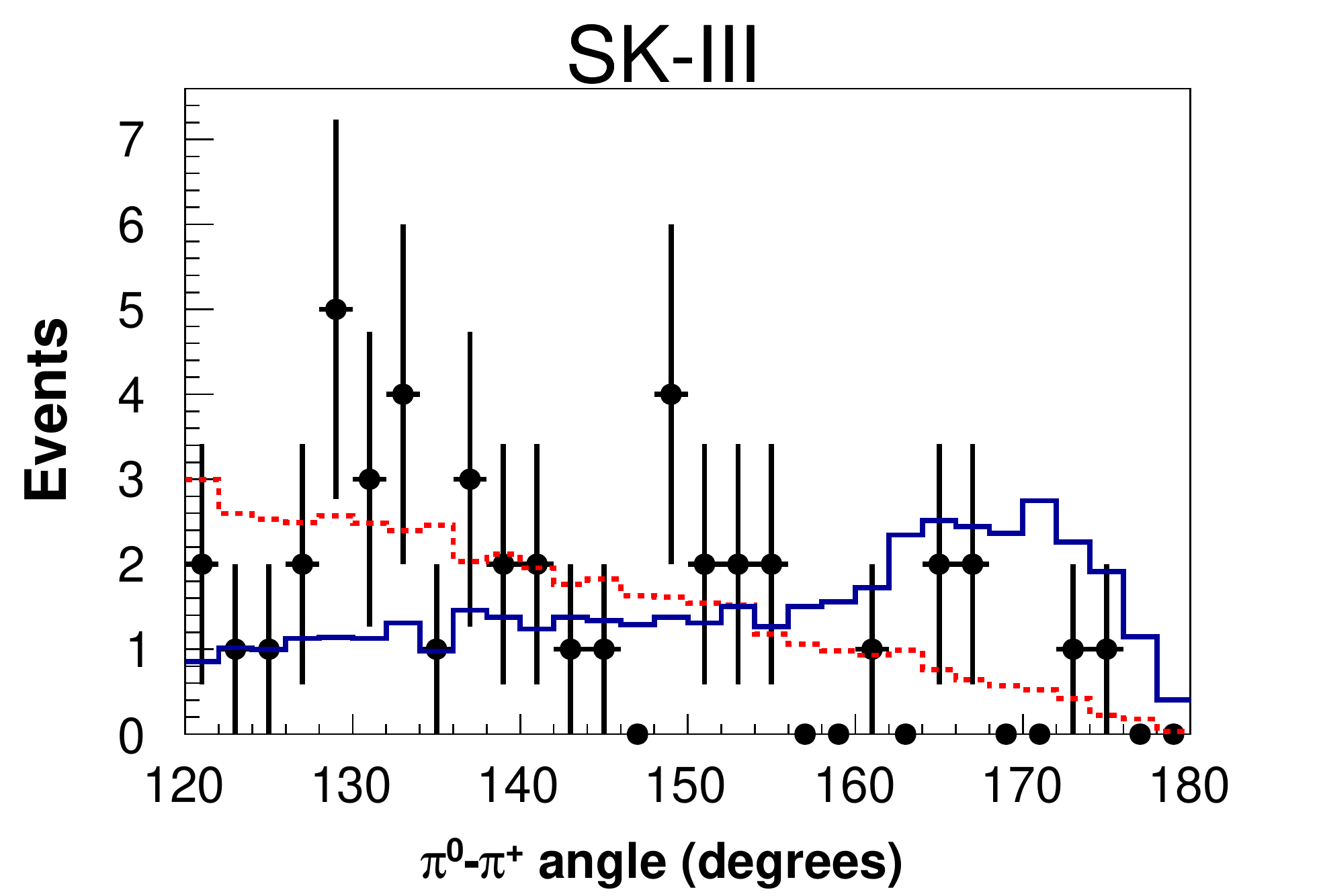}}
\subfigure{\includegraphics[width=80mm]{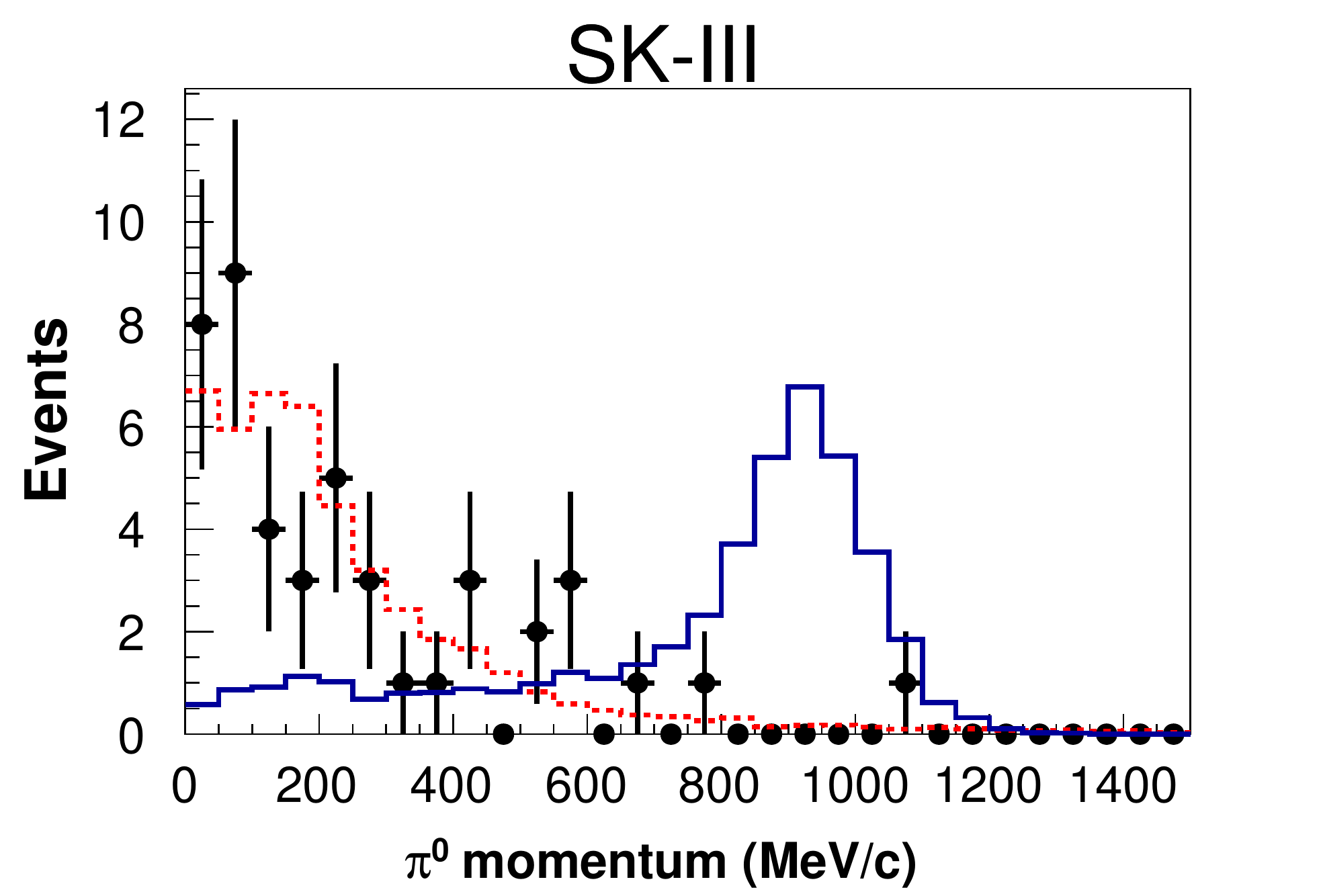}}
\subfigure{\includegraphics[width=80mm]{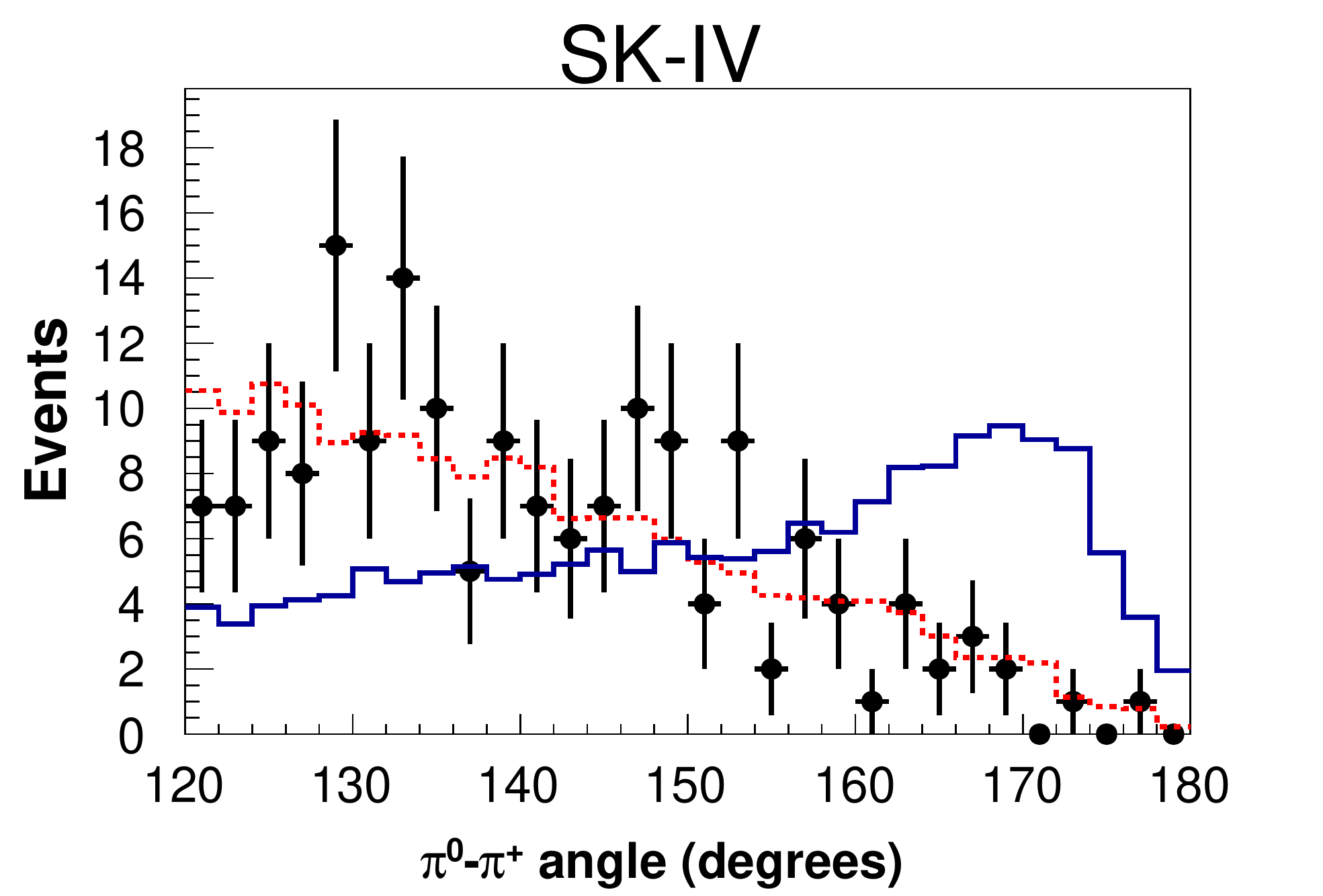}}
\subfigure{\includegraphics[width=80mm]{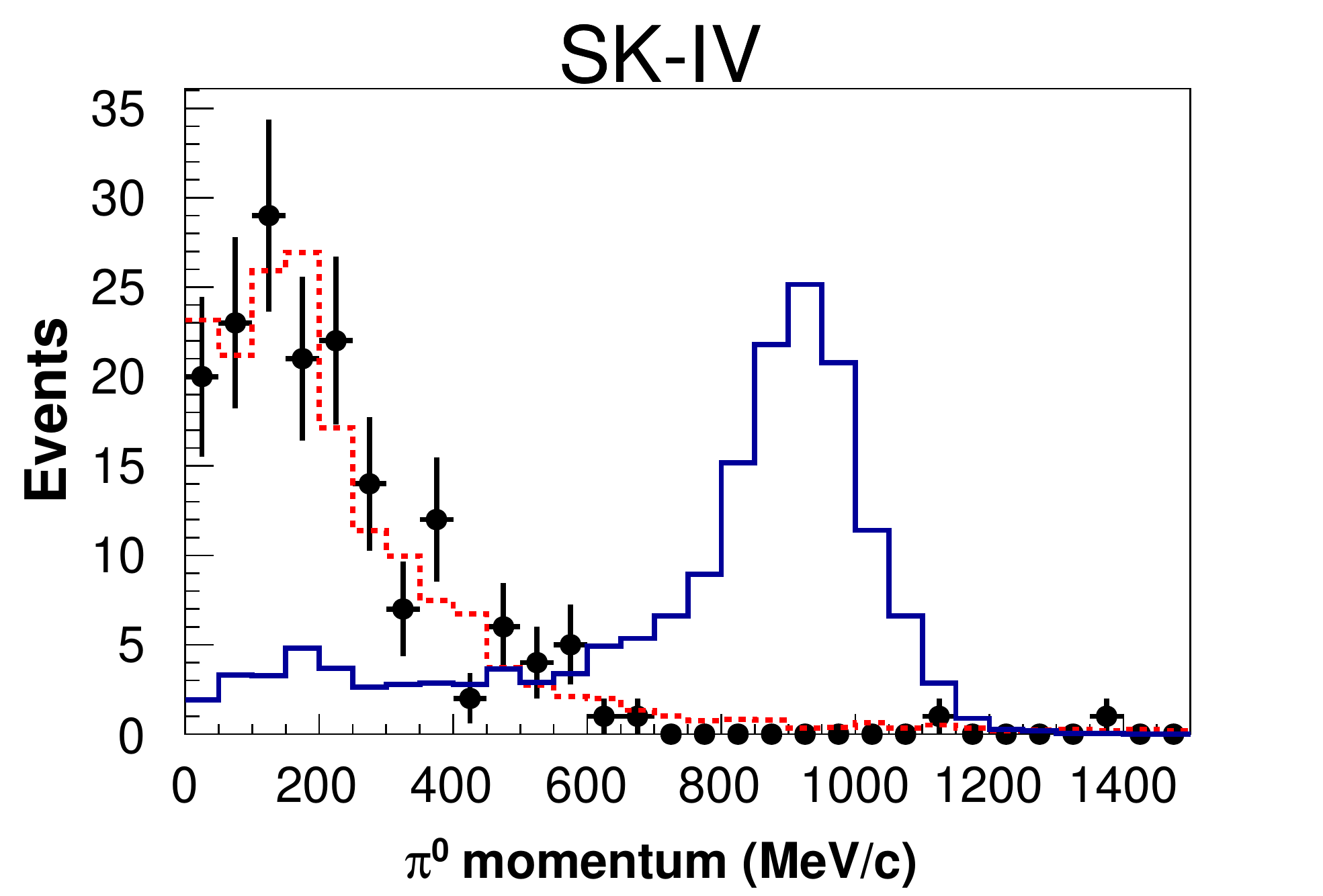}}
\caption{The $\pi^{0}$-$\pi^{+}$ angle and $\pi^{0}$ momentum for the $pn\rightarrow\pi^{+}\pi^{0}$ search. Signal and atmpospheric neutrino MC (solid and dashed histograms) are normalized to the number of events in the data (crosses) for each SK period. Details of each variable are described in the text.}
\label{figure:pn_bestvars}
\end{figure*}

Variables (b5), (b6), and (b7) were described for the $pp\rightarrow\pi^{+}\pi^{+}$ search.
The ranking of these variables is shown in Table~\ref{pn_ranking}. The two most discriminating input variables are shown in Fig.~\ref{figure:pn_bestvars}. The $\pi^{0}$-$\pi^{+}$ angle is highly discriminating for the same reason as the angle between $\pi^{+}$ candidates in the $pp\rightarrow\pi^{+}\pi^{+}$ search. The $\pi^{0}$ momentum is also a highly discriminating variable, since the peak near 1 GeV/$c$ for the signal $\pi^{0}$ momentum, expected for back-to-back pions emitted with initial energy $\sim$ 2 GeV, is absent in the background. The data and atmospheric neutrino MC are in good agreement.

\begin{table}
\setlength{\tabcolsep}{6pt}        
  \begin{center}
    \begin{tabular}{lr}
      \hline \hline
      Variable &  Importance \\
      \hline 
      $\pi^{0}$ candidate momentum & 0.19\\
      \specialcell{Angle between $\pi^{0}$ and \\ $\pi^{+}$ candidates} & 0.17\\
      $\pi^{+}$ candidate momentum & 0.16\\
      $\pi^{0}$ candidate invariant mass & 0.15\\       
      \specialcell{Ratio of charge carried by \\ most energetic ring} & 0.14 \\
      Visible energy & 0.14\\
      Number of Michel electrons & 0.058 \\
      
      \hline \hline
    \end{tabular}
  \caption{\protect \small 
Relative importance of each variable in the  \\ $pn\rightarrow \pi^{+}\pi^{0}$ search, averaged across SK periods.
}
  \label{pn_ranking}
  \end{center}
\end{table}

\begin{figure*}[t!]

\subfigure{\includegraphics[width=80mm]{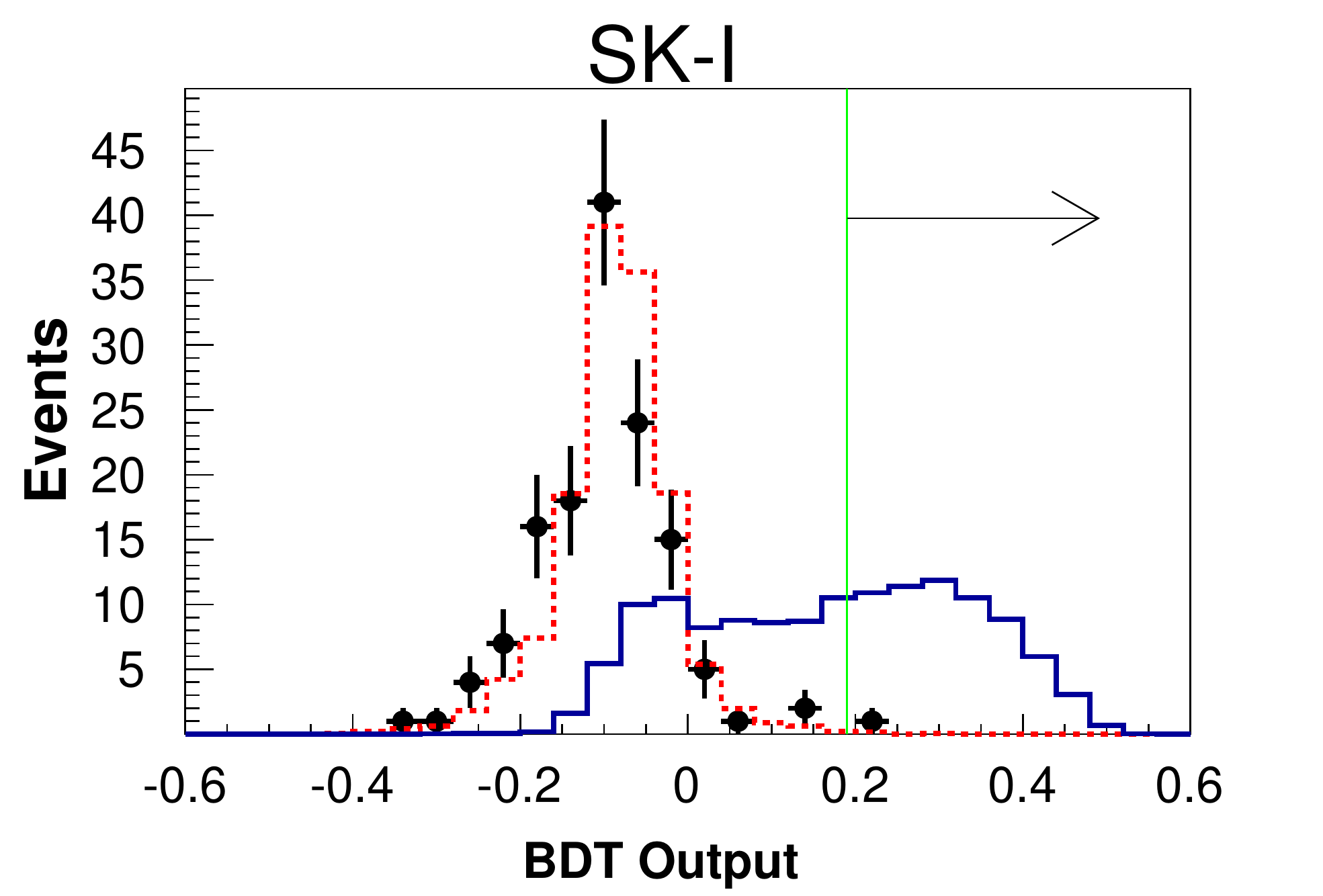}}
\subfigure{\includegraphics[width=80mm]{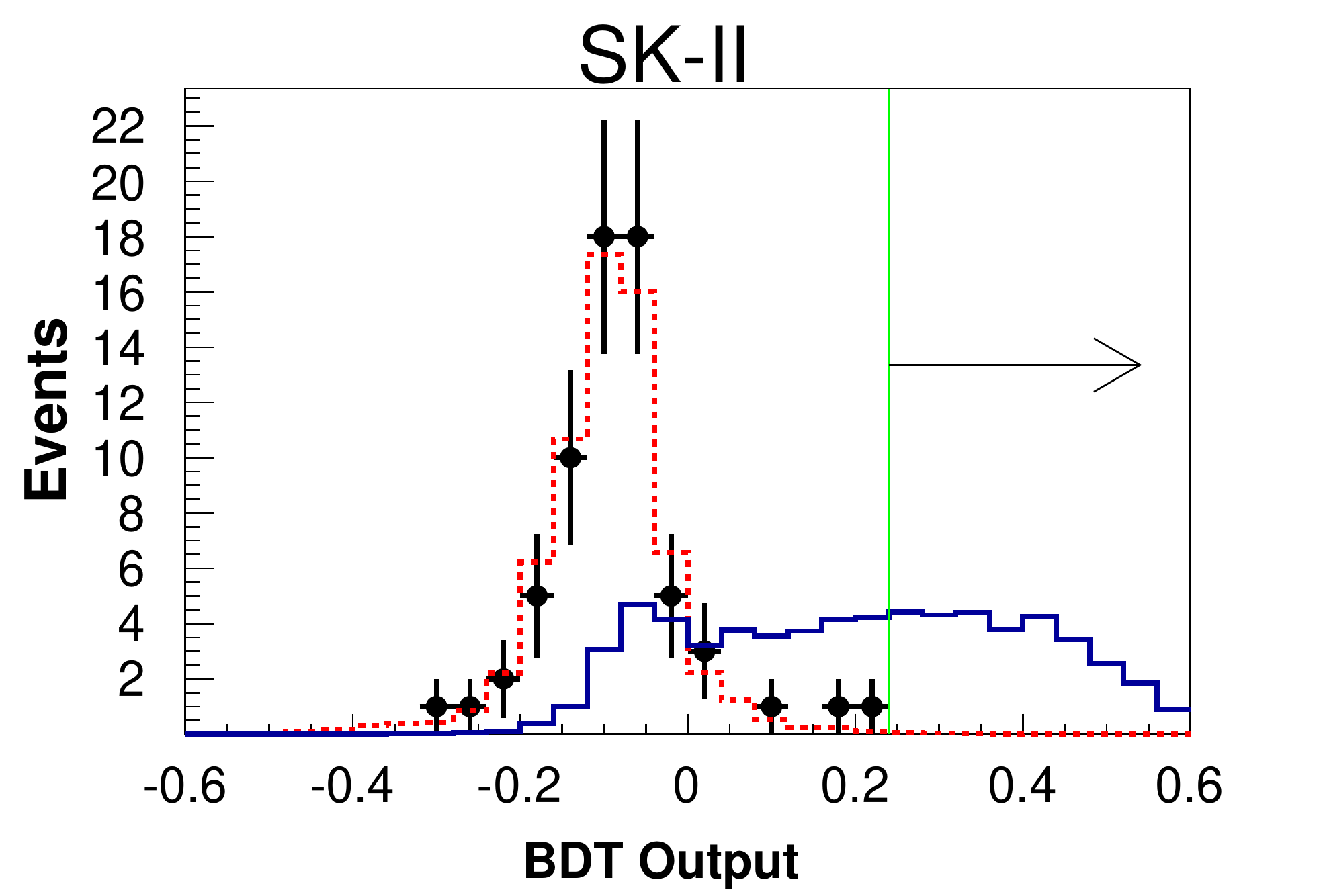}}
\subfigure{\includegraphics[width=80mm]{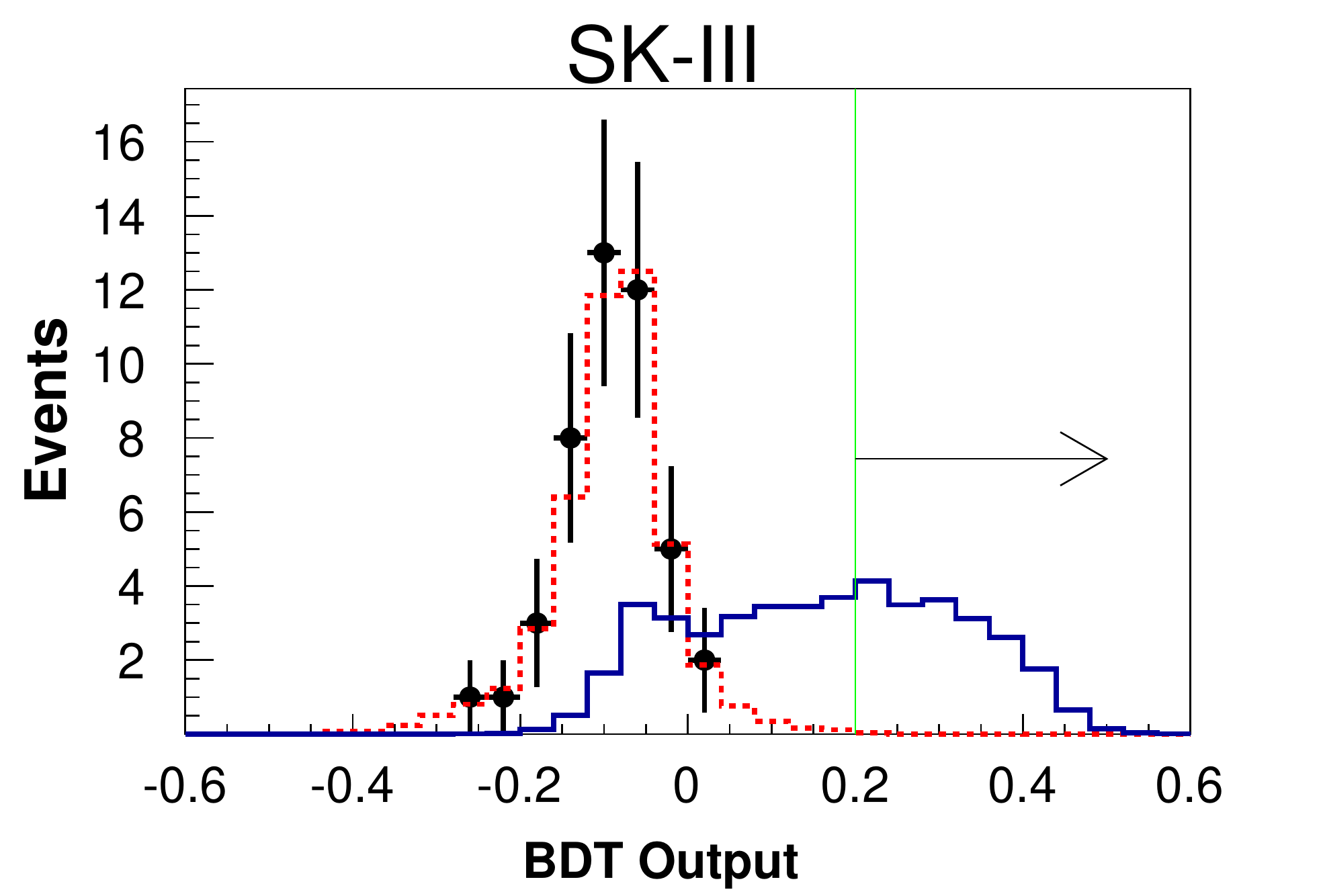}}
\subfigure{\includegraphics[width=80mm]{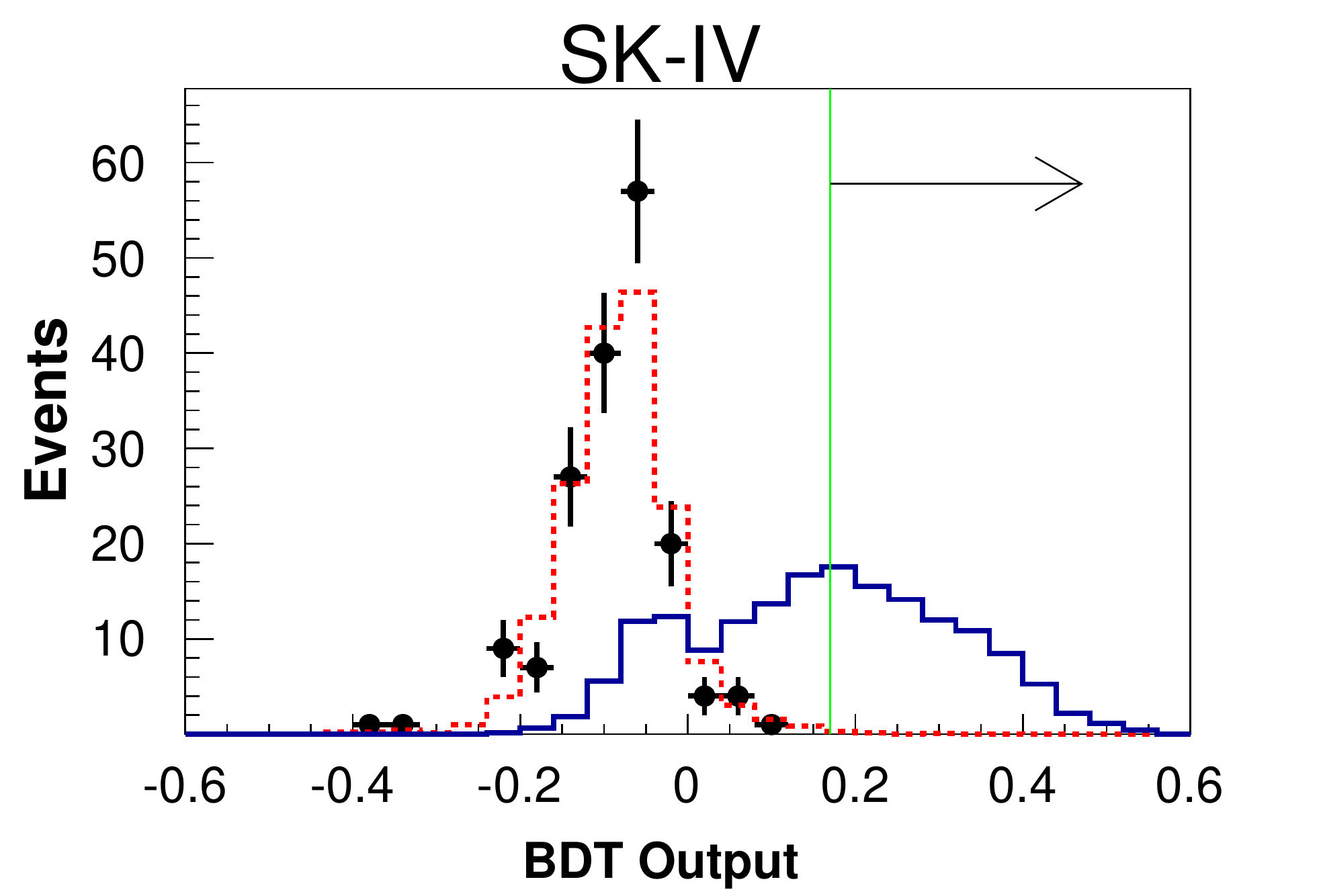}}

\caption{BDT output for $pn\rightarrow\pi^{+}\pi^{0}$ signal MC (solid), atmospheric neutrino background MC (dashed), and data (crosses). The green vertical line indicates the BDT cut value, and the arrow indicates that only events to the right of the cut are kept. One candidate event can be seen in the SK-I data distribution, just to the right of the cut at 0.19.}

\label{figure:BDT}
\end{figure*}

\begin{figure}[t]
\centering

\setlength\fboxsep{0pt}
\setlength\fboxrule{0pt}

\fbox{\includegraphics[trim ={7mm 65mm 10mm 3mm},clip=true, width=80mm]{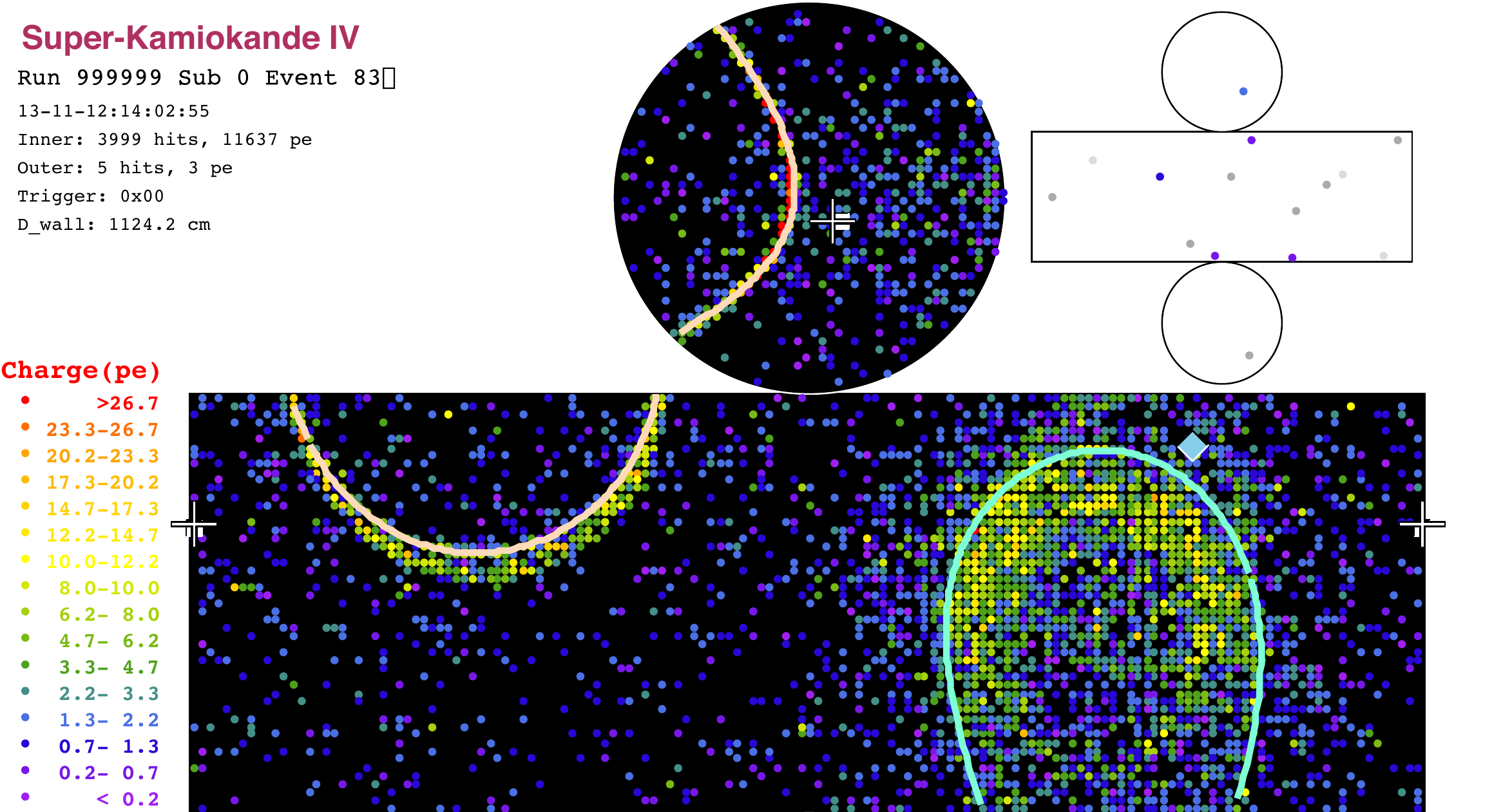}}
\linebreak
\fbox{\hspace{1.5em}\includegraphics[trim ={100mm 3mm 88mm 132mm},clip=true, width=22mm]{sk4_pn_pippi0_file1_ev83_notru.pdf}}

\caption{A $pn\rightarrow \pi^{+}\pi^{0}$ MC event. The $\mu$-like ring corresponds to a true $\pi^{+}$. The $e$-like ring contains two overlapping $\gamma$'s from a $\pi^{0}$. The angle between the $\mu$- and $e$-like rings is 165 degrees. The $e$-like ring has a reconstructed momentum of 893 MeV/$c$, while the original $\pi^{0}$ had momentum 876 MeV/$c$.}
\label{figure:pn_MC_disp}
\end{figure}

\begin{figure}[h!]
\centering

\setlength\fboxsep{0pt}
\setlength\fboxrule{0pt}

\fbox{\includegraphics[trim ={7mm 65mm 10mm 3mm},clip=true, width=80mm]{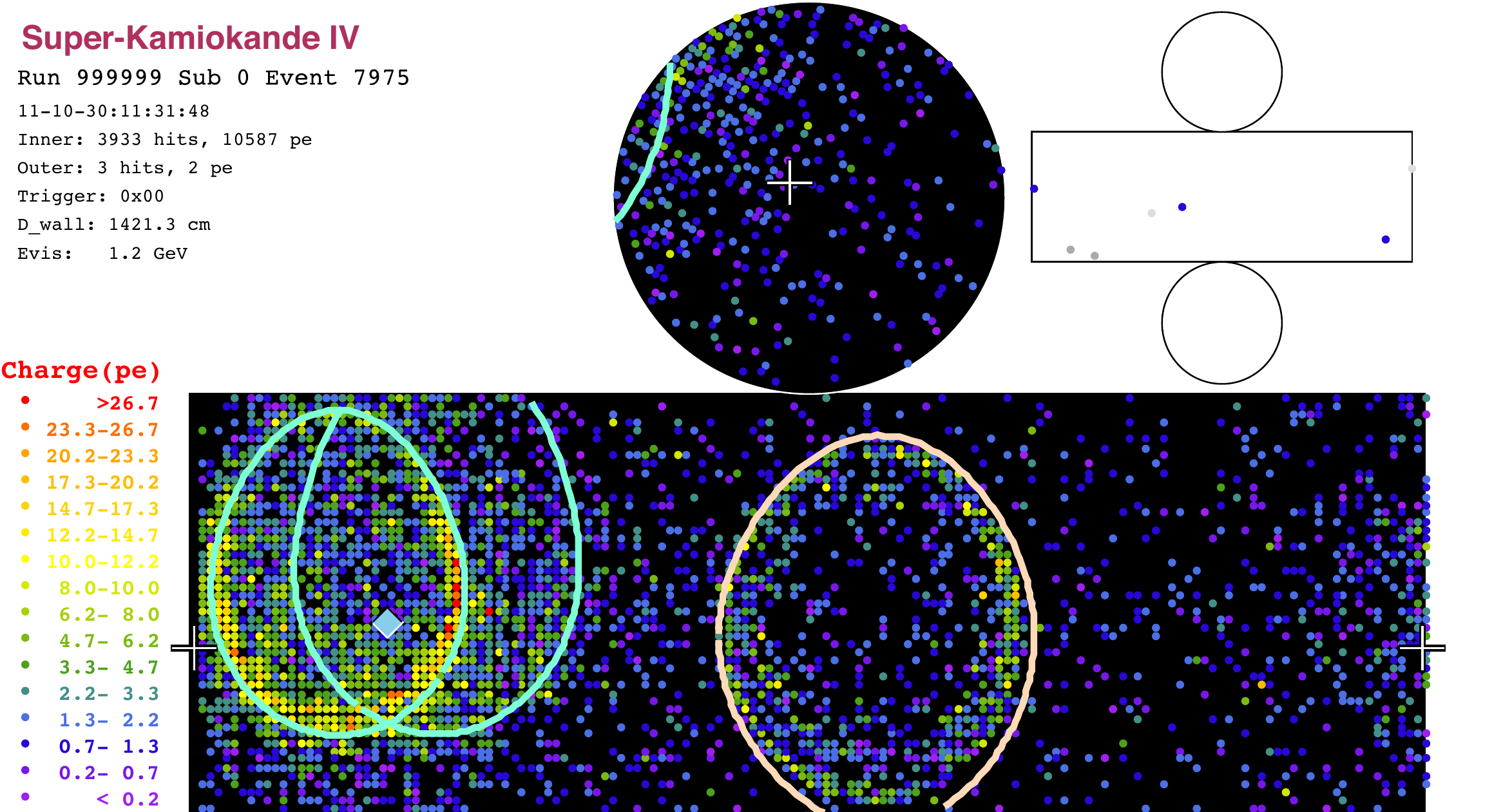}}
\linebreak
\fbox{\hspace{1.5em}\includegraphics[trim ={100mm 3mm 88mm 132mm},clip=true, width=22mm]{pn_pippi0_bkg_sk4_file499_ev7975.pdf}}

\caption{ A surviving atmospheric neutrino MC event for the $pn\rightarrow\pi^{+}\pi^{0}$ search. The interaction type is CC$1\pi^{0}$, $\nu_{\mu}n \rightarrow \mu^{-}\pi^{0}p$. The two overlapping $e$-like rings are photons from a $\pi^{0}$, with total momentum 939 MeV/$c$. The $\mu$-like ring corresponds to the $\mu^{-}$, with momentum 410 MeV/$c$. The angle between them is 151 degrees, thereby appearing back-to-back.}
\label{figure:atm_surv_pn}
\end{figure}

\begin{figure}[h!]
\centering

\setlength\fboxsep{0pt}
\setlength\fboxrule{0pt}

\fbox{\includegraphics[trim ={7mm 65mm 10mm 3mm},clip=true, width=80mm]{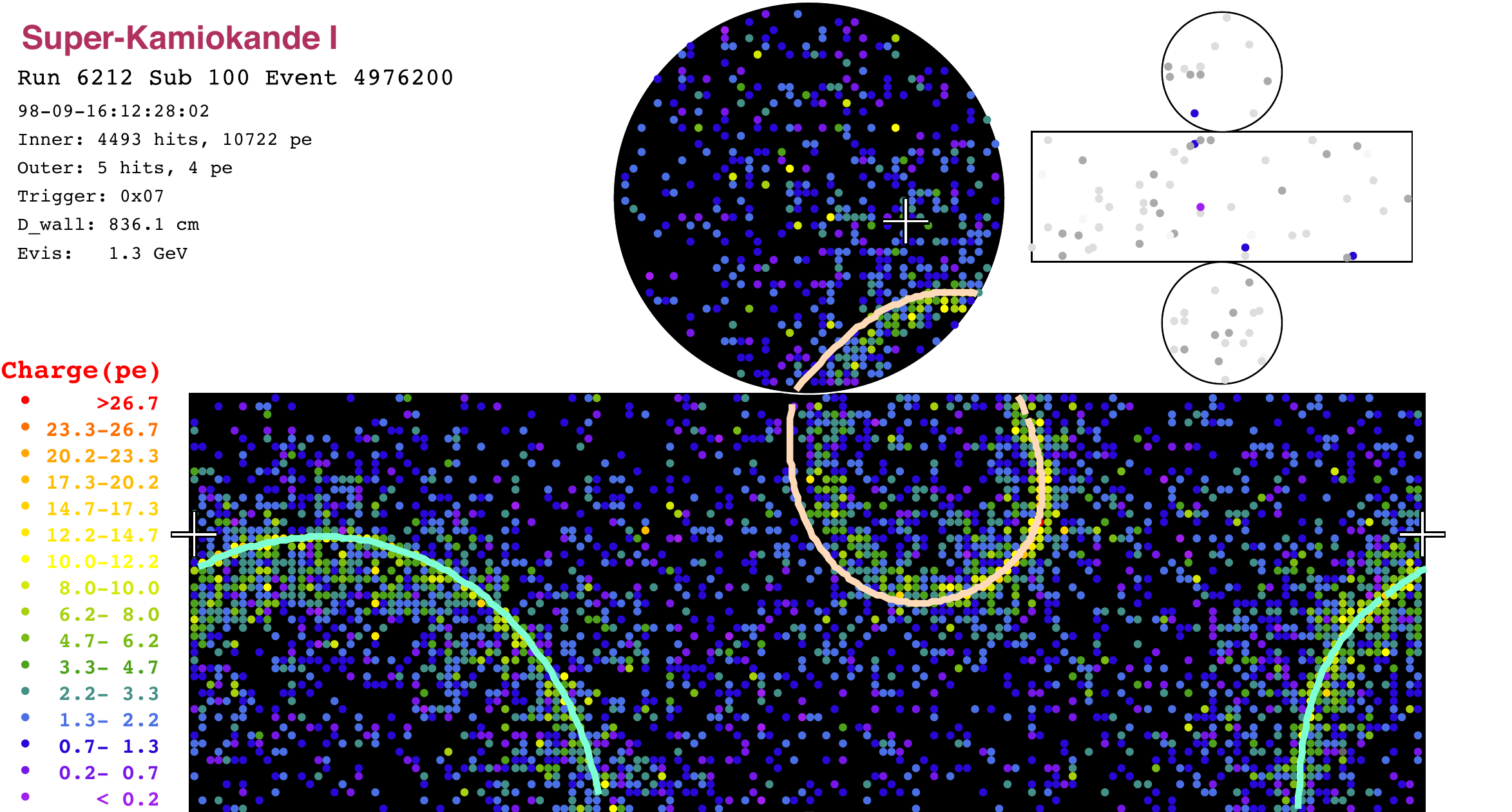}}
\linebreak
\fbox{\hspace{1.5em}\includegraphics[trim ={100mm 3mm 88mm 132mm},clip=true, width=22mm]{sk1_pn_pippi0_candidate.pdf}}

\caption{The $pn\rightarrow\pi^{+}\pi^{0}$ candidate in SK-I. The $e$-like and $\mu$-like rings are at an angle of 140 degrees. The $e$-like ring has a high momentum of 987 MeV/$c$, but its fit $\pi^{0}$ mass is low, at 10 MeV/$c^2$. The $\mu$-like ring has a momentum of 460.0 MeV/$c$. The total momentum and invariant mass are 700.1 MeV/$c$ and 1266.4 MeV/$c^2$, respectively.}
\label{figure:pn_cand_sk1}
\end{figure}

\begin{table}[h]
\setlength{\tabcolsep}{4pt}
\begin{center}
\begin{tabular}{c | r r r r } \hline\hline
  & SK-I & SK-II & SK-III & SK-IV\\ \hline
Cut & 0.19 & 0.24 & 0.20 & 0.17\\ 
Eff. (\%) & 10.2$\pm$0.2 & 10.0$\pm$0.2 & 9.4$\pm$0.2 & 10.4$\pm$ 0.2\\
Bkg. (MT-yr) & 2.7$\pm$0.7 & 2.3$\pm$0.7 & 2.2$\pm$0.7 & 2.9$\pm$0.8 \\
Bkg. (SK live.) & 0.25 & 0.11 & 0.07 & 0.32 \\
Candidates & 1 & 0 & 0 & 0\\ 

\hline \hline
\end{tabular}
\end{center}
\caption{Final BDT cut value, efficiency, expected background, and candidate data events for the SKI-IV $pn\rightarrow\pi^{+}\pi^{0}$ search, with statistical uncertainties shown for efficiency and expected background. Background rates are quoted for the appropriate SK livetime, and per megaton-year.} 
\label{table:pn_stats}
\end{table}

\begin{table}[h!]
\setlength{\tabcolsep}{6pt}
\begin{center}        

\begin{tabular}{r | r r r r} \hline\hline

mode & SK-I & SK-II & SK-III & SK-IV\\ \hline

CC1$\pi$  & 33$\pm$17\% & 29$\pm$17\% & 43$\pm$19\% & 45$\pm$18\%\\ 
CCDIS  & 38$\pm$17\% & 46$\pm$19\% & 31$\pm$18\% & 40$\pm$16\%\\
NCDIS  & 8$\pm$8\% & 20$\pm$14\% & 21$\pm$15\% & 8$\pm$8\%\\ 
CCQE  & 8$\pm$8\% & 5$\pm$5\% & 5$\pm$5\% & 8$\pm$8\%\\ 

\hline \hline
\end{tabular}
\caption{Neutrino interaction mode percentages for remaining background in the $pn\rightarrow \pi^{+}\pi^{0}$ search for SKI-IV. The percentages agree across SK periods within statistical fluctuations, which are large due to the low number of remaining events.}
\label{table:pn_bkg}
\end{center}
\end{table}

\par The final BDT outputs are shown in Fig. \ref{figure:BDT}. For the $pn\rightarrow \pi^{+}\pi^{0}$ search, an adaptive boost method \cite{Hocker:2007ht} was chosen.

\par The final cut on the BDT output was designed so that less than one background event was expected across all SK periods, with contributions from each SK period ordered by their livetime. It is determined with MC only. The total expected background is 0.75 events, consistent with the one candidate event found in SK-I. Unlike the $pp\rightarrow\pi^{+}\pi^{+}$ search, the signal distributions are not sharply peaked, so that the efficiency does not change drastically as the cut value is shifted.
\par The cut on the final BDT output, efficiencies, background rates, and candidates are shown in Table ~\ref{table:pn_stats}. CCDIS and CC1$\pi$ are the main components of the remaining background, detailed in Table~\ref{table:pn_bkg}. 

\begin{table}[h!]
\begin{center}
\begin{tabular}{l | c c c } \hline\hline

Variable & Candidate & Signal & Background  \\ \hline

$\pi^{0}$ momentum (MeV/$c$) & 986.5 & 756.0 & 233.7 \\
$\pi^{0}$-$\pi^{+}$ angle (degrees) & 142.8 & 152.9 & 140.4 \\
$\pi^{+}$ momentum (MeV/$c$) & 459.5 & 470.3 & 394.5 \\
$\pi^{0}$ mass (MeV/$c^2$) & 10.0 & 143.0 & 102.2 \\
Charge ratio (dimensionless) & 0.77 & 0.56 & 0.58 \\
Visible energy (MeV) & 1171.6 & 1098.6 & 495.0 \\
Michel electrons (dimensionless) & 0 & 0.40 & 0.51 \\

\hline \hline
\end{tabular}
\end{center}
\caption{Candidate variable values, and mean variable values for signal and background distributions, for each of the variables used in the BDT for the $pn\rightarrow \pi^{+}\pi^{0}$ search. Letters for each variable are assigned in the text. Values are given for SK-I, for which a single candidate event was found.}
\label{table:pn_var_breakdown}
\end{table}

\par An example $pn\rightarrow\pi^{+}\pi^{0}$ MC event display is shown in Fig.~\ref{figure:pn_MC_disp}, and an example background MC event is shown in Fig.~\ref{figure:atm_surv_pn}. One candidate is found in SK-I, shown in Fig.~\ref{figure:pn_cand_sk1}. Table~\ref{table:pn_var_breakdown} shows the values of the BDT input variables for the candidate event, compared with the means of the signal and background distributions. While the candidate event appears closer to the signal mean values for energy-related variables, it appears more background-like for the $\pi^{0}-\pi^{+}$ angle and the reconstructed $\pi^{0}$ mass. There are no candidate events in SKII-IV data. Although the expected background in SK-I is much lower than one event, it is consistent within about $1\sigma$ with the expected background of 0.75 events in the entire SKI-IV dataset. The probability of observing one or more background event in SK-I, without incorporating systematic uncertainties, is 22\%, and therefore the event is consistent with background expectations. Further, the low $\pi^{0}$ mass of the candidate event, as reconstructed by the specialized algorithm discussed in Section~\ref{sec:redrec}, suggests that it is not a $pn\rightarrow \pi^{+}\pi^{0}$ signal. The $e$-like ring is more likely to be an electron from, for instance, a CC$\nu_{e}$ 1$\pi$ event. Thus, we conclude that there is no evidence for $pn\rightarrow\pi^{+}\pi^{0}$ in the SK-I-IV dataset.


\subsection*{$\mathbf{nn}\boldsymbol{\rightarrow}\boldsymbol{\pi^{0}}\boldsymbol{\pi^{0}}$}

The $nn\rightarrow\pi^{0}\pi^{0}$ search is distinct from the other two searches, in that no multivariate method is used. Similar to other nucleon decay searches at SK~\cite{Nishino:2012ipa, Regis:2012sn, Kobayashi:2005pe}, the total momentum and invariant mass are highly discriminatory variables, so that the search proceeds with a simple set of selection criteria. The criteria are similar to those used for the $p\rightarrow e^{+}\pi^{0}$ analysis~\cite{Nishino:2012ipa}. 
\begin{enumerate}
\item[(C1)] $2\leq$ number of Cherenkov rings $\leq 4$;
\item[(C2)] all rings are $e$-like;
\item[(C3)] no Michel electrons;
\item[(C4)] $P_{tot} \leq$ 600 MeV/$c$;
\item[(C5)] $1600 \: MeV/c^{2} \leq M_{inv} \leq 2000 \: MeV/c^{2}$.
\end{enumerate}


\begin{table}[h!]
\setlength{\tabcolsep}{4pt}
\begin{center}
\begin{tabular}{c | r r r r} \hline\hline
  & SK-I & SK-II & SK-III & SK-IV \\ \hline

 & & & & \\
\textbf{FCFV} & & & & \\
Eff. (\%) & 94.1$\pm$0.7 & 96.6$\pm$0.7 & 94.4$\pm$0.7 & 94.1$\pm$0.7 \\
Bkg. & 11846$\pm$11.1 & 6389$\pm$5.8 & 4187$\pm$3.7 & 14214$\pm$12.7 \\
Data & 12299 & 6610 & 4355 & 14444\\ 

 & & & & \\
\textbf{(C1)} & & & & \\
Eff. (\%) & 76.7$\pm$0.6 & 76.4$\pm$0.6 & 76.6$\pm$0.6 & 76.8$\pm$0.6 \\
Bkg. & 3434$\pm$5.6 & 1846$\pm$3.0 & 1204$\pm$2.0 & 4135$\pm$6.7 \\
Data & 3558 & 1969 & 1239 & 4215\\ 

 & & & & \\
\textbf{(C2)} & & & & \\
Eff. (\%) & 61.0$\pm$0.5 & 59.1$\pm$0.5 & 59.6$\pm$0.5 & 60.1$\pm$0.5 \\
Bkg. & 2131$\pm$4.3 & 1175$\pm$2.3 & 728$\pm$1.5 & 2502$\pm$5.0 \\
Data & 2249 & 1268 & 782 & 2623\\ 

 & & & & \\
\textbf{(C3)} & & & & \\
Eff. (\%) & 54.4$\pm$0.5 & 52.6$\pm$0.5 & 52.8$\pm$0.5 & 52.1$\pm$0.5 \\
Bkg. & 1441$\pm$3.4 & 799$\pm$1.9 & 497$\pm$1.2 & 1543$\pm$1.1 \\
Data & 1492 & 837 & 529 & 1627\\ 

 & & & & \\
\textbf{(C4)} & & & & \\
Eff. (\%) & 30.1$\pm$0.4 & 27.0$\pm$0.4 & 29.1$\pm$0.4 & 29.3$\pm$0.4 \\
Bkg. & 837$\pm$2.6 & 470$\pm$1.4 & 290$\pm$0.9 & 940$\pm$3.0 \\
Data & 853 & 503 & 310 & 1021\\ 

 & & & & \\
\textbf{(C5)} & & & & \\
Eff. (\%) & 22.1$\pm$0.3 & 18.8$\pm$0.3 & 20.9$\pm$ 0.3 & 21.4$\pm$ 0.3\\ 
Bkg. & 0.05$\pm$0.02 & 0.04$\pm$0.01 & 0.03$\pm$0.01 & 0.02$\pm$0.01 \\
Data & 0 & 0 & 0 & 0\\ 


\hline \hline
\end{tabular}
\end{center}
\caption{Efficiency, expected background, and data events for the SKI-IV $nn\rightarrow\pi^{0}\pi^{0}$ search, for each analysis selection criterion. Statistical errors are shown. The criteria are described in the text.} 
\label{Table:effs}
\end{table}

\begin{figure}[h!]
\centering

\setlength\fboxsep{0pt}
\setlength\fboxrule{0pt}

\subfigure{\includegraphics[trim ={7mm 65mm 10mm 3mm},clip=true, width=80mm]{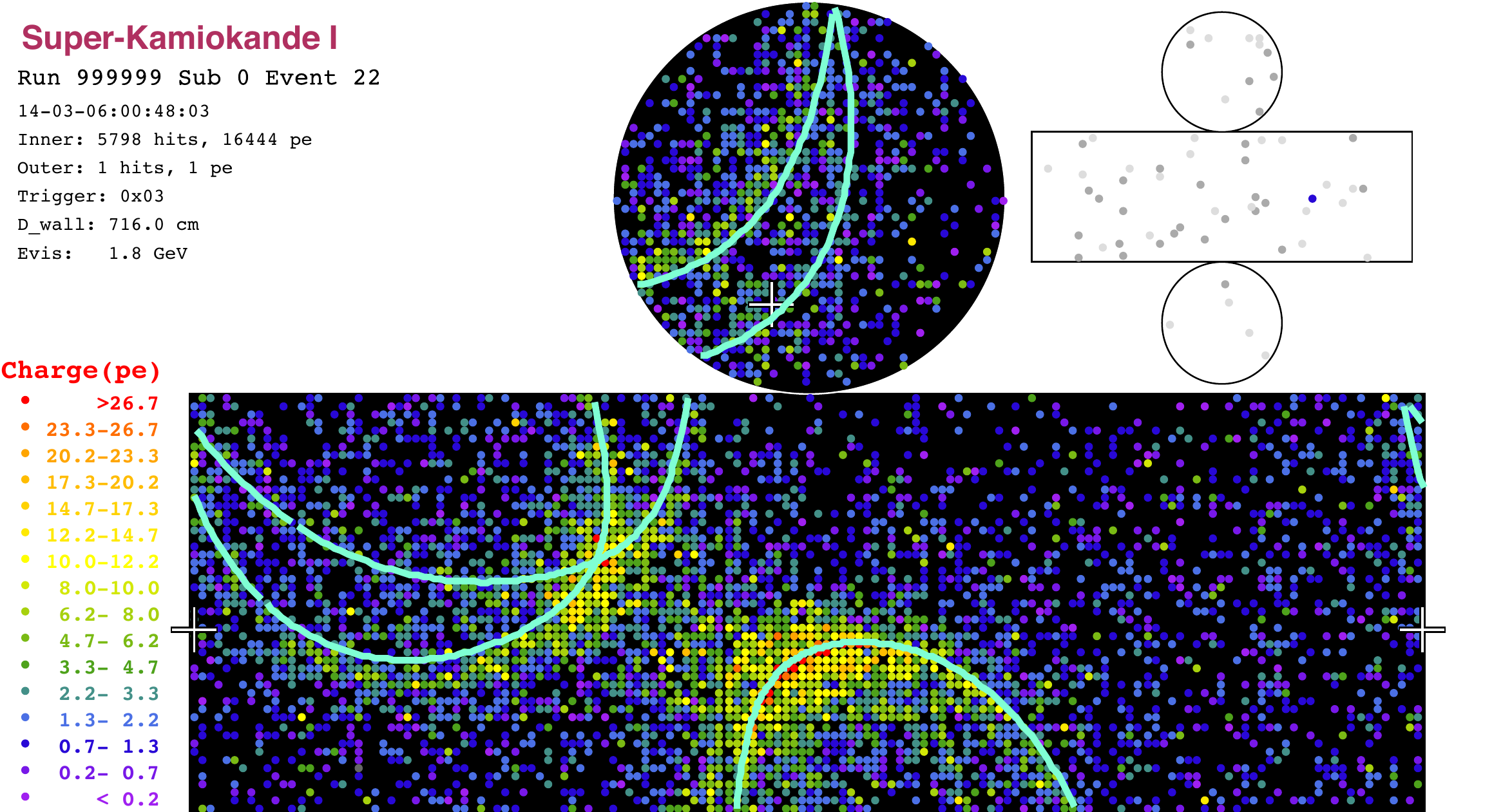}}
\linebreak
\subfigure{\hspace{1.5em}\includegraphics[trim ={100mm 3mm 88mm 132mm},clip=true, width=22mm]{nn_pi0pi0_file1_ev22.pdf}}

\caption{A $nn\rightarrow\pi^{0}\pi^{0}$ MC event. The two overlapping $e$-like rings on the top left correspond to two $\gamma$'s from a $\pi^{0}$ decay, and the other ring also corresponds to a $\pi^{0}$. The true momentum of the $\pi^0$ corresponding to the two-ring fit is 872.7 MeV/$c$, and the true momentum of the $\pi^0$ corresponding to the one-ring fit is 936.7 MeV/$c$. The total reconstructed momentum of this event is 219 MeV/$c$, and the reconstructed invariant mass is 1793 MeV/$c^2$.}
\label{figure:nn_MC}
\end{figure}

\begin{figure}[h!]
\centering

\setlength\fboxsep{0pt}
\setlength\fboxrule{0pt}

\subfigure{\includegraphics[trim ={7mm 65mm 10mm 3mm},clip=true, width=80mm]{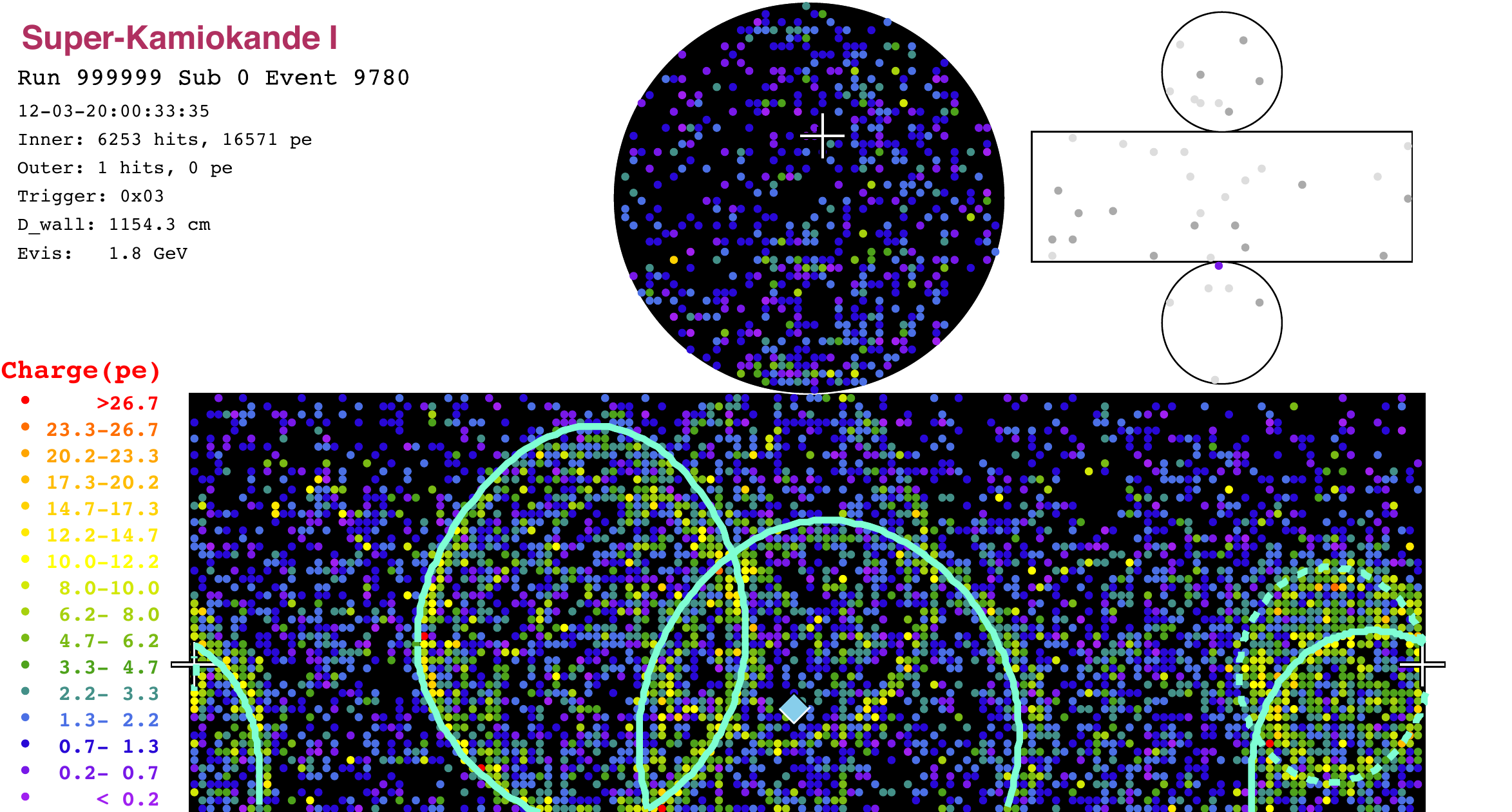}}
\linebreak
\subfigure{\hspace{1.5em}\includegraphics[trim ={100mm 3mm 88mm 132mm},clip=true, width=22mm]{sk1_nn_pi0pi0_bkg_file388_ev9780.pdf}}

\caption{Surviving atmospheric neutrino MC event for the $nn\rightarrow \pi^{0}\pi^{0}$ search. The interaction is $\nu_{\mu}$ neutral current deep-inelastic scattering. In particular, the interaction produced a $\pi^{0}$ and an $\eta$ meson, from which four $\gamma$'s are produced. All but the left-most ring correspond to true photons from the $\pi^{0}$ and $\eta$ meson. Many additional charged pions are also produced, concentrated near the center of the display, thereby complicating reconstruction. The left-most ring corresponds to a charged pion. There is a ring near the center from multiple, closely overlapping charged pion tracks, which is not found by the reconstruction program. The total reconstructed momentum is 553 MeV/$c$, and the reconstructed invariant mass is 1763 MeV/$c^2$.}
\label{figure:nn_atm_surv}
\end{figure}

\begin{figure*}[t!]
\centering        
\subfigure{\includegraphics[width=80mm]{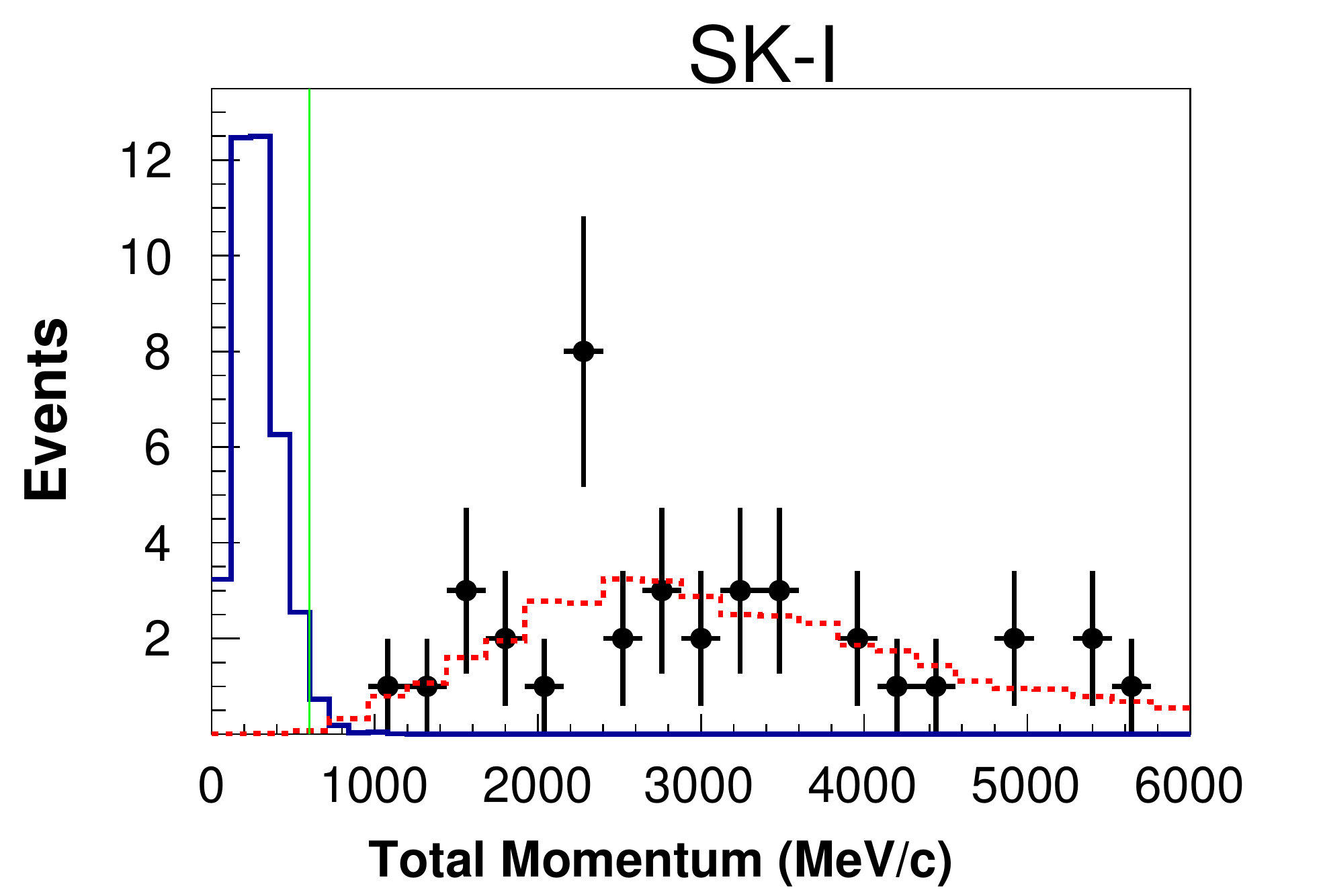}}
\subfigure{\includegraphics[width=80mm]{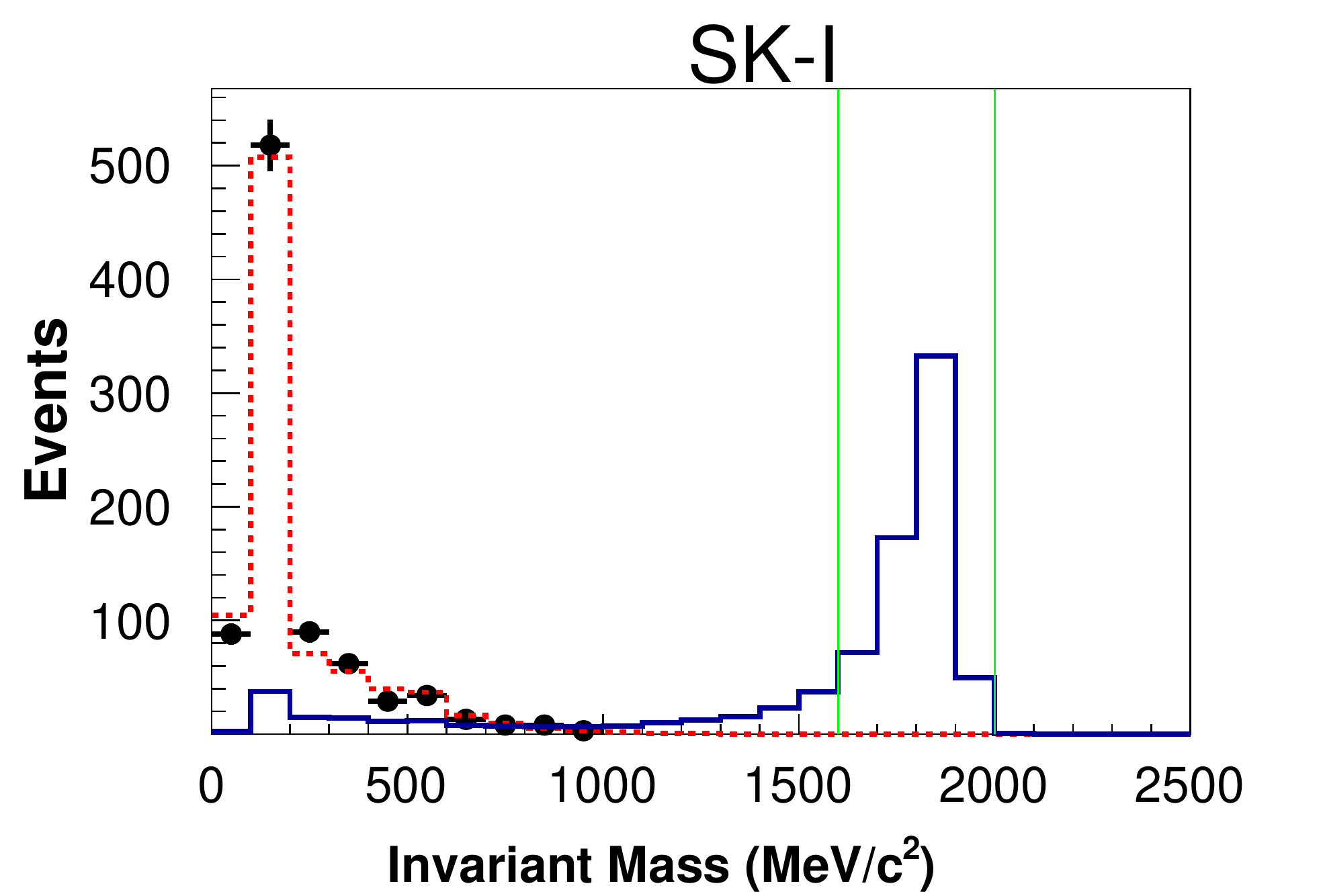}}
\subfigure{\includegraphics[width=80mm]{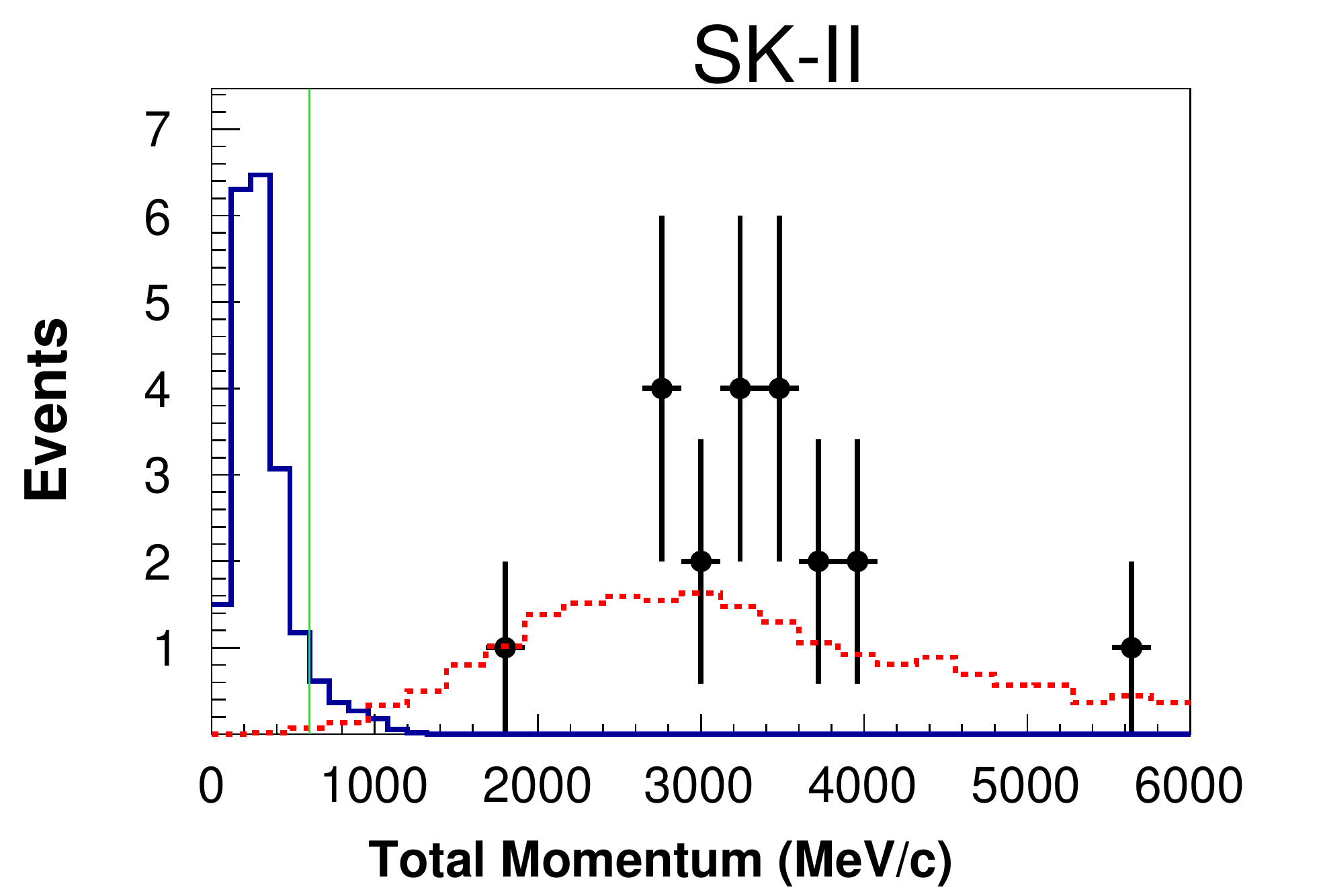}}
\subfigure{\includegraphics[width=80mm]{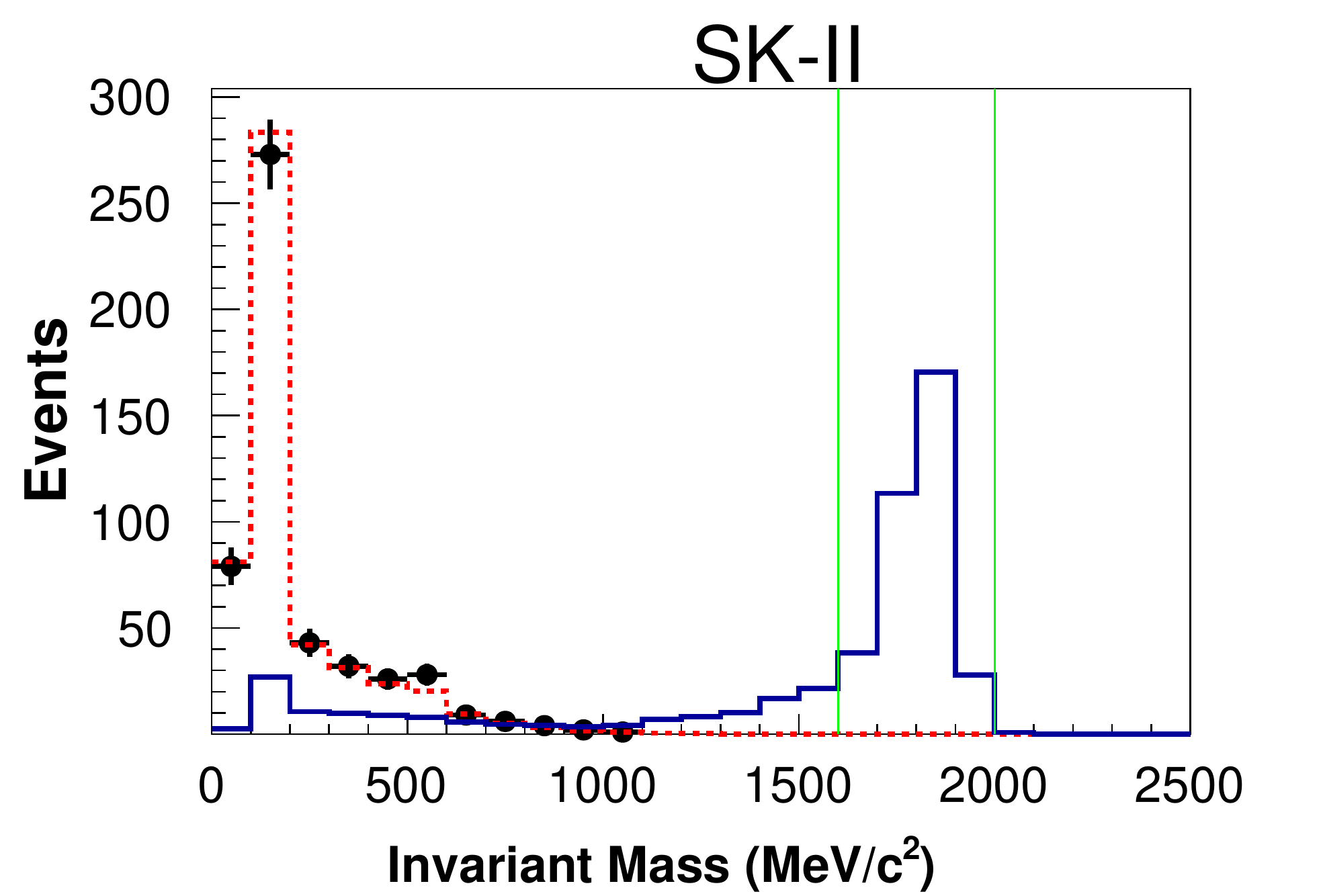}}
\subfigure{\includegraphics[width=80mm]{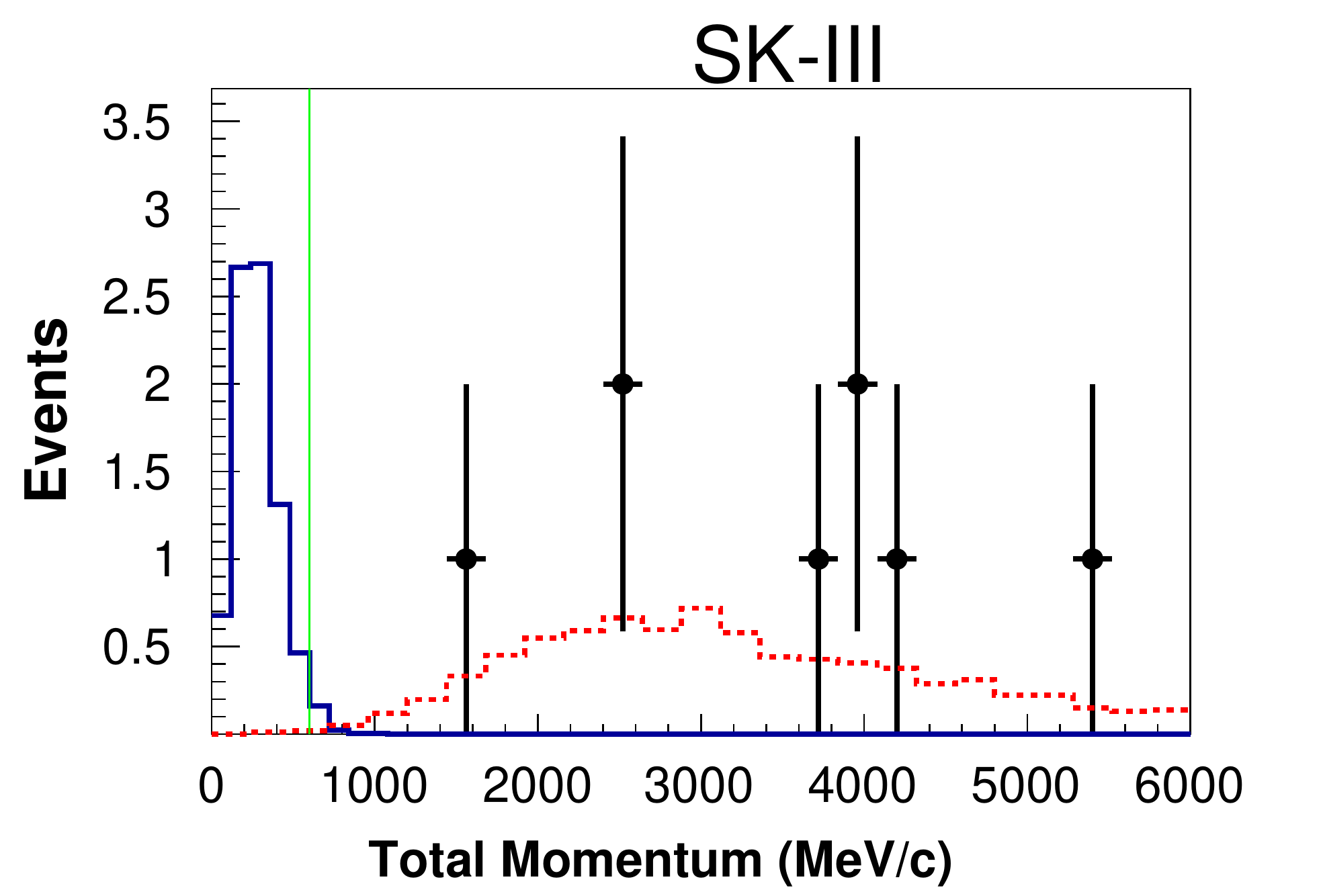}}
\subfigure{\includegraphics[width=80mm]{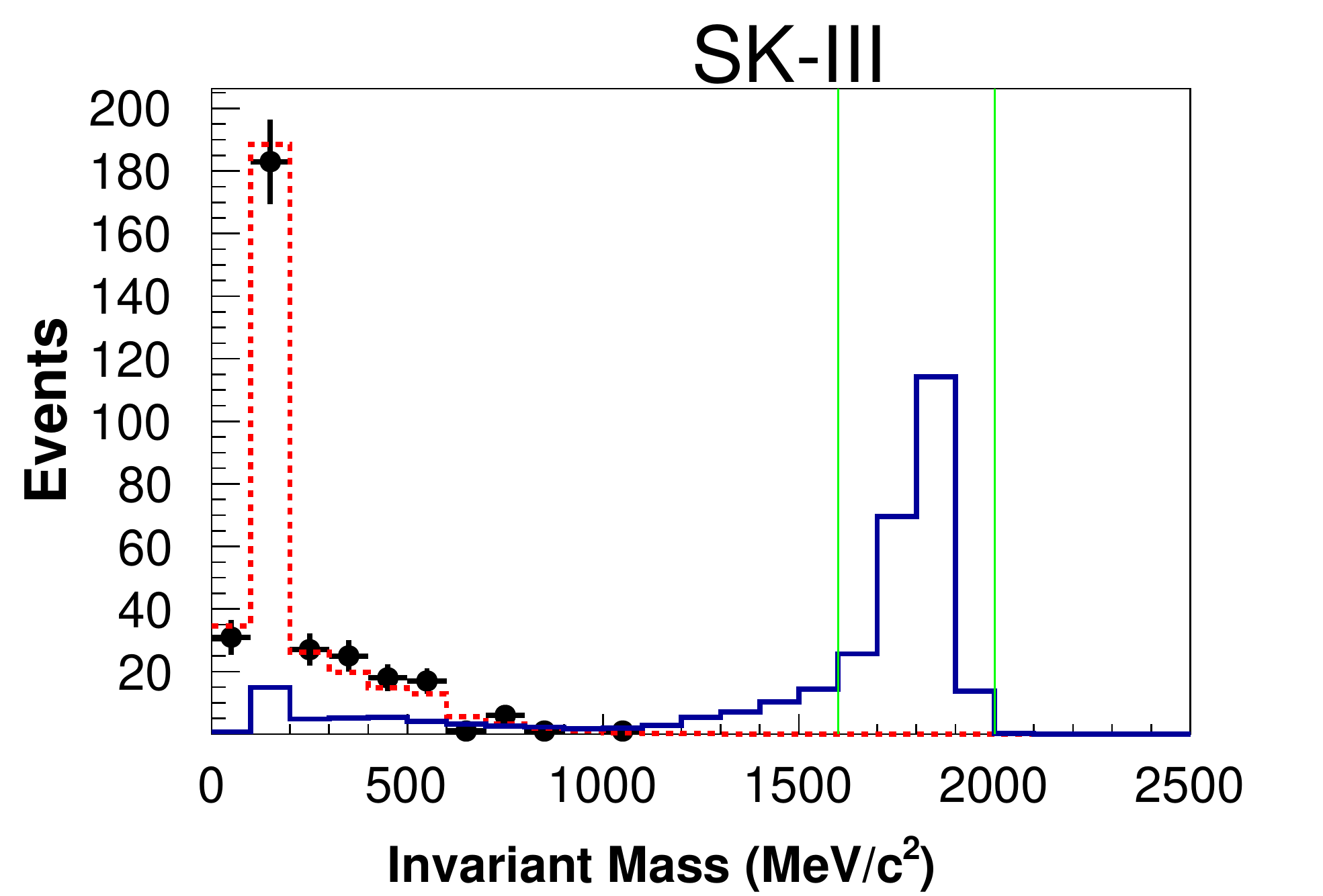}}
\subfigure{\includegraphics[width=80mm]{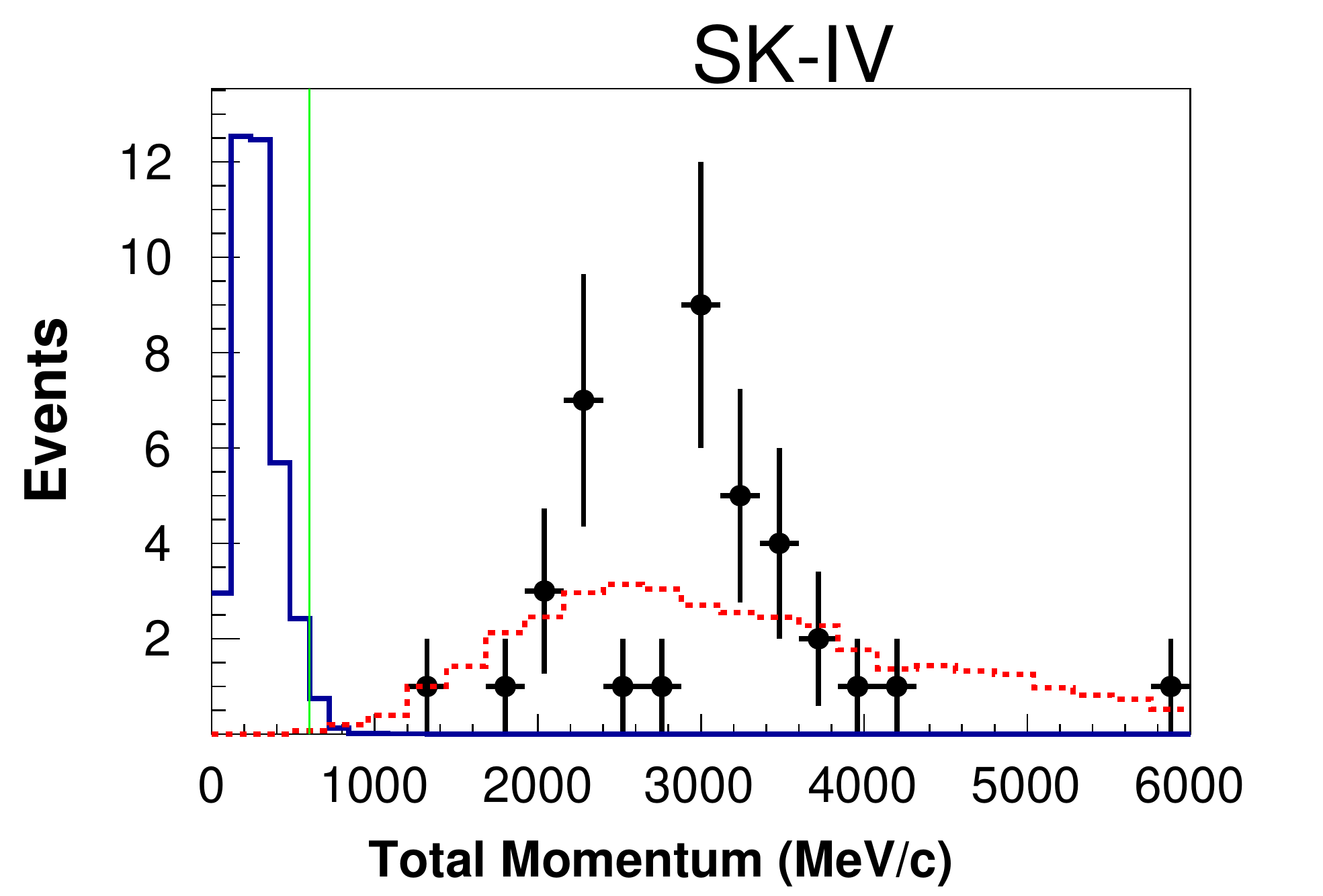}}
\subfigure{\includegraphics[width=80mm]{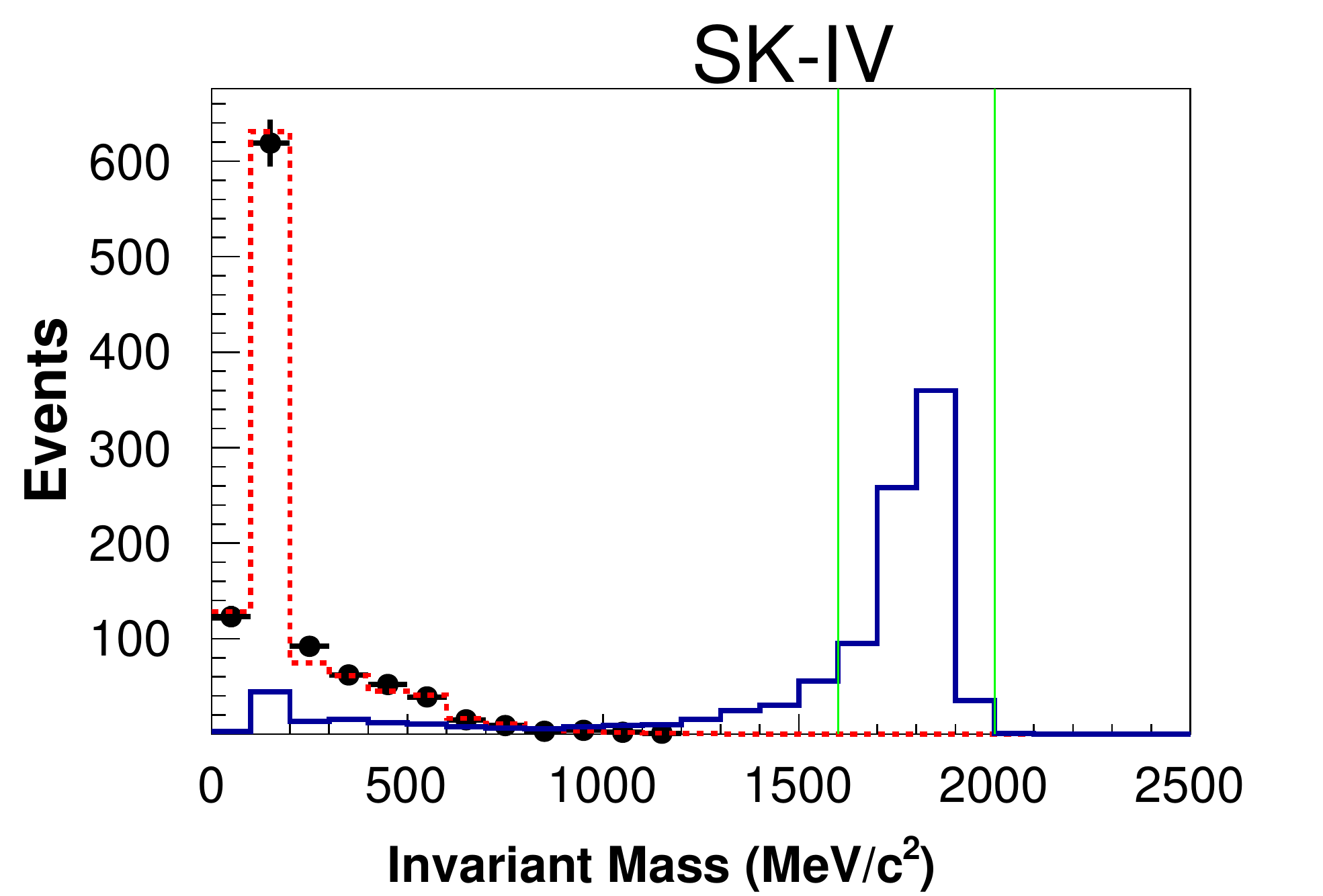}}
\caption{Total momentum (left) and invariant mass (right) for data (crosses), $nn\rightarrow \pi^{0}\pi^{0}$ MC (solid histogram), and atmospheric neutrino MC (dashed histogram). Criteria (C1)-(C3) have been applied to all distributions. The invariant mass criterion (C5) has been applied to the total momentum, and the total momentum criterion (C4) to the invariant mass. The thin green vertical lines indicate the locations of the final cuts for each variable.}
\label{figure:nn_1D_plots}
\end{figure*}

\begin{figure*}[t!]

   \subfigure{\includegraphics[width=160mm]{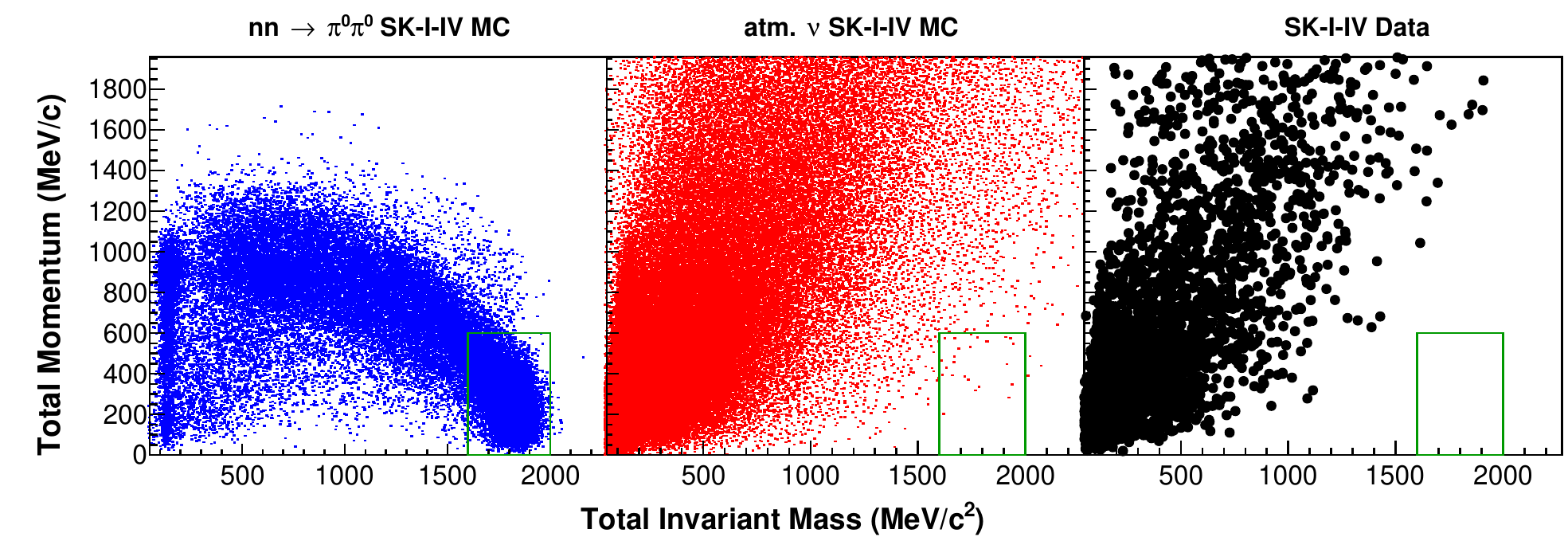}}

\caption{The total momentum and invariant mass distributions for $nn\rightarrow\pi^{0}\pi^{0}$ MC (left), atmospheric neutrino MC (center), and data (right) for SK I-IV. As seen in Fig.~\ref{figure:nn_1D_plots}, there is no major structural difference for SK-I-IV distributions, so events from all periods are combined. Events have passed criteria (C1)-(C3). The solid box indicates the total momentum and invariant mass cuts ((C4) and (C5) in the text).}
\label{figure:boxcut}
\end{figure*}

The total momentum $P_{tot}$ is defined as $P_{tot} = |\sum_{i=1}^{n_{ring}}{\vec{p}_{i}}|$, where $n_{ring}$ is the number of reconstructed rings, $\vec{p}_{i}$ is the 3-momentum of the $i^{th}$ ring, and the invariant mass is defined as $M_{inv} = \sqrt{E_{tot}^{2}-P_{tot}^2}$, where $E_{tot} = \sum_{i=1}^{n_{ring}}{|\vec{p}_{i}|}$ (the rings are assumed to come from photons, so there is no rest-mass energy). A $nn\rightarrow\pi^{0}\pi^{0}$ MC event is shown in Fig.~\ref{figure:nn_MC}, and a background MC event surviving all selection criteria is shown in Fig.~\ref{figure:nn_atm_surv}.

\par Criterion (C1) allows for both one- and two-ring $\pi^{0}$ fits. Criteria (C2) and (C3) are designed to eliminate muons and charged pions. Criterion (C4) is chosen to ensure a low total momentum, and to reject cases where only one $\pi^{0}$ is reconstructed, as each $\pi^{0}$ from $nn\rightarrow\pi^{0}\pi^{0}$ has a momentum that peaks closer to 1 GeV/$c$. Criterion (C5) is chosen to cover the $\sim 2M_{n}$ region as tightly as possible without getting too close to the peak (Fig.~\ref{figure:nn_1D_plots}). This eliminates nearly all background, which peaks sharply near $M_{\pi^{0}}$. \par The signal efficiency, background, and data events after each selection criteria are shown in Table~\ref{Table:effs}. The final efficiencies and background rates are summarized in Table~\ref{table:nn_stats}. The remaining background is mostly neutral current deep-inelastic scattering (Table~\ref{table:nn_bkg}). The interaction mode percentages agree across SK periods within statistical fluctuations, which are large due to the very low number of remaining events. The lower background in SK-IV is due to improved Michel electron tagging relative to SK-I, II, and III. 

\begin{table}[h!]
\setlength{\tabcolsep}{4pt}
\begin{center}
\begin{tabular}{c | r r r r} \hline\hline
  & SK-I & SK-II & SK-III & SK-IV\\ \hline
Efficiency (\%) & 22.1$\pm$0.3 & 18.8$\pm$0.3 & 20.9$\pm$ 0.3 & 21.4$\pm$ 0.3\\ 
bkg. (MT-yr.) & 0.6$\pm$0.2 & 0.9$\pm$0.3 & 1.0$\pm$0.3 & 0.2$\pm$0.1 \\  
bkg. (SK live.) & 0.05 & 0.04 & 0.03 & 0.02 \\  
candidates & 0 & 0 & 0 & 0\\ 

\hline \hline
\end{tabular}
\end{center}
\caption{Efficiency, expected background, and candidate data events for the SKI-IV $nn\rightarrow\pi^{0}\pi^{0}$ search, with statistical uncertainties. Background is quoted both for each SK livetime and in megaton-years.} 
\label{table:nn_stats}
\end{table}

\begin{table}[h!]
\setlength{\tabcolsep}{6pt}
\begin{center}
\begin{tabular}{r | r r r r} \hline\hline
mode & SK-I & SK-II & SK-III & SK-IV \\ \hline


NCDIS  & 63$\pm$32\% & 30$\pm$18\% & 67$\pm$25\% & 50$\pm$50\%\\
CCDIS  & 15$\pm$16\% & 50$\pm$22\% & 24$\pm$14\% & 0+50\%\\ 
CC1$\pi$  & 21$\pm$15\% & 20$\pm$14\% & 9$\pm$9\% & 51$\pm$51\%\\ 


\hline \hline
\end{tabular}
\caption{neutrino interaction mode percentages for the remaining background in the $nn\rightarrow \pi^{0}\pi^{0}$ search for SKI-IV.}

\label{table:nn_bkg}
\end{center}
\end{table}

\par Fig.~\ref{figure:boxcut} shows the final total momentum and invariant mass distributions for signal and background MC, and the data. No candidate events are found for this mode, consistent with the near-zero expected background for all SK periods. 

\subsection*{Systematic Uncertainties}

Systematic uncertainties for the dinucleon decay searches are comprised of MC simulation uncertanties specific to the signal and background MC, and common signal and background uncertainties related to event reconstruction and BDT bias (for modes which use a BDT).
\par The largest systematic uncertainties for the dinucleon decay signal come from the simulation stage. The dominant systematic uncertainty for all modes come from pion final state interaction (FSI) uncertainties. To estimate this, we generated separate sets of dinucleon decay MC for 24 parameter variations consisting of 1-sigma changes in pion-nucleon cross-sections, corresponding to uncertainties determined from fits to pion scattering data. This is the same set of variations used in the T2K $\nu_{e}$ appearance result~\cite{Abe:2013xua}. The variations affect the cross section for both low-energy (charge exchange, absorption, quasi-elastic scattering) and high-energy (hadron production) interactions. These variations were propagated through all stages of the analysis. For the $pp\rightarrow\pi^{+}\pi^{+}$ and $pn\rightarrow\pi^{+}\pi^{0}$ modes, each set of MC was processed with the trained, unvaried BDT used for the data and analysis MC. The systematic error is conservatively taken to be the largest percentage change in efficiency across all variations upon applying all of the analysis selection criteria. Other significant simulation systematic uncertainties come from Fermi motion ($\sim$20\%) and correlated decay ($\sim$10\%). These are not passed through the BDT, but estimated through a simple re-weighting method of the final event sample. For Fermi motion, the re-weighting comes from differences between the Fermi momentum simulation for dinucleon decay and the Fermi gas model. For correlated decay, the effect is not well understood, and so a conservative 100\% uncertainty is placed on the detection efficiency of events with a correlated decay, which is 10\% at the simulation stage and still about 10\% in the final event sample.
\par The largest overall contributions to background systematic uncertainties are related to neutrino interactions. In particular, for $pp\rightarrow\pi^{+}\pi^{+}$ and $pn\rightarrow\pi^{+}\pi^{0}$, the uncertainties associated with charged current single pion production are dominant. This is mainly due to a 40\% uncertainty in the $\pi^{0}/\pi^{+}$ production ratio. Other simulation systematic uncertainties due to pion FSI and neutrino flux are also considered.

\begin{table}[h!]
\setlength{\tabcolsep}{6pt}
\centering

\begin{tabular}{c | c c c c } \hline\hline

& \multicolumn{4}{c}{\bf{$pp\rightarrow\pi^{+}\pi^{+}$}}\\ \hline
Signal (\%) & SK-I & SK-II & SK-III & SK-IV \\ \hline
Simulation & 35.2 & 35.1 &33.6 & 38.5 \\
Reconstruction & 6.0 & 8.6 & 4.0 & 3.2 \\ 
BDT & 3.6 & 2.2 & 4.4 & 2.0  \\ 
\bf{Total} & 35.9 & 36.2 & 34.1 & 38.7 \\ \hline
Background (\%) & SK-I & SK-II & SK-III & SK-IV \\ \hline
Simulation & 29.1 & 29.1 & 35.8 & 26.5\\ 
Reconstruction & 6.1 & 8.1 & 4.1 & 3.2\\ 
BDT & 6.8 & 1.0 & 4.3 & 1.4\\
\bf{Total} & 30.5 & 30.3 & 36.4 & 26.8\\ 

\hline \hline
\end{tabular}
\caption{Systematic uncertainties for the $pp\rightarrow\pi^{+}\pi^{+}$ mode.}
\label{Table:pp_sys}
\end{table}

\begin{table}[h!]
\setlength{\tabcolsep}{6pt}
\begin{tabular}{c | c c c c } \hline\hline

& \multicolumn{4}{c}{\bf{$pn\rightarrow\pi^{+}\pi^{0}$}}\\ \hline
Signal (\%) & SK-I & SK-II & SK-III & SK-IV \\ \hline
Simulation & 33.3 & 32.2 & 28.4 & 35.0 \\ 
Reconstruction & 3.3 & 1.7 & 5.6 & 5.6 \\ 
BDT & $<$1 & 1.6 & $<$1 & $<$1 \\ 
\bf{Total} & 33.4 & 32.3 & 28.9 & 35.3 \\ \hline

Background (\%) & SK-I & SK-II & SK-III & SK-IV \\ \hline
Simulation & 22.1 & 19.9 & 24.0 & 27.8\\ 
Reconstruction & 1.8 & 1.8 & 3.3 & 3.8\\ 
BDT & 6.3 & 7.4 & 10.3 & 11.3\\ 
\bf{Total} & 23.1 & 21.3 & 26.3 & 28.6\\ 

\hline \hline
\end{tabular}
\caption{Systematic uncertainties for the $pn\rightarrow\pi^{+}\pi^{0}$ mode.}
\label{Table:pn_sys}
\end{table}

\begin{table}[h!]
\setlength{\tabcolsep}{6pt}
\begin{tabular}{c | c c c c } \hline\hline

& \multicolumn{4}{c}{\bf{$nn\rightarrow\pi^{0}\pi^{0}$}}\\ \hline
Signal (\%) & SK-I & SK-II & SK-III & SK-IV \\ \hline
Simulation & 31.1 & 34.4 & 37.3 & 33.1 \\
Reconstruction & 1.5 & 1.7 & 4.0 & 3.6 \\
\bf{Total} & 31.2 & 34.4 & 37.6 & 33.3 \\ \hline 

Background (\%) & SK-I & SK-II & SK-III & SK-IV \\ \hline
Simulation & 13.6 & 15.5 & 14.5 & 13.9\\
Reconstruction & 10.9 & 18.1 & 28.9 & 24.3\\ 
\bf{Total} & 17.5 & 24.0 & 32.3 & 28.0\\ 

\hline \hline
\end{tabular}

\caption{Systematic uncertainties for the $nn\rightarrow\pi^{0}\pi^{0}$ mode.}
\label{Table:nn_sys}
\end{table}

\par The most important reconstruction systematic uncertainty for both dinucleon decay and atmospheric neutrinos comes from the energy scale. The energy scale uncertainty is determined from comparing the absolute momenta of data and MC for several control samples, such as cosmic ray muons, their resulting decay electrons, and $\pi^{0}$'s from neutral current interactions. Time variation in the reconstructed momenta of the control samples is also taken into account in the uncertainty calculation. The energy scale uncertainty is estimated to be 1.1\%, 1.7\%, 2.7\%, and 2.3\% for SK-I, II, III, and IV, respectively \cite{Regis:2012sn}~\cite{Abe:2013xua}. This uncertainty is propagated into all energy-dependent variables, and is propagated through the trained, unvaried BDT used for the data and analysis MC for $pp\rightarrow\pi^{+}\pi^{+}$ and $pn\rightarrow\pi^{+}\pi^{0}$. The energy scale uncertainty is the dominant systematic uncertainty for the atmospheric neutrino background in the $nn\rightarrow \pi^{0}\pi^{0}$ search. Other reconstruction-related uncertainties taken into account are ring separation, particle identification, and the ring directional resolution. 
\par Finally, for the multivariate method, an additional systematic uncertainty for the training bias of the BDT is considered. This is estimated by assessing the difference in the testing and analysis distributions of the BDT output for both signal and background. Tables~\ref{Table:pp_sys}-\ref{Table:nn_sys} break down the systematic uncertainties for each mode and each SK period.

\section{Results and Discussion}

Since no significant excess was observed in the data, we set a lower limit on the lifetime of each dinucleon decay mode. This is done using a Bayesian method, which incorporate systematic uncertainties as follows~\cite{Barnett:1996hr}:

 $$P\left(\Gamma | n \right) = A\iiint \frac{e^{-\left(\Gamma\lambda\epsilon + b\right)}\left(\Gamma\lambda\epsilon + b\right)^{n}}{n!}$$ 
$$\times P\left(\Gamma\right)P\left(\lambda\right)P\left(\epsilon\right)P\left(b\right)d\lambda d\epsilon db, $$

\noindent where $\Gamma$ is the true dinucleon decay rate, $n$ is the number of candidate events, $\lambda$ is the exposure, $\epsilon$ is the efficiency, $b$ is the background rate, and $A$ is a normalization constant that ensures $\int_0^{\infty}P\left(\Gamma | n\right) d\Gamma = 1$. 

\begin{table}[h]
\begin{center}
\begin{tabular}{c | c c} \hline\hline
 \rule{0pt}{4ex}mode & Frejus limit ($^{56}$Fe) & \bf{This analysis ($^{16}$O)}\\ \hline
 \rule{0pt}{4ex}$pp\rightarrow\pi^{+}\pi^{+}$ & $7.0 \times 10^{29}$ yrs & $7.22 \times 10^{31}$ yrs \\ 
\rule{0pt}{4ex}$pn\rightarrow\pi^{+}\pi^{0}$ & $2.0 \times 10^{30}$ yrs & $1.70 \times 10^{32}$ yrs \\ 
\rule{0pt}{4ex}$nn\rightarrow\pi^{0}\pi^{0}$ & $3.4 \times 10^{30}$ yrs & $4.04 \times 10^{32}$ yrs \\ 

\hline \hline
\end{tabular}
\end{center}
\caption{Limits on each mode, compared with previous results in \cite{Berger:1991fa}.}
\label{table:limits}
\end{table}

The distributions $P\left(\lambda\right)$, $P\left(\epsilon\right)$, and $P\left(b\right)$ are taken to be Gaussian, while $P\left(\Gamma\right)$ is uniform for $\Gamma > 0$. The uncertainty on $\lambda$ is assumed to be negligible. The exposure is $^{16}$O-years, not nucleon-years as in most proton decay analyses. This is consistent with the exposure used in \cite{Litos:2014} and \cite{Berger:1991fa}, which used total $^{16}$O and $^{56}$Fe nuclei, respectively.
\par Finally, the confidence level is calculated as: $$C.L.=\int_{0}^{\Gamma_{limit}}P\left(\Gamma | n \right) d\Gamma $$ and the lower lifetime limit is just the reciprocal of $\Gamma_{limit}$. Table~\ref{table:limits} shows the lifetime limits at 90\% confidence level, compared with the previous limits set by the Frejus collaboration in ref. \cite{Berger:1991fa}.

\par The interpretation of our results depends on the underlying physics model. For instance, our limit for $nn \rightarrow \pi^{0}\pi^{0}$ is the most stringent experimental limit for dinucleon decay to light mesons, including \cite{Litos:2014}, which measured $\tau_{pp \rightarrow K^{+}K^{+}} > 1.77 \times 10^{32}$ years. However, \cite{Litos:2014} was motivated by supersymmetric models that forbid couplings between quarks of the same generation. If, as with the models considered in \cite{Csaki:2013jza}, such couplings are not forbidden, and no preference is given for couplings to first- or second-generation quarks, our result for $nn \rightarrow \pi^{0}\pi^{0}$ provides the strongest general constraint on dinucleon decay to mesons.
\par As stated in Section I, there is a connection between neutron-antineutron oscillation (n-nbar) and dinucleon decay to pions, if the same operator is dominant for both processes. This is discussed further in Appendix A. Our results provide a useful cross-check to n-nbar searches. By isolating a single final state for each mode, as opposed to the many non-negligible final states of n-nbar~\cite{Abe:2011ky}, these analyses are simple and robust.

\section{Conclusion}
We have performed searches for dinucleon decay to pions via the processes $pp\rightarrow \pi^{+}\pi^{+}$, $pn\rightarrow\pi^{+}\pi^{0}$, and $nn\rightarrow\pi^{0}\pi^{0}$ with 282.1 kiloton-years of data with the Super-Kamiokande detector. This is the first search for dinucleon decay to pions done in a water Cherenkov detector. We do not observe any significant event excess for any of the modes, and the data remain consistent with the atmospheric neutrino background. Therefore, partial lifetime limits of $\tau_{pp\rightarrow\pi^{+}\pi^{+}} > 7.22 \times 10^{31}$ years, $\tau_{pn\rightarrow\pi^{+}\pi^{0}} > 1.70 \times 10^{32}$ years, and $\tau_{nn\rightarrow\pi^{0}\pi^{0}} > 4.04 \times 10^{32}$ years per $^{16}$O nucleus have been set. The limits per nucleus are about two orders of magnitude more stringent than previous limits set in \cite{Berger:1991fa}, and constrain models that predict $\Delta B = 2$ processes.

\section*{Acknowledgements}
We gratefully acknowledge the cooperation of the Kamioka Mining and Smelting Company. The Super-Kamiokande experiment has been built and operated from funding by the Japanese Ministry of Education, Culture, Sports, Science and Technology, the United States Department of Energy, and the U.S. National Science Foundation. This work was partially supported by the Research Foundation of Korea (BK21 and KNRC), the Korean Ministry of Science and Technology, the National Science Foundation of China, the European Union FP7 (DS laguna-lbno PN-284518 and ITN invisibles GA-2011-289442), the National Science and Engineering Research Council (NSERC) of Canada, and the Scinet and West-grid consortia of Compute Canada.

\appendix*
\section{A: Dinucleon decay lifetimes}

Here we discuss the lifetime formula for dinucleon decay to pions. We assume that the same operator which mediates neutron-antineutron oscillation (n-nbar) also mediates dinucleon decay to pions, which follows from crossing symmetry.

\par Reference \cite{Csaki:2013jza} discusses both n-nbar and dinucleon decay with the same operator. This operator takes the general form: 

$$ O_{\Delta B = 2} = \frac{1}{\Lambda_{sup}^5}O_{6Q} $$ where $O_{6Q}$ is a six quark operator containing two up-type and four down-type quarks, and $\Lambda_{sup}$ is a suppression scale, with the fifth power following from dimensional analysis. Indices related to quark generation, color, etc., are suppressed. For all first-generation quarks, this operator leads to both n-nbar and dinucleon decay. The (free) n-nbar lifetime is given by:

\begin{equation} \tau_{n\bar{n}} \approx \frac{\Lambda_{sup}^{5}}{\Lambda_{QCD}^{6}}, \label{eq:taunnbar}\end{equation} where $\Lambda_{QCD}$ is a hadronic matrix element near the QCD scale, estimated to be 200 MeV. The dinucleon decay lifetime $\tau_{NN}$ is given by \cite{Goity:1994dq, Csaki:2013jza}
\begin{equation} \tau_{NN} \approx \frac{\pi M_{N}^{2}}{8\rho_{N}}\frac{\Lambda_{sup}^{10}}{\Lambda_{QCD}^{10}}, \label{eq:tauNN} \end{equation} where $M_{N}$ is the nucleon mass and $\rho_{N} \approx 0.25$ $fm^{-3}$ is the average nuclear density. From equations \ref{eq:taunnbar} and \ref{eq:tauNN}, 

\begin{equation} \tau_{NN} \approx \frac{\pi M_{N}^{2}}{8\rho_{N}}\Lambda_{QCD}^{2}\tau_{n\bar{n}}^{2} = T_{nuc}\tau_{n\bar{n}}^{2}. \label{eq:connection} \end{equation} Equation~\ref{eq:connection} is mathematically similar to the formalism for comparing free and bound n-nbar lifetimes, although in the latter case the factor $T_{nuc}$ comes from different physics related to the difference in neutron and antineutron nuclear potentials. For dinucleon decay, $T_{nuc} \approx 1.1 \times 10^{25}$ $s^{-1}$, while for n-nbar $T_{nuc}$ is two orders of magnitude less, with \cite{Friedman:2008es} calculating it to be $0.517 \times 10^{23}$ $s^{-1}$. The current lower limit on the n-nbar lifetime, from Super-Kamiokande, is $\tau_{n\bar{n}} > 2.7 \times 10^{8}$  s~\cite{Abe:2011ky}. This translates to a lower lifetime for dinucleon decay to pions of $\tau_{NN \rightarrow \pi\pi} \approx 2.5 \times 10^{34}$ years, about two orders of magnitude beyond the limits obtained in this paper. Thus, our results do not place any new constraints on n-nbar.

\end{document}

%% file: authors-20140520.tex
\newcommand{\AFFicrr}{\affiliation{Kamioka Observatory, Institute for Cosmic Ray Research, University of Tokyo, Kamioka, Gifu 506-1205, Japan}}
\newcommand{\AFFkashiwa}{\affiliation{Research Center for Cosmic Neutrinos, Institute for Cosmic Ray Research, University of Tokyo, Kashiwa, Chiba 277-8582, Japan}}
\newcommand{\AFFipmu}{\affiliation{Kavli Institute for the Physics and
Mathematics of the Universe (WPI), Todai Institutes for Advanced Study,
University of Tokyo, Kashiwa, Chiba 277-8582, Japan }}
\newcommand{\AFFmad}{\affiliation{Department of Theoretical Physics, University Autonoma Madrid, 28049 Madrid, Spain}}
\newcommand{\AFFubc}{\affiliation{Department of Physics and Astronomy, University of British Columbia, Vancouver, BC, V6T1Z4, Canada}}
\newcommand{\AFFbu}{\affiliation{Department of Physics, Boston University, Boston, MA 02215, USA}}
\newcommand{\AFFbnl}{\affiliation{Physics Department, Brookhaven National Laboratory, Upton, NY 11973, USA}}
\newcommand{\AFFuci}{\affiliation{Department of Physics and Astronomy, University of California, Irvine, Irvine, CA 92697-4575, USA }}
\newcommand{\AFFcsu}{\affiliation{Department of Physics, California State University, Dominguez Hills, Carson, CA 90747, USA}}
\newcommand{\AFFcnm}{\affiliation{Department of Physics, Chonnam National University, Kwangju 500-757, Korea}}
\newcommand{\AFFduke}{\affiliation{Department of Physics, Duke University, Durham NC 27708, USA}}
\newcommand{\AFFfukuoka}{\affiliation{Junior College, Fukuoka Institute of Technology, Fukuoka, Fukuoka 811-0295, Japan}}
\newcommand{\AFFgifu}{\affiliation{Department of Physics, Gifu University, Gifu, Gifu 501-1193, Japan}}
\newcommand{\AFFgist}{\affiliation{GIST College, Gwangju Institute of Science and Technology, Gwangju 500-712, Korea}}
\newcommand{\AFFuh}{\affiliation{Department of Physics and Astronomy, University of Hawaii, Honolulu, HI 96822, USA}}
\newcommand{\AFFkek}{\affiliation{High Energy Accelerator Research Organization (KEK), Tsukuba, Ibaraki 305-0801, Japan }}
\newcommand{\AFFkobe}{\affiliation{Department of Physics, Kobe University, Kobe, Hyogo 657-8501, Japan}}
\newcommand{\AFFkyoto}{\affiliation{Department of Physics, Kyoto University, Kyoto, Kyoto 606-8502, Japan}}
\newcommand{\AFFmiyagi}{\affiliation{Department of Physics, Miyagi University of Education, Sendai, Miyagi 980-0845, Japan}}
\newcommand{\AFFnagoya}{\affiliation{Solar Terrestrial Environment Laboratory, Nagoya University, Nagoya, Aichi 464-8602, Japan}}
\newcommand{\AFFpol}{\affiliation{National Centre For Nuclear Research, 00-681 Warsaw, Poland}}
\newcommand{\AFFsuny}{\affiliation{Department of Physics and Astronomy, State University of New York at Stony Brook, NY 11794-3800, USA}}
\newcommand{\AFFokayama}{\affiliation{Department of Physics, Okayama University, Okayama, Okayama 700-8530, Japan }}
\newcommand{\AFFosaka}{\affiliation{Department of Physics, Osaka University, Toyonaka, Osaka 560-0043, Japan}}
\newcommand{\AFFregina}{\affiliation{Department of Physics, University of Regina, 3737 Wascana Parkway, Regina, SK, S4SOA2, Canada}}
\newcommand{\AFFseoul}{\affiliation{Department of Physics, Seoul National University, Seoul 151-742, Korea}}
\newcommand{\AFFshizuokasc}{\affiliation{Department of Informatics in
Social Welfare, Shizuoka University of Welfare, Yaizu, Shizuoka, 425-8611, Japan}}
\newcommand{\AFFskk}{\affiliation{Department of Physics, Sungkyunkwan University, Suwon 440-746, Korea}}
\newcommand{\AFFtokyo}{\affiliation{The University of Tokyo, Bunkyo, Tokyo 113-0033, Japan }}
\newcommand{\AFFtoronto}{\affiliation{Department of Physics, University of Toronto, 60 St., Toronto, Ontario, M5S1A7, Canada }}
\newcommand{\AFFtriumf}{\affiliation{TRIUMF, 4004 Wesbrook Mall, Vancouver, BC, V6T2A3, Canada }}
\newcommand{\AFFtokai}{\affiliation{Department of Physics, Tokai University, Hiratsuka, Kanagawa 259-1292, Japan}}
\newcommand{\AFFtsinghua}{\affiliation{Department of Engineering Physics, Tsinghua University, Beijing, 100084, China}}
\newcommand{\AFFuw}{\affiliation{Department of Physics, University of Washington, Seattle, WA 98195-1560, USA}}

\AFFicrr
\AFFkashiwa
\AFFmad
\AFFbu
\AFFubc
\AFFbnl
\AFFuci
\AFFcsu
\AFFcnm
\AFFduke
\AFFfukuoka
\AFFgifu
\AFFgist
\AFFuh
\AFFkek
\AFFkobe
\AFFkyoto
\AFFmiyagi
\AFFnagoya
\AFFsuny
\AFFokayama
\AFFosaka
\AFFregina
\AFFseoul
\AFFshizuokasc
\AFFskk
\AFFtokai
\AFFtokyo
\AFFipmu
\AFFtoronto
\AFFtriumf
\AFFtsinghua
\AFFuw

\author{J.~Gustafson}
\AFFbu

\author{K.~Abe}
\AFFicrr
\AFFipmu
\author{Y.~Haga}
\AFFicrr
\author{Y.~Hayato}
\AFFicrr
\AFFipmu
\author{M.~Ikeda}
\AFFicrr
\author{K.~Iyogi}
\AFFicrr 
\author{J.~Kameda}
\author{Y.~Kishimoto}
\author{M.~Miura} 
\author{S.~Moriyama} 
\author{M.~Nakahata}
\AFFicrr
\AFFipmu 
% ...

\collaboration{The Super-Kamiokande Collaboration}
\noaffiliation